\documentclass[12pt,a4paper]{article}
\usepackage[T2A]{fontenc}
\usepackage[utf8]{inputenc}
\usepackage{amsmath,amssymb}
\usepackage{graphicx}
\usepackage{indentfirst}
\usepackage{setspace}
\usepackage{cite}
\usepackage[left=4cm,right=2cm,top=2cm,bottom=2cm,bindingoffset=0cm]{geometry}
\usepackage{mathtext}
\usepackage{mathrsfs} 
\usepackage{hyperref}

\usepackage{color}

\DeclareMathOperator{\sign}{sign}
\DeclareMathOperator{\Pf}{Pf}
\DeclareMathOperator{\diag}{diag}
\DeclareMathOperator{\Tr}{Tr}

\begin{document}

\doublespacing

\title{\Large \textbf{Topological superconductivity and \\
		Majorana states in low-dimensional systems}}

\author{V.\,V. Val'kov$^{a}$, M.\,S. Shustin$^{a,\square}$, S.\,V.~Aksenov$^{a,\circ}$, A.\,O.~Zlotnikov$^{a,\ast}$, \\ A.\,D.~Fedoseev$^{a}$, V.\,A.~Mitskan$^{a}$, M.\,Yu.~Kagan$^{b,c}$}

\date{}

\maketitle
\begin{center}
	\textit{\small $^{a)}$Kirensky Institute of Physics,
		 660036 Krasnoyarsk, Russia\\
		$^{b)}$National Research University Higher School of Economics, 101000 Moscow, Russia\\
		$^{c)}$Kapitza Institute for Physical Problems RAS, 119334 Moscow, Russia}\\
	E-mail: $^{\square}$ \verb"mshustin@yandex.ru",
	$^{\circ}$ \verb"asv86@iph.krasn.ru",
	$^{\ast}$ \verb"zlotn@iph.krasn.ru"
\end{center}

We discuss the properties of topologically nontrivial superconducting phases and the conditions for their realization in condensed matter, the criteria for the appearance of elementary Majorana-type excitations in solids, and the corresponding principles and experimental methods for identifying Majorana bound states (MBSs). Along with the well-known Kitaev chain and superconducting nanowire (SW) models with spin-orbit coupling in an external magnetic field, we discuss models of quasi-two-dimensional materials in which MBSs are realized in the presence of noncollinear spin ordering. For finite-length SWs, we demonstrate a cascade of quantum transitions occurring with a change in the magnetic field, accompanied by a change in the fermion parity of the ground state. The corresponding anomalous behavior of the magnetocaloric effect can be used as a tool for identifying MBSs. We devote considerable attention to the analysis of the transport characteristics of devices that contain topologically nontrivial materials. The results of studying the conductance of an Aharonov-Bohm ring whose arms are connected by an SW are discussed in detail. An important feature of this device is the appearance of Fano resonances in the dependence of conductance on the magnetic field when the SW is in a topologically nontrivial phase. We establish a relation between the characteristics of such resonances and the spatial structure of the lowest-energy SW state. The conditions for the occurrence of an MBS in the phase of the coexistence of chiral $d+id$ superconductivity and 120-degree spin ordering are determined in the framework of the $t-J-V$ model on a triangular lattice. We take electron-electron interactions into account in discussing the topological invariants of low-dimensional superconducting materials with noncollinear spin ordering. The formation of Majorana modes in regions with an odd value of a topological $\mathbb{Z}$ invariant is demonstrated. The spatial structure of these excitations in the Hubbard fermion ensemble is determined.


\tableofcontents

\section{\label{sec1}Introduction}

Topological superconductors are defined as materials with a nontrivial topological index, or topological invariant (TI), of the superconducting phase. This invariant, well known from the theory of the quantum Hall effect \cite{haldane-81, haldane-81b, niu-85, volovik-09}, describes nonlocal characteristics of the many-body wave function of the electron ensemble.

Under consideration are both the materials in which topological superconductivity is uniform in the bulk due to internal interactions and external fields, and systems in which this phase can be induced in a bounded region of a solid-state structure by external fields and the mutual effect of materials brought into contact (for example, by the proximity effect).

In this review, we mainly focus on systems whose spectrum of elementary excitations has a gap in the entire Brillouin zone in the case of periodic boundary conditions (such a spectrum is often called the bulk spectrum). We emphasize that there is another class of materials with nodal points, lines, or surfaces in the Brillouin zone with gapless excitations, for which a TI has also been found (for more details, see \cite{volovik-09}). Nontrivial TI values indicate the formation of surface or edge states in a system with open boundary conditions. The excitation energies corresponding to the realization of such states then lie in a range of values below the bulk gap, down to zero.
The topological classification is useful due to two factors. The first is the possibility of predicting edge states in a material based on the results of relatively simple calculations. The second, and more important one, is that the TI value does not change under perturbations and variations in the parameters unless the gap in the excitation spectrum vanishes and the symmetry of the Hamiltonian changes. As a result, the surface states in the materials under consideration are topologically stable, and topological transitions with a change in the TI are realized only if the bulk spectrum becomes gapless at the transition point.
It follows from the foregoing that an analogy exists between topological superconductors and topological insulators \cite{hasan-10, lozovik-12, tarasenko-18, pankratov-18}. In both classes of materials, the gap is realized in the bulk excitation spectrum, and conducting surface states exist at the boundaries in the absence of an electric current in the bulk. Moreover, disregarding electron-electron interactions, all topological materials can be divided into 10 classes, in accordance with different combinations of three symmetries: time reversal symmetry, electron-hole symmetry, and chiral symmetry \cite{schnyder-08}.

At the same time, there are differences among these classes of topological materials, for example, due to the mechanism of the appearance of a gap in the excitation spectrum, the existence of screening superconducting currents in any superconductors, and the structure of edge excitations. In topological superconductors, they are Majorana bound states (MBSs) formed by a pair of Majorana modes (MMs). These states are singled out by their specific properties, such as spatial nonlocality, quantum entanglement, and non-abelian exchange statistics. These features motivate interest in topological superconductivity from the standpoint of the prospects of creating qubits robust against to local perturbations \cite{nayak-08}.

A necessary condition for the implementation of topological insulators is the presence of a band structure with band inversion and spin-orbit coupling. For some topological
superconductors with MMs, an important role is played by the superconducting pairing of electrons with the same spin projections, i.e., triplet pairing \cite{read-00, kitaev-01}. Because there are a limited number of candidates for triplet-pairing materials, the search for conditions for the realization of MMs in materials with a singlet Cooper pairing of fermions has become important. It turns out that, under certain conditions (see Sections \ref{sec3} and \ref{sec4}), introducing an interaction that induces a mixing of states with different spin projections is capable of initiating the occurrence of MMs. In such topological superconductors, a nontrivial topology can be caused by both internal interactions and interactions that appear as a result of growing heterostructures.

We note that the concept of topological superconducting systems was largely formed in previous studies in the field of superfluid quantum liquids, such as studies of the A and B superfluid phases of $^3$He. Historically, the possibility of a triplet $p$-pairing was first predicted there \cite{volovik-09, kagan-19}. Majorana modes were also discovered at the boundary of superfluid $^3$He-B \cite{volovik-09b}.

Topological superconductors can be divided into the following groups.

\begin{enumerate}
	\item Chiral superconductors, in which the onset of the superconducting phase is accompanied by the setting in of a $d_{x^2-y^2} + id_{xy}$-type order parameter symmetry \cite{laughlin-94, volovik-97}. Such a phase can form in $d_{x^2-y^2}$ superconductors with broken time reversal symmetry \cite{balatsky-98}, in materials with triangular layers \cite{baskaran-03}, and in graphene \cite{nandkishore-12, nandkishore-14, kagan-15}.
	
	\item Superconductors with the $p$ or chiral $p_x+ip_y$ symmetry type, which include Sr$_2$RuO$_4$ \cite{mackenzie-03, sarma-06} (a discussion of the type of symmetry in this compound is ongoing \cite{pustogow-19, suzuki-20}) and uranium superconductors UGe$_2$, UCoGe, URhGe \cite{sau-12, mineev-17}.
	
	\item Heterostructures based on topological insulators and conventional (nontopological) superconductors \cite{fu-08, snelder-15}.
	
	\item Noncentrosymmetric superconductors in which triplet pairings can be induced due to spatial inversion symmetry breaking \cite{sato-09}.
	
	\item Doped topological insulators in which superconductivity occurs, such as Cu$_x$Bi$_2$Se$_3$ \cite{hor-10, sasaki-15}. This group also includes iron-based superconductors, such as FeTe$_{x}$Se$_{1-x}$ \cite{zhang-18}, in which topological surface states form in the normal nonsuperconducting phase.
	
	\item Hybrid structures containing semiconducting materials with strong spin-orbit coupling and conventional superconductors \cite{sau-10, lutchyn-10, oreg-10, stanescu-11}, such as InAs or InSb nanowires in contact with niobium-titanium nitride or aluminum \cite{mourik-12, deng-12}.
	
	\item Superconducting systems with inhomogeneous (noncollinear or helicoidal) magnetic ordering, including chains of magnetic nanoparticles or atoms on a superconducting substrate \cite{choy-11} and magnetic superconductors \cite{martin-12}.
\end{enumerate}

In each of these groups except the first one, the formation of MMs is possible. There are several studies describing the main ideas and results on topological superconductivity and MMs in condensed media in detail \cite{alicea-12, beenakker-13, elliot-15, sato-17}.

This review is devoted to a detailed description of the conditions for the formation of MMs and to an analysis of the properties of these edge states in quasi-one-dimensional and two-dimensional (2D) systems that can be explored in modern experiments. We discuss methods for identifying MMs based on the use of transport characteristics of devices involving materials in the topological superconducting phase. Much attention is devoted to the most interesting cases where MMs are caused by spin-orbit coupling (item 6 above) and inhomogeneous magnetic ordering (item 7). In particular, we show that an important manifestation of a topologically nontrivial phase in various superconducting materials is oscillations of the fermion parity (FP) of the ground state, which can be observed experimentally. Current theoretical and experimental problems in dealing with MMs in topological superconductors are also discussed.

This review is organized as follows. In Section \ref{sec2}, we consider the minimal model for describing MMs in topological superconductors, the Kitaev chain model. In Section \ref{sec3}, we describe the properties of MMs and observable effects in semiconducting nanowires with induced superconductivity and also consider quantum transport through various devices with nanowires in the topological superconducting phase. Section \ref{sec4} is devoted to superconducting systems with inhomogeneous magnetic ordering, which have a deep connection to materials with spin-orbit coupling in a uniform field.

\section{\label{sec2}Majorana modes in Kitaev chains}
\subsection{\label{sec2.1} General properties of Majorana modes}

One of the best studied models predicting the existence of MMs in solid systems is the Kitaev chain model. Its popularity is due to its simplicity, which allows describing many effects analytically or in the most transparent form. In this model, an ensemble of spinless fermions on a chain is considered, capable of one-dimensional (1D) motion by hopping between nearest-neighbor sites. In addition, the chain is assumed to be brought into contact with a massive superconductor. Due to the proximity effect, intersite $p$-symmetry Cooper pairing is induced in it. This type of symmetry corresponds to irreducible representation in which the superconducting order parameter transforms under rotations perpendicular to the chain. Because the chain is in material contact with a thermostat, the grand canonical ensemble is used to describe the statistical properties of the system. The chemical potential is therefore included as an additional independent parameter.

\begin{figure}[h!]
	\begin{center}
		\includegraphics[width=0.7\textwidth]{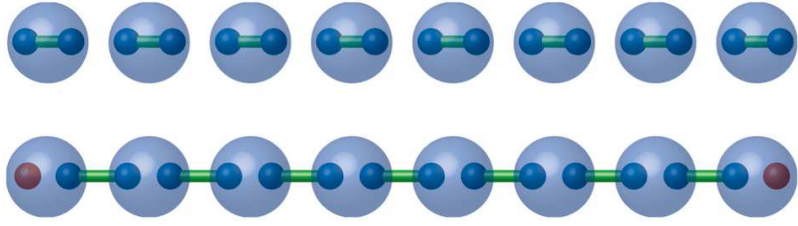}
		\caption{Schematic representation of terms in Hamiltonian (4) at (top) $t=|\Delta|=0$, $\mu\neq0$ (trivial case) and (bottom) $\mu=0$, $t=|\Delta|$ (nontrivial case). Large dots corresponds to sites $l$, and small ones, to operators $\gamma_{Al}$ and $\gamma_{Bl}$. Pairs of operators entering (4) with nonvanishing coefficients are connected by horizontal lines.}\label{fig1}
	\end{center}
\end{figure}

The Hamiltonian of the Kitaev chain model has the form
\begin{eqnarray}
\label{HamiltonianKitaev1}
H_{K} & = & \sum_{l=1}^{N} (\epsilon - \mu)
a_{l}^{\dag}a_{l} -
t\sum_{l=1}^{N-1} \left( a_{l}^{\dag}a_{l+1} + a_{l+1}^{\dag}a_{l} \right)
+\sum_{l=1}^{N-1} \left( \Delta a_{l}a_{l+1} + \Delta^* a_{l+1}^{\dag}a_{l}^{\dag} \right).
\end{eqnarray}
The first two terms of the Hamiltonian describe the on-site energy of fermions $\epsilon$, measured from the chemical potential $\mu$, and the hops of fermions between nearest-neighbor sites. The hopping rate is determined by the parameter $t$. The superconducting order parameter is written as $\Delta=|\Delta|e^{i\theta}$. In Hamiltonian (\ref{HamiltonianKitaev1}), following [12], we move to the Majorana operators
\begin{eqnarray}
\label{Majorana_gamma1}
\gamma_{lA} =  e^{i\theta/2}a_l + e^{-i\theta/2}a_l^{\dag},~~
\gamma_{lB} =  i\left(e^{-i\theta/2}a_l^{\dag}-e^{i\theta/2}a_l\right),
\end{eqnarray}
which satisfy the anticommutation relations
\begin{eqnarray}
\label{Gamma_comm}
\{\gamma_{lA},\gamma_{l'B}\}=0;~\{\gamma_{lA},\gamma_{l'A}\}=\{\gamma_{lB},\gamma_{l'B}\} = 2\delta_{ll'}.
\end{eqnarray}
In terms of the Majorana operators, Hamiltonian (1), up to constant terms, is given by the expression
\begin{eqnarray}
\label{HamiltonianKitaev0}
{H}_{K} = \frac{i}{2}(\epsilon - \mu)\sum_{l=1}^{N} \gamma_{lA}\gamma_{lB} + \frac{i}{2} \sum_{l=1}^{N-1} \left[ \left(t+|\Delta|\right) \gamma_{lB}\gamma_{l+1A} \right. - \left. \left(t-|\Delta|\right) \gamma_{lA}\gamma_{l+1B} \right].
\end{eqnarray}
Figure \ref{fig1} illustrates the single-site and inter-site coupling between Majorana operators. With the parameters chosen $\epsilon - \mu = 0$, $|t| = |\Delta|$,only one of the two terms in the square brackets in the second summand remains nonzero in expression (4): the first for $t>0$ and the second for $t<0$.

In this section, we restrict ourself to the case $t>0$. Hamiltonian (4) then takes the simple form
\begin{eqnarray}
\label{HamiltonianKitaev_esp_point}
&~&H_{K} = it \sum_{l=1}^{N-1} \gamma_{lB}\gamma_{l+1A}.
\end{eqnarray}
It is essential that this not involve operators  $\gamma_{1A}$ and $\gamma_{NB}$. Introducing the Fermi operators
\begin{eqnarray}
\label{fer_maj}
\alpha_{0} &=& \frac{1}{2}(\gamma_{NB} + i\gamma_{1A});~~~\alpha_{m} = \frac{1}{2}(\gamma_{mB} + i\gamma_{m+1A}),~~m=1,\ldots, N-1.
\end{eqnarray}
allows diagonalizing the Hamiltonian of the chain:
\begin{eqnarray}
\label{Ham_Kit_esp_diag}
{H}_{K} &=& -t\left(N-1\right) + \varepsilon_0\alpha_0^{\dag}\alpha_0 + 2t\sum_{m=1}^{N-1}\alpha_m^{\dag}\alpha_m;~~\varepsilon_{0} = 0.
\end{eqnarray}
It hence follows that $N$ excited states of a chain with one quasiparticle are described by the Hilbert space vectors
\begin{eqnarray}
\label{exite_states}
|\,\tilde{0}\,\rangle = \alpha^+_0 |\,0\,\rangle,~~
|\,m\,\rangle = \alpha^+_m |\,0\,\rangle,~~~m=1,2,...,N-1.
\end{eqnarray}
where the vector $|0\rangle$ corresponds to the many-body ground state of the chain ($\alpha_0|0\rangle=0|0\rangle$, $\alpha_m|0\rangle=0|0\rangle$). The operators $\alpha_m^+$ create quasiparticles with an energy of $2t$. The energy of such quasiparticles is independent of their number because of the localization of excitations on a pair of neighboring sites of the chain.

The state $|\,\tilde{0}\,\rangle$ deserves special attention: its energy coincides with the ground state energy $E_{\tilde{0}}=E_0$ because $\varepsilon_{0} = 0$. The energy value $\varepsilon_{0} = 0$ splits off from the energy spectrum of other quasiparticles because of the special structure of the $|\widetilde{0}\rangle$ state. To demonstrate this, we introduce the structural characteristics $F_{\pm}(l)$ of the $|\,\tilde{0}\,\rangle$ state, defined as site-dependent superpositions of the probability amplitudes of electron and hole creation in the course of the transition of the system from the ground state to the considered excited state:
\begin{eqnarray}
\label{F-factor}
F_{\pm}(l) = \langle \, \tilde{0} \, | (a_{l}^{+} \pm a_{l}) |\, 0 \, \rangle.
\end{eqnarray}
Recalling that $\gamma_{NB} = \alpha^{+}_{0} + \alpha_{0}$,
$\gamma_{1A} = i\left(\alpha^{+}_{0} - \alpha_{0}\right)$, we find that the introduced structural characteristics have a delta-like dependence on the site number:
\begin{eqnarray}
\label{w_z_ampl}
F_{+}(l) = i\delta_{l1}, ~~~ F_{-}(l) = -i\delta_{lN}.
\end{eqnarray}
This dependence suggests that, when the state $|\,\tilde{0}\,\rangle$ is formed, the structure of the many-body function of the ground state $|\,0\,\rangle$ changes only at the first and last sites of the chain. We note that this localized transformation is generated by the Majorana operators $\gamma_{1A}$ and $\gamma_{NB}$.

It is therefore common to say that the considered excited state can be associated with MMs whose wave functions are localized at the edges of the chain. Because these modes are realized simultaneously, we speak of the occurrence of an MBS for the state described by the vector $|\,\tilde{0}\,\rangle$. We now discuss this in more detail.

To represent Fermi excitations in terms of MMs and
MBSs, we introduce self-conjugate operators $b'_{m} = (b'_{m})^{+}$ and $b''_{m} = (b''_{m})^{+}$ and write their decompositions in terms of the Majorana operators $\gamma_{lA}$ and $\gamma_{lB}$:
\begin{eqnarray}
\label{b'_b''_1}
b_{m}' = \alpha_{m}^{+} + \alpha_{m} = \sum^{N}_{l=1} w_{lm}\gamma_{lA};~~~
b_{m}'' = i\left(\alpha_{m}^{+} - \alpha_{m}\right) = \sum^{N}_{l=1} z_{lm}\gamma_{lB}.
\end{eqnarray}
It is easy to see how the decomposition coefficients $w_{lm}$ and $z_{lm}$ are related to the $u-v$ coefficient of the Bogoliubov transformation:
\begin{eqnarray}
\label{w_z_def}
\alpha_{m} = \sum_{l=1}^{N}\left( u_{lm}a_{l} + v_{lm}a_{l}^{+}\right);~~~
w_{lm} = |u_{lm}| + |v_{lm}|;~~z_{lm} = |u_{lm}| - |v_{lm}|.
\end{eqnarray}
The Hamiltonian can be expressed in terms of the $b'_{m}$ and $b''_{m}$ operator as
\begin{eqnarray}
\label{Ham_b'_b''}
{H}_{K}=\frac{1}{2}\sum_{m=1}^{N}\varepsilon_{m}+
\frac{i}{2}\sum_{m=1}^{N}\varepsilon_{m}b_{m}'b_{m}''.
\end{eqnarray}
Because the coefficients $u_{lm}$ and $v_{lm}$ describe the spatial behavior of the electron- and hole-like wave functions of Fermi quasiparticles, we can assume, by analogy, that the coefficients $w_{lm}$ and $z_{lm}$ characterize the spatial distribution of the wave functions of the modes associated with the respective operators $b'_{m}$ and $b''_m$. These coefficients are determined from the system of equations
\begin{eqnarray}
\label{wz_syst2}
&~&-\mu z_{lm} - (t+|\Delta|)z_{l-1,m} - (t-|\Delta|)z_{l+1,m}=\varepsilon_{m} w_{lm},
\nonumber \\
&~&\hphantom{-}\mu w_{lm} + (t-|\Delta|) w_{l-1,m} + (t+|\Delta|)w_{l+1,m}=\varepsilon_{m} z_{lm}.
\end{eqnarray}
Let us analyze the important case where the Fermi excitation
generated by the operator  $\alpha_0=b'_{0} + \,i\,b''_{0}$ has spatially nonoverlapping distributions of $w_{l0}$ and $z_{l0}$ localized near the respective left and right edges of the chain. In other words, we suppose that there is a set of sites $\{\, l_{L} \, \}$ $\left(\{\, l_{R} \, \}\right)$ near the left (right) edge of the chain such that $\{\, l_{L}:~w_{l_L,0}\neq 0,~z_{l_L,0}= 0 \, \}$ and $\{\, l_{R}:~z_{l_R,0}\neq 0,~w_{l_R,0}= 0 \, \}$. Hence, in particular, it follows that $w_{l0}z_{l0}= 0$ for each $l$ site. In this case, the Fermi operator $\alpha_0=b'_{0} + \,i\,b''_{0}$ is said to describe an MBS made of a pair of MMs described by the operators $b'_{0}$ and $b''_{0}$ whose wave functions do not overlap. It can be deduced from system (14) that such a solution can be realized only for a mode with zero excitation energy, $\varepsilon_{0}=0$. In particular, for the parameters $\mu = 0$ and $t = \Delta$, we see by comparing (7) and (11) that $w_{l0} = \delta_{l,1}$ $z_{l0} = \delta_{l,N}$ for the zero mode. This is a mathematical manifestation of the fact that MMs are localized strictly at the edge sites of the chain. In view of the bounded nature of localization, we say such a point is "special" in the parameter space of the Kitaev chain. For it, MMs with a strictly zero excitation energy are realized for an open chain of an arbitrary length. Under insignificant variations in the parameters relative to the special point, the edge modes are still realized in the system. Their excitation energy, being nonzero in general, decreases exponentially as the chain length increases, e0 $\varepsilon_{0} \sim e^{-N}$. With an increase in the number of sites $N$, the overlap of the $w_{l0}$ and $z_{l0}$ distributions diminishes exponentially, disappearing in the limit as $N\to \infty$ and leading to the formation of a pair of MMs and one MBS. The criterion for the realization of such excitations is discussed in Section \ref{sec2.2}.

The localization of MM wave functions at opposite ends of the chain automatically leads to the MBS energy tending to zero. As a result, the energies of many-body states of the system that belongs to Hilbert-space sectors with different FPs tend to degenerate. Here and hereafter, we assume the eigenvalues $P$ of the FP operator $\textbf{P}$ of a many-body state to be positive (negative), $P=1(-1)$, if this state is described by a superposition of partial contributions with an even (odd) number of fermions. From the standpoint of quantum calculations, it is important that the condition $w_{l0}z_{l0}=0$ guarantees the stability of the wave function of a qubit constructed on such states,
$$|\,\Psi\,\rangle=c_0|\,0\,\rangle + c_1|\,1\,\rangle;~~
\textbf{P} |\,0\,\rangle=|\,0\,\rangle;~~\textbf{P} |\,1\,\rangle=-|\,1\,\rangle,$$
under local external perturbations described  by the one-particle operator $f = \sum_{l=1}^{N}f_{l}\,a^{+}_{l}a_{l}$. The condition that local perturbations not change the relative phase of the qubit states can then be represented as
$\delta f = \langle \, 0 \,| \, f \,|\,0\,\rangle - \langle \, 1 \, | \, f \, |\,1\,\rangle = \sum_{l=1}^{N}f_{l}\,w_{l0}z_{l0} = 0$.
It can be seen that the spatial separation condition for MMs automatically leads to the realization of robustness against to decoherence in the system. This is an important factor in considering nanowires with MMs as promising objects for the elemental base of quantum computing. We note that, in the case $f_{l}= const$, the relative phase stability condition $\delta {f} = 0$ coincides with the electroneutrality condition for MBSs.

It is often claimed that MMs are described by self-conjugate operators $b'$ and $b''$ with Majorana anticommutation relations, while the Fermi operator $\alpha_{0} = b' + ib''$ describes an MBS. For an MBS, $\varepsilon_0=0$ and $[\,H,\alpha_{0}\,]=0$, and hence each operator $b'$ and $b''$ , which is determined by system (14) independently, commutes with the Hamiltonian. Moreover, due to the single-particle nature of the MMs, their operators anticommute with the FP operator $\textbf{P}$.

The foregoing defines the properties of the Majorana operators $b'$ and $b''$,
\begin{eqnarray}
\label{b_pecular}
b^2 = 1;~[\, b\, , \, H\,]=0;~\{\, b\, , \, \textbf{P}\,\}=0;~\{\, b'\, , \, b''\,\}=0,
\end{eqnarray}
which are often postulated as basic ones in describing MMs in Fermi systems with electron-electron interaction, when the structure of elementary excitations becomes more complex [46-49] (also see Section \ref{sec2.5}).

For systems described by Hamiltonians that are quadratic in Fermi operators, the problem of finding MBSs reduces to finding the zero eigenvalue of the Bogoliubov-de Gennes matrix, which is related to the Hamiltonian of the system as
\begin{eqnarray}
\label{BdG_gen}
&~&{H} =\Psi^{+}\cdot \hat{H}_{BG} \cdot \Psi;
~~\Psi = (a_{1},a_{2},...,a_{\tilde{N}},a^+_{1},a^+_{2},...,a^+_{\tilde{N}})^{T} \\
&~&{\hat{H}_{BG}} = \left( {\begin{array}{*{20}{c}}
	\hat{A}&\hat{B}\\
	{ - \hat{B}^*}&{ - \hat{A}^*}
	\end{array}} \right);~ \hat{A} = \hat{A}^{+};~\hat{B} = - \hat{B}^{T}. \nonumber
\end{eqnarray}

We adopt the convention that letters under a circumflex denote the matrices of operators defined in the space of single-particle states and matrices of the Green's functions. In (16), the $\hat{A}$ and $\hat{B}$ matrices have the size $\tilde{N} = N\cdot n_{0}$, where $N$ is the number of sites in the chain and $n_{0}$ is the number of internal degrees of freedom associated with one site.

Fermions in the Kitaev model are spinless, and therefore $\tilde{N}=N$. The nonzero matrix elements of $\hat{A}$ and $\hat{B}$ are
\begin{eqnarray}
\label{A_B_Kitaev}
\hat{A}_{n,n} = \epsilon - \mu;~\hat{A}_{n,n+1} = -t;~\hat{B}_{n,n+1} = -\Delta.
\end{eqnarray}
The Bogolubov-de Gennes matrix $\hat{H}_{BG}$ has the property
\begin{eqnarray}
\label{H_BdG_particle_hole}
\{ \hat{H}_{BG},~ \hat{C} \} = 0;~\hat{C} = \hat{\Lambda} \cdot {K};~
\hat{\Lambda} =  \left({\begin{array}{*{20}{c}}
	\hat{0}    &     \hat{I}\\
	\hat{I}    &     \hat{0}  \end{array}}\right),
\end{eqnarray}
where $\hat{I}$ is the unit $\tilde{N} \times \tilde{N}$ matrix and ${K}$ is the operator of a complex conjugation. A direct corollary of property (18) is that, if a vector $\psi_m = \left( u,~v \right)^T_m$ is the eigenvector of the matrix $\hat{H}_{BG}$ with an eigenvalue $\varepsilon_{m}$, then the conjugate vector $\bar{\psi}_m = \hat{C}\psi =\left( v^*,~u^*\right)^T_m$ is also an eigenvector of $\hat{H}_{BG}$, but corresponds to the eigenvalue $-\varepsilon_{m}$. The components of the $\psi_m$ vectors coincide with the $u-v$ Bogoliubov coefficients (12),
\begin{eqnarray}
\label{psi_to_uv}
(\psi_m)_n = u_{nm},~~n = 1,\ldots , \tilde{N};~~~
(\psi_m)_n = v_{nm},~~n = \tilde{N}+1,\ldots , 2\tilde{N},
\end{eqnarray}
which describe the electron- and hole-like properties of Fermi excitations. Positive eigenvalues of $\hat{H}_{BG}$ determine the energies $\varepsilon_{m}$ of Fermi excitations. In accordance with the foregoing, property (18) determines the electron-hole symmetry of Hamiltonian (16).

The Kitaev chain model (1) has an effective time reversal symmetry and belongs to the BDI class of the well-known symmetry classification. This class of symmetries allows several pairs of MMs to be realized in the system: those described by the operator $\alpha_{j}=b'_{j}+ib''_{j}$ of quasiparticle excitations for which the operators $b'_{j}$ and $b''_{j}$ correspond to MMs with nonoverlapping wave functions. It was shown in [50-52] that, in order to realize several MM pairs, hops and superconducting pairing beyond the nearest-neighbor sites must be taken into account in the Kitaev chain. At the same time, it was found that the generalization of the system to the case where the hopping amplitudes and anomalous pairings decrease in accordance with a power law can result in violating the topological classification, weakening the bulk-boundary correspondence (Section \ref{sec2.2}) and inducing new quasiparticles, called Dirac fermions [53-55].

\subsection{\label{sec2.2} Conditions for the realization of Majorana modes in one-dimensional systems. Topological indices}

Analyzing the conditions for the formation of MMs in an open chain based on the solution to the eigenvalue problem of Bogoliubov-de Gennes matrix (16) is mathematically difficult due to its large size. A useful criterion was proposed in [12], according to which the bulk-boundary correspondence is established between the conditions ensuring the realization of MMs in open chains and the characteristics of chains in the closed geometry, for which not only numerical but also analytic calculations are much simpler.

The idea of the proposed criterion, revealing its physical meaning, is as follows. We consider an open chain of length $N=N_{1}+N_{2}$, large enough for a pair of MMs located at opposite ends to be realized in it. We then impose periodic boundary conditions on this chain and calculate the FP index of its ground state, conventionally written as $P = P\left( {{H}}(N_{1}+N_{2}) \right)$. Next, we divide this chain into two closed chains of lengths $N_{1}$ and $N_{2}$ and calculate the FP index of the ground state of the new system, which factors through the indices of the FPs of individual chains: $P=P\left( {{H}}(N_{1}) \right)P\left( {{H}}(N_{2}) \right)$. Now, if MMs were realized in the initial chain, then the FP of the ground state would change; in the absence of MMs in the original system, it would remain the same. Mathematically, this criterion is formulated by introducing the Majorana number ${\mathscr{M}}$ taking the values ${\mathscr{M}}=\pm 1$:
\begin{eqnarray}
\label{M_numb_def}
P\left( {{H}}(N_{1}+N_{2}) \right) = {\mathscr{M}}P\left( {{H}}(N_{1}) \right)P\left( {{H}}(N_{2}) \right).
\end{eqnarray}
If ${\mathscr{M}}=-1$, then MMs are realized in long wires. It follows from the above definition that, for a nanowire with an even number of sites, the Majorana number and the FP index coincide. We note that it was assumed in deriving (20) that the lengths $N_{1}$ and $N_{2}$ of the open chains obtained by splitting the original one are large enough for MMs with nonoverlapping wavefunctionstoberealizedinthem.However, as shown below, the values of $N_{1}$ and $N_{2}$ do not appear in specific calculations of ${\mathscr{M}}$.

In fact, the Majorana number ${\mathscr{M}}$ is a $\mathbb{Z}_2$ TI of 1D superconducting systems with broken time reversal invariance (quadratic Hamiltonians of the symmetry class D). In the presence of such a symmetry described by an antiunitary operator (class BDI, the case of model (1)), the first Chern number ${\mathscr{C}}$ acts as a TI, taking integer values and being $\mathbb{Z}$-invariant. The index ${\mathscr{M}}$ allows predicting the existence of MMs in an open system of the symmetry class D based on its analysis in the closed geometry. Similarly, the ${\mathscr{C}}$ index allows determining the existence and the number of MBSs in systems of the symmetry class BDI. Moreover, ${\mathscr{M}}=(-1)^{\mathscr{C}}$ in the last case [56-58].

To calculate Majorana number (20) and demonstrate the last relation, it is quite useful to move from the coordinate representation to the energy one in Bogoliubov-de Gennes Hamiltonian (16). This can be done by a unitary transformation given by a matrix $\hat{\text{U}}$ in the space of single-particle states
\begin{eqnarray}
\label{U_matrix}
\left( \begin{array}{*{20}{c}} {a} \\ {~a^+} \end{array} \right) \to
\left( \begin{array}{*{20}{c}} {\alpha} \\ {~\alpha^+} \end{array} \right) =
{\hat{\rm{U}}} \cdot \left( \begin{array}{*{20}{c}} {a} \\ ~{a^+} \end{array} \right),~~~
{\hat{\rm{U}}}{\hat{H}}{\hat{\rm{U}}}^{+} =  \left({\begin{array}{*{20}{c}}
	{\hat{\varepsilon}}    &     ~\hat{0}\\
	~~\hat{0}              &     ~{-\hat{\varepsilon}}  \end{array}}\right).
\end{eqnarray}
Here, $\hat{\varepsilon}$ is a diagonal $\tilde{N}\times \tilde{N}$ matrix whose positive elements give the energies of Fermi excitations $\varepsilon_{0},\ldots\varepsilon_{\tilde{N}-1}$. The column vectors of the unitary transformation matrix $\hat{\rm{U}}$ that correspond to the energies $\varepsilon_{m}$ are given by ${\bf \psi_{m}}$ in (19). A similar transition has to be performed in the representation of the Majorana operators
${\gamma_{nA} = e^{-i\theta/2}a_{n}^{+} + e^{i\theta/2}a_{n}}$, \\
${\gamma_{nB} = i(e^{-i\theta/2}a_{n}^{+} - e^{i\theta/2}a_{n})}$,\\
$b_{m}' = \alpha_{m}^{+} + \alpha_{m}$ и $b_{m}'' = i(\alpha_{m}^{+} - \alpha_{m})$,\\
defined in the respective site and energy representations,
\begin{eqnarray}
\label{from_gamma_to_b}
&~&{\left( \begin{array}{*{20}{c}}
	\gamma_{A} \\ ~\gamma_{B}\end{array} \right) \to
	\left( \begin{array}{*{20}{c}} b' \\ ~b''\end{array} \right) =
	{\hat{W}} \cdot \left( \begin{array}{*{20}{c}} \gamma_{A} \\ ~\gamma_{B}\end{array} \right);} \nonumber\\
&~&{\hat{W}=\hat{R}^{+}(\theta=0)\cdot\hat{\rm{U}}\cdot\hat{R}(\theta);~~\hat{R}(\theta) = \frac{1}{\sqrt{2}}\left({\begin{array}{*{20}{c}}
		\hat{I}\cdot e^{-i\theta/2}    &     \hphantom{-}i\hat{I}\cdot e^{-i\theta/2}\\
		\hat{I}\cdot e^{i\theta/2}    &    -i\hat{I}\cdot e^{i\theta/2}  \end{array}}\right),}
\end{eqnarray}
where $\hat{I}$ is still the unit $\tilde{N} \times \tilde{N}$ matrix, the matrix ${\hat{W}}$ is real and orthogonal, and ${\hat{W}^{-1}}={\hat{W}}^{T}$. The FP index $P$ of the ground state of many-body system can be evaluated as \cite{kitaev-01}
\begin{eqnarray}
\label{Parity_W_U}
P = \sign\left(\det({\hat{W}})\right) = \sign\left(\det(\hat{\rm{U}})\right).
\end{eqnarray}
We note that formula (23) is applicable for calculating the FP index of chains with arbitrary boundary conditions. The definition of $\mathscr{M}$ in (20) involves periodic boundary conditions, and, to calculate it, it is therefore useful to move to the quasimomentum representation:
\begin{eqnarray}
\label{ak_gammak}
a_{k} = \frac{1}{\sqrt N} \sum_{l=1}^{N} a_{l}e^{-ikl};~k &=& \frac{2\pi s}{N};~~s=0,\ldots,N-1.
\end{eqnarray}
\begin{eqnarray}
\label{Hk_BdG}
&~&H = \sum_{k}H(k);~~~H(k) = \frac{1}{2}\left( {a_k^+}\,, \,
{a_{-k}} \right)
{\hat{H}_{BG}}(k)\left( \begin{array}{*{20}{c}} {a_{k}} \\ {~{a}_{-k}^{+}}\end{array} \right); \\
&~&{\hat{H}_{BG}}(k) = \left( \begin{array}{*{20}{c}}
\hat{A}_{k}&\hat{B}_{k}\\
- \hat{B}_{-k}^*& - \hat{A}_{-k}^*
\end{array} \right);
\begin{array}{*{20}{c}}
\hat{A}_{k} = \hat{A}_{k}^{+}\\
~~~~\hat{B}_{k} = - \hat{B}_{-k}^{T}.
\end{array}\nonumber
\end{eqnarray}
We note a number of general properties of single-particle Hamiltonians, with their manifestations in specific models to be discussed later. The Bogoliubov-de Gennes matrix ${\hat{H}_{BG}}(k)$ in the quasimomentum representation has the property
\begin{eqnarray}
\label{Hk_BdG_sym}
\hat{\Lambda}\cdot{\hat{H}_{BG}}(k)\cdot\hat{\Lambda} = -{\hat{H}^{*}_{BG}}(-k),
\end{eqnarray}
where the matrix $\hat{\Lambda}$ is formally the same as in (\ref{H_BdG_particle_hole}). The matrices ${\hat{\rm{U}}}_{k}$ that diagonalize ${\hat{H}_{BG}}(k)$ can therefore be
represented as
\begin{eqnarray}
\label{Uk_matrix}
{\hat{\rm{U}}_{k}}^{+}{\hat{H}_{BG}}(k){\hat{\rm{U}}}_{k} &=&
{\left({\begin{array}{*{20}{c}}
		~~\hat{\varepsilon}_{k}  &     ~\hat{0}\\
		~~\hat{0}              &  ~~-\hat{\varepsilon}_{-k}      \end{array}}\right)};~~~
{\hat{\rm{U}}}_{k}  =
{\left({\begin{array}{*{20}{c}}
		\hat{u}_{k}    &     \hat{v}_{-k}^{*}\\
		\hat{v}_{k}    &     ~\hat{u}_{-k}^{*} \end{array}}\right)}.
\end{eqnarray}
It follows from (\ref{Uk_matrix}) that ${\hat{\rm{U}}_{k} = \hat{\Lambda}\cdot{\hat{\rm{U}}}^{*}_{-k}\cdot\hat{\Lambda}}$ and therefore ${\det({\hat{\rm{U}}}_{k})=\left( \det({\hat{\rm{U}}}_{-k}) \right)^{*}}$. Hence, the expression
\begin{eqnarray}
\label{detU_closed}
\det({\hat{\rm{U}}})=\prod_{-\pi<k\leq\pi}\det({\hat{\rm{U}}}_{k})
\end{eqnarray}
entering (\ref{Parity_W_U}) (for an even number of sites in the chain), can be split into two parts. The first one is the product of the factors $\det({\hat{\rm{U}}}_{k})$ for which $k \neq -k$, and the second is determined by the product of factors for the quasimomentum values, such that $k = -k \doteq K$ (symmetric points of the Brillouin zone).

For chains with an even number of sites, there is a pair of symmetric points $K = 0,~\pi$. Therefore, taking the equality $\det({\hat{\rm{U}}}_{k})=\left( \det({\hat{\rm{U}}}_{-k}) \right)^{*}$ into account, we see that the product $\prod_{k \neq -k}\det({\hat{\rm{U}}}_{k})$ is real and positive. Hence, the Majorana number can be represented as
\begin{eqnarray}
\label{M_numb_1}
{\mathscr{M}}=\sign\left(\frac{\det({\hat{\rm{U}}}_{0})}{\det({\hat{\rm{U}}}_{\pi})}\right)=
\frac{\det({\hat{\rm{U}}}_{0})}{\det({\hat{\rm{U}}}_{\pi})}.
\end{eqnarray}
This form allows relating the Majorana number of a quantum wire to the Zak-Berry phase [59, 60] for band states, which has a topological origin. For this, we consider the known relation
\begin{eqnarray}
\label{to_Berri_Connection}
\frac{d}{dk}\ln\left( \det({\mathbf{U}}_{k}) \right) &=& -i\cdot {{\Tr}}\left( {\hat{\rm{O}}}_k \right);
~~{\hat{\rm{O}}}_{k} = {\hat{\rm{U}}}^{-1}_{k}\frac{d}{dk}{\hat{\rm{U}}}_{k}; \Rightarrow\nonumber\\
\Rightarrow\frac{\det({\hat{\rm{U}}}_{0})}{\det({\hat{\rm{U}}}_{\pi})} &=&
\exp\left(-i\int_{0}^{\pi}{\Tr}\left( {\hat{\rm{O}}}_k \right)dk\right).
\end{eqnarray}
Using the definition of the matrix ${\mathbf{U}}_{k}$ in (27) and the property ${\mathbf{U}}^{-1}_{k}={\mathbf{U}}^{+}_{k}$, after several simple transformations, we can represent ${\hat{\Tr}}\left( {\hat{\rm{O}}}_k \right)$ as
\begin{eqnarray}
\label{Berri_Connection}
{\Tr}\left( \hat{\rm{O}}_k \right) =
\sum_{m}i\left[ \left(\phi^*_m(k),\frac{d}{dk}\phi_m(k)\right) +
\left(k \to -k\right) \right].
\end{eqnarray}
where $\phi_m(k)$ is the eigenvector of ${\mathbf{U}}_{k}$ corresponding to the energy $\varepsilon_{m}(k)\geq 0$, from the quasiparticle branch of the excitation spectrum. The sum over $m$ in (31) involves all functions $\phi_m(k)$ that cannot be transformed into one another by the electron-hole symmetry operation.

Comparing relations (29)-(31), we obtain the sought expression for the Majorana number:
\begin{eqnarray}
\label{M_numb_2}
{\mathscr{M}}=\exp\left(i\int_{-\pi}^{\pi} dk \cdot \sum_{m} \left(\phi^*_m(k),\frac{d}{dk}\phi_m(k)\right)
\right)\doteq\exp\left(i\pi\cdot{\mathscr{C}}\right).
\end{eqnarray}
The points $k=-\pi$ and $k=\pi$ are identified, and therefore the exponent involves the integral of a differential 1-form, the well-known Berry connection, along a closed curve. The topological nature of the Majorana number is thus established. In what follows, the set of parameters corresponding to the conditions for a negative value of $\mathscr{M}$ to be realized (and therefore the conditions for MMs to be realized in open wires) is referred to as the topologically nontrivial parameter domain (TNPD). Moreover, we say that the system is in a topologically nontrivial phase.

To conclude this Section, we also present another formula for the Majorana number, which is more convenient in specific calculations in some cases \cite{kitaev-01}:
\begin{eqnarray}
\label{M_numb_3}
{\mathscr{M}}=\sign\left[ {\Pf}({\hat{X}}(k=0))\cdot {\Pf}({\hat{X}}(k=\pi)) \right].
\end{eqnarray}
Here, ${\Pf}({\hat{X}}(K))$ denotes the Pfaffian of the skew-symmetric matrix ${\hat{X}}(k)$, the matrix of the Bogoliubov-de Gennes Hamiltonian written in terms of the Majorana operators $a_{k} = \gamma_{kA}+i\gamma_{kB}$; $a^{+}_{k} = \gamma_{-k,A}-i\gamma_{-k,B}:$
\begin{eqnarray}
\label{Hk_BdG_Maj_repr}
{\hat{X}}(k) &=& \frac{1}{2}\left( {\begin{array}{*{20}{c}}
	{ {\hat{A}}_{k+} + {\hat{B}}_{k-}} & {~~~{\hat{A}}_{k+} - {\hat{B}}_{k+}} \\
	{ {\hat{A}}_{k+} + {\hat{B}}_{k+}} & {~~~{\hat{A}}_{k-} - {\hat{B}}_{k-}}\end{array}} \right);\nonumber\\
{\hat{A}}_{k\pm} &=& {\hat{A}}_{k} \pm {\hat{A}}^{*}_{-k};~~~{\hat{B}}_{k\pm} = {\hat{B}}_{k} \pm {\hat{B}}^{*}_{-k}.
\end{eqnarray}
Using the properties of ${\hat{A}}_{k}$ and ${\hat{B}}_{k}$, it can be easily verified that the matrix  $\hat{X}(k)$ is real and antisymmetric at the $k = K = 0,~\pi$ points of the Brillouin zone.

\subsection{\label{sec2.3} Topological quantum transitions and the ground state structure}

We use formulas (29)-(33) to evaluate the Majorana number in the Kitaev model. For this, we use expression (32) and find the Zak-Berry phase in the closed geometry. The Bogoliubov-de Gennes matrix ${\hat{H}_{BG}}(k)$ in quasimomentum representation (25) has the form
\begin{eqnarray}
\label{Hk_BdG_Kitaev}
&~&\hat{H}_{BG}(k) = \left({\begin{array}{*{20}{c}}
	\xi_k                        &     ~~~~\Delta_k e^{i\theta\,'}\\
	\Delta_k e^{-i\theta\,'}        & ~~-\xi_k      \end{array}}\right);\nonumber\\
&~&\theta\,' = \theta + \frac{\pi}{2};~\xi_k = -\mu - 2t\cos k;~\Delta_k = 2|\Delta|\sin k.
\end{eqnarray}
We proceed to the new representation in terms of the matrix $\hat{R}$:
\begin{eqnarray}
\label{Hk_BdG_Kitaev2}
&~&\hat{R} = \frac{1}{\sqrt{2}}\left( {\begin{array}{*{20}{c}}
	{ e^{-i\theta\,'/2} } & {\hphantom{-}ie^{i\theta\,'/2}} \\
	{ e^{-i\theta\,'/2} } & {-ie^{i\theta\,'/2}} \end{array}} \right);\nonumber\\
&~&{\hat{H}_{BG}}(k) \to {\hat{{\tilde{H}}}_{BG}}(k) = \hat{R}\cdot{\hat{H}_{BG}}(k)\cdot\hat{R}^{+}  = \left( {\begin{array}{*{20}{c}}
	{ 0 } & {~~~\xi_{k} + i\Delta_{k}} \\
	{ \xi_{k} - i\Delta_{k}} & {~~~0}\end{array}} \right).
\end{eqnarray}
In this representation, the electron and hole components of the Bloch amplitudes $\phi_+(k)$ corresponding to the positive branch of the spectrum of Bogoliubov quasiparticles with the energy $\varepsilon_{k} = \sqrt{\xi_{k}^{2} + \Delta_{k}^{2}}$ are given by
\begin{eqnarray}
\label{spectr_Bloch_Kitaev}
\phi_+(k) = \frac{1}{\sqrt{2}\cdot\varepsilon_{k}}\left( {\begin{array}{*{20}{c}}
	{ \xi_{k} + i\Delta_{k} }  \\
	{ \varepsilon_{k} } \end{array}} \right) =
\frac{1}{\sqrt{2}}\left( {\begin{array}{*{20}{c}}
	{ e^{i\beta_k}}   \\
	{ 1 } \end{array}} \right).
\end{eqnarray}
In (37), we introduce the angle $\beta_k$ between the effective vector ${\bf h} = (h_{x},~h_{y})$ and the unit vector of the 2D Cartesian coordinate system:
\begin{eqnarray}
\label{theta_angle}
\beta_k = \arctan\left( \frac{h_{y}}{h_{x}} \right);~~~
h_{x}=\frac{\xi_{k}}{2t}=-\frac{\mu}{2t} - \cos k;~~h_{y}=\frac{\Delta_{k}}{2t}=\frac{|\Delta|}{t}\sin k.
\end{eqnarray}
The change in the Zak-Berry phase $\Delta\Phi_{ZB}$ under adiabatic evolution of the system (such that the quasimomentum varies along a closed contour in the first Brillouin zone) is determined by the variation in that angle:
\begin{eqnarray}
\label{Berri_Phase_Kitaev}
\Delta\Phi_{ZB} &=& i\int_{-\pi}^{\pi} dk \cdot \left(\phi^*_+(k),\frac{d}{dk}\phi_+(k)\right)=\frac{1}{2}\oint d\beta_k=\nonumber\\
&=&\frac{1}{2|h|^{2}}\int_{-\pi}^{\pi}\left( h_{y}\frac{dh_{x}}{dk} - h_{x}\frac{dh_{y}}{dk} \right) = \nu/2.
\end{eqnarray}

\begin{figure}[h!]
	\begin{center}
		\includegraphics[width=0.8\textwidth]{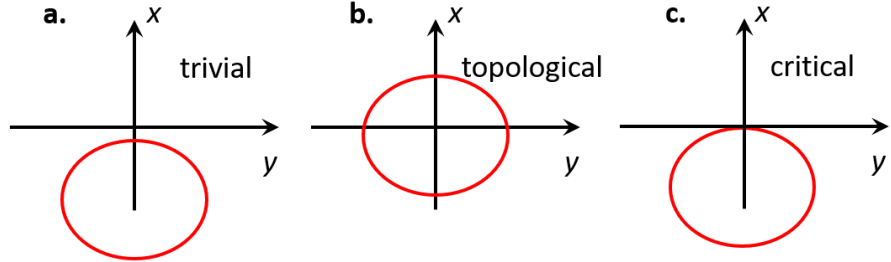}
		\caption{Schematic representation of the Zak-Berry phase variations in different cases: (a) $\nu=0$ corresponds to trivial phase of the Kitaev chain, (b) $\nu=1$ corresponds to the topological phase, (c) $\nu$ is undefined and corresponds to a topological quantum phase transition point.}\label{fig2}
	\end{center}
\end{figure}

For completeness, we here introduce the winding number $\nu$. Inspecting expressions (38) and (39) readily shows what the conditions are for the phase change $\Delta\Phi_{ZB}$ to be nonzero under adiabatic evolution (these are the conditions for the realization of a topologically nontrivial phase). The components of the vector ${\bf h}$ satisfy the equation for an ellipse,
\begin{eqnarray}
\label{h_circle}
\left(h_{x} + \frac{\mu}{2t}\right)^2 + \left(\frac{t}{|\Delta|}\cdot h_{y}\right)^2 = 1;
\end{eqnarray}
with a unit semi-axis along the $x$-axis. The center of the ellipse is displaced by $-\mu/2t$ along the $x$-axis (Fig. \ref{fig2}). We see that $\Delta\Phi_{ZB} = \pi$ if the origin is in the domain bounded by the ellipse, and $\Delta\Phi_{ZB} = 0$ otherwise. The first case is realized under the condition $|\mu|<2t$. Substituting the result of this calculation into (32), we obtain a condition for the realization of a topologically nontrivial phase in the Kitaev chain model:
\begin{eqnarray}
\label{M_number_Kitaev}
{\mathscr{M}} = \left\{ {\begin{array}{*{20}{c}}
	{{e^{i\pi }} =  - 1;~~~\left| \mu  \right| < 2t}\\
	{{e^{i0}} = ~1;~~~\left| \mu  \right| > 2t}
	\end{array}} \right.
\end{eqnarray}
Both equalities are satisfied in the case $|\Delta| \neq 0$. For $|\Delta| = 0$, the ellipse degenerates into a straight line, and the winding number becomes undefined. We note that $\nu$ defines the degree of the mapping $\mathbb{R}^{2}\setminus \{0\} \to \mathbf{S}^{1}$ of the 2D plane $\{h_{x}(k),~h_{y}(k)\}$ with punctured origin (the bulk spectrum being gapless for $h_{x}=h_{y}=0$) onto a 1D sphere. The homotopy classes of such a map are characterized by the integers $\mathbb{Z}$ that can be related to the values of $\nu/2\pi=\mathscr{C}$. In this case, the Majorana number is determined by the parity of $\nu/2\pi$ and is therefore $\mathbb{Z}_{2}$ -invariant.

Thus, in the Kitaev chain, a topologically nontrivial phase is realized for the parameter values $-2t<\mu<2t$, $|\Delta| \neq 0$. In this parameter domain, MMs exist in a very long open chain. From the expression for the bulk spectrum,
\begin{eqnarray}
\label{spectrum_Kitaev}
\varepsilon_{k} = \sqrt{(-2t\cos(k) - \mu)^2 + 4|\Delta|^2\sin(k)^{2}},
\end{eqnarray}
it is clear that, at the points $\{\,|\mu| = 2t\,;\,k=0,~\pi\,\}$, bulk spectrum closes. Therefore, the transition from a topologically nontrivial to a trivial phase and the reverse transition are quantum transitions with a change in the $\mathbb{Z}_{2}$ topological index ${\mathscr{M}}$. As noted, the FP index of the ground state of the system changes under such a transition.

To consider the structure of the ground state and its change under topological quantum transitions, we write the Hamiltonian of the system in the quasimomentum representation by isolating the terms corresponding to the symmetric quasimomentum values $k=0,~\pi$:
\begin{eqnarray}
\label{Hk_Kitaev}
{H}_{K} = (-2t - \mu)a^{+}_{0}a_{0} + (2t-\mu)a^{+}_{\pi}a_{\pi}+\sum_{-\pi<k<\pi}\left(
\xi_{k}a^{+}_{k}a_{k} + \left(\frac{{\Delta}_{k}}{2}e^{i\theta\,'}a^{+}_{k}a^{+}_{-k} + h.c.\right) \right).
\end{eqnarray}
The terms corresponding to Brillouin zone points that are symmetric under the electron-hole symmetry operation are isolated, because they initially have a diagonal form. To diagonalize the other terms, we introduce the unitary operator
\begin{eqnarray}
\label{U_op}
{U} &=& \exp \left(\sum_{{-\pi<k<\pi}}\frac{\beta_{k}}{2}\left( e^{-i\tilde{\theta}}a^{+}_{-k}a^{+}_{k} - e^{i\tilde{\theta}}a_{k}a_{-k}  \right) \right)=\nonumber\\
&=& \prod_{-\pi<k<\pi}\Big[ 1 + \sin\frac{\beta_k}{2}\left( e^{-i\tilde{\theta}}a^{+}_{-k}a^{+}_{k} - e^{i\tilde{\theta}}a_{k}a_{-k}  \right) + \nonumber\\
&+& \left( \cos\frac{\beta_k}{2}-1\right)\left( 1 - n_{k} - n_{-k} + n_{k}n_{-k} \right) \Big],
\end{eqnarray}
where $\tilde{\theta}=\theta + \frac{\pi}{2}\left(2 - \sign(\Delta_k)\right)$.
The Hamiltonian  ${H}$ can be then be brought to the form
\begin{eqnarray}
\label{H_eff_diag}
{H}_{K} \to  {UH_{K}U^{+}} &=& (-2t-\mu)a^{+}_{0}a_{0} + (2t-\mu)a^{+}_{\pi}a_{\pi} + \sum_{0<k<\pi} \varepsilon_{k}\alpha_{k}^{+}\alpha_{k} + E_{0},
\end{eqnarray}
where
\begin{eqnarray}
\label{H_eff_diag_param}
&~&\alpha_{k}= {U}\cdot a_{k} \cdot {U}^{+}=\cos\frac{\beta_k}{2}\cdot a_{k} + e^{-i\tilde{\theta}}\sin\frac{\beta_k}{2}\cdot a^{+}_{-k};\\
&~&\cos\frac{\beta_k}{2}= \sqrt{\frac{1}{2}\left( 1 + \frac{\xi_{k}}{\varepsilon_{k}} \right)};~
\sin\frac{\beta_k}{2}= \sqrt{\frac{1}{2}\left( 1 - \frac{\xi_{k}}{\varepsilon_{k}} \right)}.\nonumber
\end{eqnarray}
In the subspace of states with a quasimomentum $0<k<\pi$, the ground state $|\Psi_{0}\rangle$ of Hamiltonian (43) can be obtained by acting with the operator $U$ on the quasiparticle vacuum $|0\rangle$:
\begin{eqnarray}
\label{H_eff_ground_state}
|\Psi_{0}\rangle = \prod_{-\pi<k<\pi}\left( \cos\frac{\beta_k}{2} + e^{-i\tilde{\theta}}\sin\frac{\beta_k}{2}a^{+}_{-k}a^{+}_{k} \right)|0\rangle.
\end{eqnarray}
To find the structure of the ground state of the system in the Hilbert space containing single-particle states with quasimomenta $k=0,\pi$, we have to consider the isolated terms in (43) corresponding to such values of $k$. If $\mu<-2t$, then the modes with quasimomenta $k=0,\pi$ are unfilled. In the range of the chemical potential values $-2t<\mu<2t$, the mode with the quasimomentum $k=0$ is filled. For $\mu > 2t$ , both energy states with the selected values of the quasimomentum are filled with fermions.

As a result of the analysis, we find that the structure of the ground state $|\Psi_{GS}\rangle$ of a closed Kitaev chain with an even number of sites can be one of three qualitatively different types, each of which is realized in its own parameter domain:
\begin{eqnarray}
\label{Kitaev_Param_areas}
&\text{I})&~|\mu|<2t:~\Delta\Phi_{ZB} = \pi;~{\mathscr{M}}=-1;~|\Psi_{GS}\rangle = a_{0}^{+}|\Psi_{0}\rangle;
\nonumber\\
&\text{II})&~~\mu>2t:~\Delta\Phi_{ZB} = 0;~{\mathscr{M}}=1;~|\Psi_{GS}\rangle = a_{\pi}^{+}a_{0}^{+}|\Psi_{0}\rangle;\nonumber\\
&\text{III})&~\mu<-2t:~\Delta\Phi_{ZB} = 0;~{\mathscr{M}}=1;~|\Psi_{GS}\rangle = |\Psi_{0}\rangle;
\end{eqnarray}
Domain I, where a negative FP index of the ground state is realized, coincides with the TNPD, for which MMs are realized in a long nanowire.

\subsection{\label{sec2.4} Finite-size effects}

We discuss how the properties of MMs depend on the length of the chain. It was noted in \cite{kitaev-01, kao-14} that, in the context of studying MBSs, open-boundary chains with a very large but finite number of sites are fundamentally different from infinite or semi-infinite chains due to the need to consider at least two MMs simultaneously. In finite open chains, the MMs localized at opposite edges of the system tend to hybridize, which in most cases leads to the disappearance of gapless excitations. The Majorana polarization $\mathscr{P}$ introduced in \cite{sedlmayr-15, sedlmayr-16} can serve as a quantitative characteristic of the degree of overlap of the MM wave functions:
\begin{equation}\label{MP}
{\mathscr{P}} = \frac{2\left|\sum\limits_{l}{'} u^{*}_{l0}v_{l0}\right|}{\sum\limits_{l}{'}\left(u_{l0}^2 + v_{l0}^2\right)} =\frac{\left|\sum\limits_{l}{'}\left(w_{l0}^2 - z_{l0}^2\right)\right|}
{\sum\limits_{l}{'}\left(w_{l0}^2 + z_{l0}^2\right)},
\end{equation}
with $w_{l0}$ and $z_{l0}$ defined in (12). Here, the summation ranges the site numbers $l$ from only the left (or only the right) half of the chain, which is indicated by the prime on the summation symbol. If ${\mathscr{P}} \cong 1$, then the MM wave functions do not overlap, leading to the formation of quasiparticles with an exponentially low excitation energy. In the case ${\mathscr{P}}<1$, the MM wave functions overlap significantly. In the case of long chains, ${\mathscr{P}}\to 1$ in all of the TNPD. As $N$ decreases, the range of parameters for which ${\mathscr{P}} \cong 1$ decreases but lies within the TNPD. Moreover, $\varepsilon_{0} \sim 1 - \mathscr{P}$. Quasiparticle excitations with ${\mathscr{P}} \cong 1$ correspond to the generally accepted concept of MBSs described in Section \ref{sec2.1}.

In [61-67], the characteristics of the boundary states of the model of an open and finite Kitaev chain were analyzed analytically, which recently resulted in obtaining a complete analytic solution of the eigenvector and eigenvalue problem for the Bogoliubov-de Gennes Hamiltonian (16), (17) of such a system. It was shown that taking the open geometry into account leads to an oscillatory dependence of the minimum energy of one-particle excitations on the parameters of the system and the length of the chain. This effect manifests itself in the TNPD of the systems and is associated with the hybridization of the MM wave functions. Analytic methods were also used in [61-67] to obtain the conditions for the destructive interference of MMs, stating that the Fermi excitation energy must strictly vanish in the case of a finite chain. It was shown that the vanishing of the excitation energy is accompanied by a quantum phase transition with a change in the FP of the ground state. This effect can manifest itself in the measured characteristics, which we discuss in Section \ref{sec2.5}. It was demonstrated that FP oscillations are stable with respect to disorder and can manifest themselves in the dynamical features of MBSs. Subsequently, it was shown that hybridization of MMs can be realized not only through their direct overlap but also as a result of virtual processes of quasiparticle tunneling into a massive superconductor brought into contact with the wire [68].

\begin{figure}[h!]
	\begin{center}
		\includegraphics[width=0.6\textwidth]{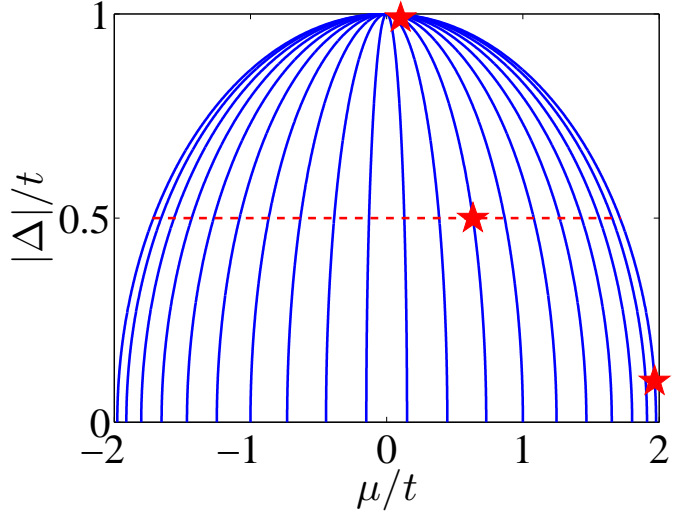}
		\caption{Lines determining values of the chemical potential and superconducting gap at which an MM exists for a Kitaev chain with $N=20$ sites.}\label{fig3}
	\end{center}
\end{figure}

We now discuss in more detail single-particle excitations whose energy strictly vanishes due to the hybridization of MMs. By analogy with MBSs, we call the edge modes with strictly zero excitation energy Majorana hybridization states (in a generalized sense to be explained below). They are realized in a parameter null set, and degenerate into MBSs in the case of long chains. But in the case of short chains, the conditions for their realization may differ. It is convenient to approach the problem of finding Majorana hybridization states in Kitaev's chain model by considering system of equations (14) with $\varepsilon_{0} = 0$:
\begin{eqnarray}
\label{wz_syst}
&~&-\mu z_{l0} - (t+|\Delta|)z_{l-1,0} - (t-|\Delta|)z_{l+1,0} = {0} ,
\nonumber \\
&~&\mu w_{l0} + (t-|\Delta|) w_{l-1,0} + (t+|\Delta|)w_{l+1,0} = {0} ,
\nonumber \\
&~&z_{00}=w_{N+1,0}=z_{N+1,0} = 0.
\end{eqnarray}
System (50) decouples into two independent subsystems for the coefficients  $w_{l0}$ and $z_{l0}$. The symmetry of the equations is such that $w_{l0}=Cz_{N-l+1,0}$. Solving this system, we find that zero-energy excitations exist on the lines (Fig. \ref{fig3})
\begin{eqnarray}
\label{mueq}
\mu = 2\sqrt{t^2-|\Delta|^2}\cos\phi_{m},~~\phi_{m} = \frac{\pi m}{N+1},~~m=1,...,N.
\end{eqnarray}
Nontrivial solutions exist in the entire parameter domain $|\Delta|/t>0$, $\mu=0$ for an odd $N$. The expression for the Fermi operator $\alpha_{0}$ corresponding to a zero mode on line $m$ then takes the form
\begin{eqnarray}
\label{d0_general}
\alpha_{0,m} = \frac{1}{S_m}\sum_{l=1}^{N} \big[ r^{l-1}\cdot\sin\phi_{m}l\cdot(\gamma_{lA} + i\gamma_{N+1-l,B}],
\end{eqnarray}
where $r = \sqrt{(t - |\Delta|)/(t + |\Delta|)}$ and $S_{m}$ is a normalization factor.

\begin{figure}[h!]
	\begin{center}
		\includegraphics[width=0.6\textwidth]{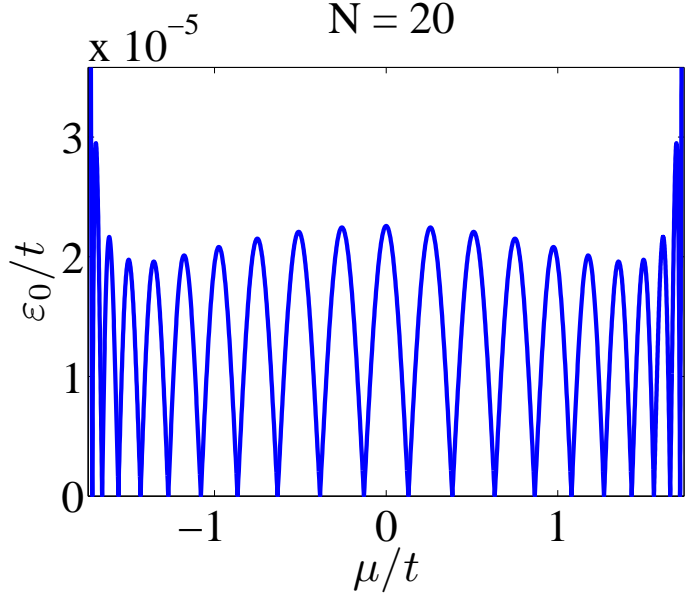}
		\caption{Dependence of the minimal energy of elementary excitation $\varepsilon_0$ in a Kitaev chain on the chemical potential $\mu$ at $N=20$ and $|\Delta|/t=0.5$ (dashed line in Fig. \ref{fig3})}\label{fig4}
	\end{center}
\end{figure}

The parameter space of the Kitaev model therefore contains $N$ lines at the points of which Fermi-type zero-energy excitations can exist. All such lines converge at the singular point $\mu = 0$, $t = |\Delta|$. In Fig. \ref{fig4}, we show the dependence of the minimum excitation energy for the parameters corresponding to the dashed line in Fig. \ref{fig3}. We can see that, at the intersection of the indicated lines of parameters at 20 points, $\varepsilon_{0}$ is strictly equal to zero, while between such points, $\varepsilon_{0}$ is exponentially small, $\varepsilon_{0} \sim r^{-N}$, but finite.

\begin{figure*}[h!]
	\begin{center}
		\includegraphics[width=0.48\textwidth]{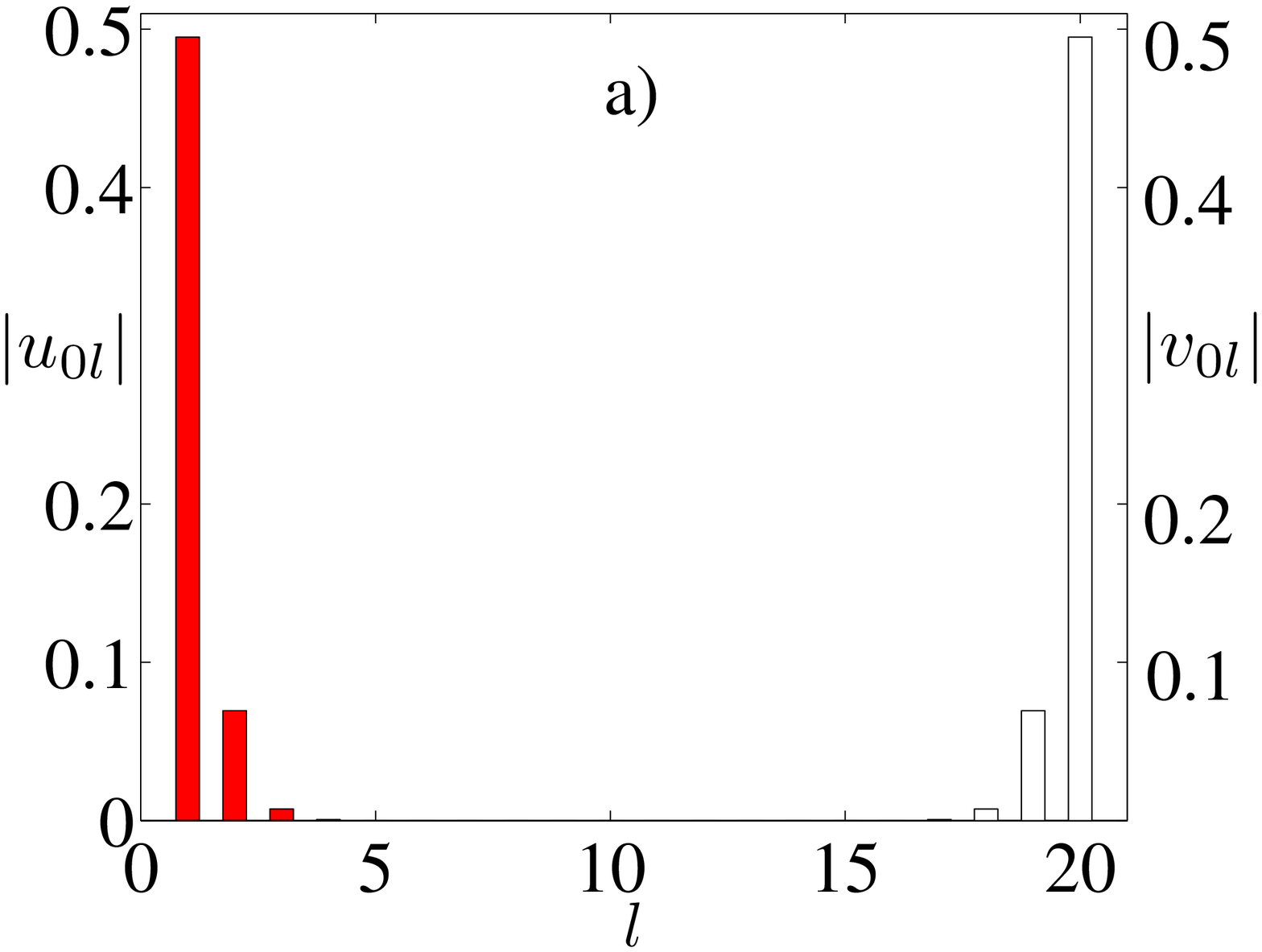}
		\includegraphics[width=0.48\textwidth]{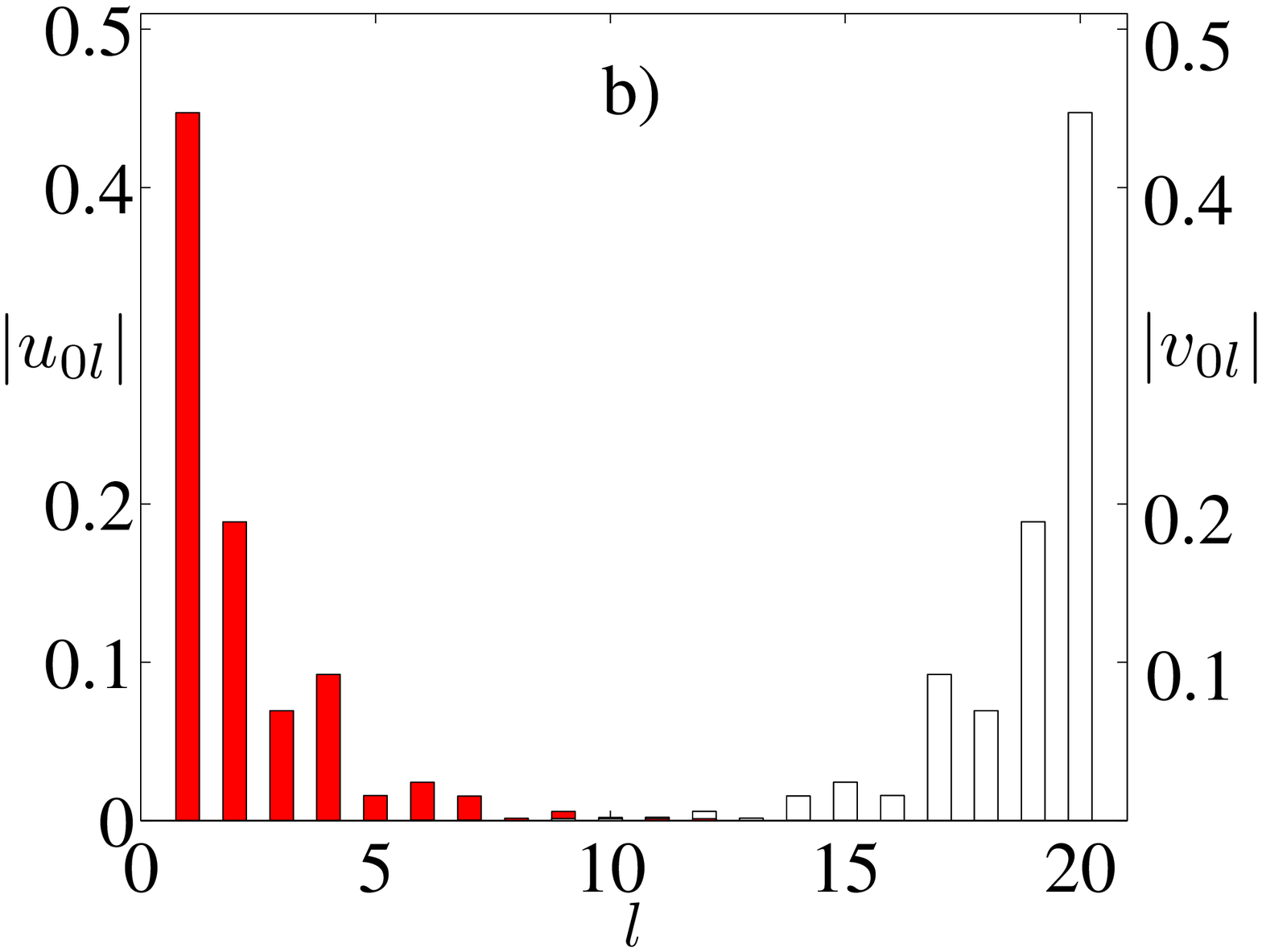}
		\includegraphics[width=0.48\textwidth]{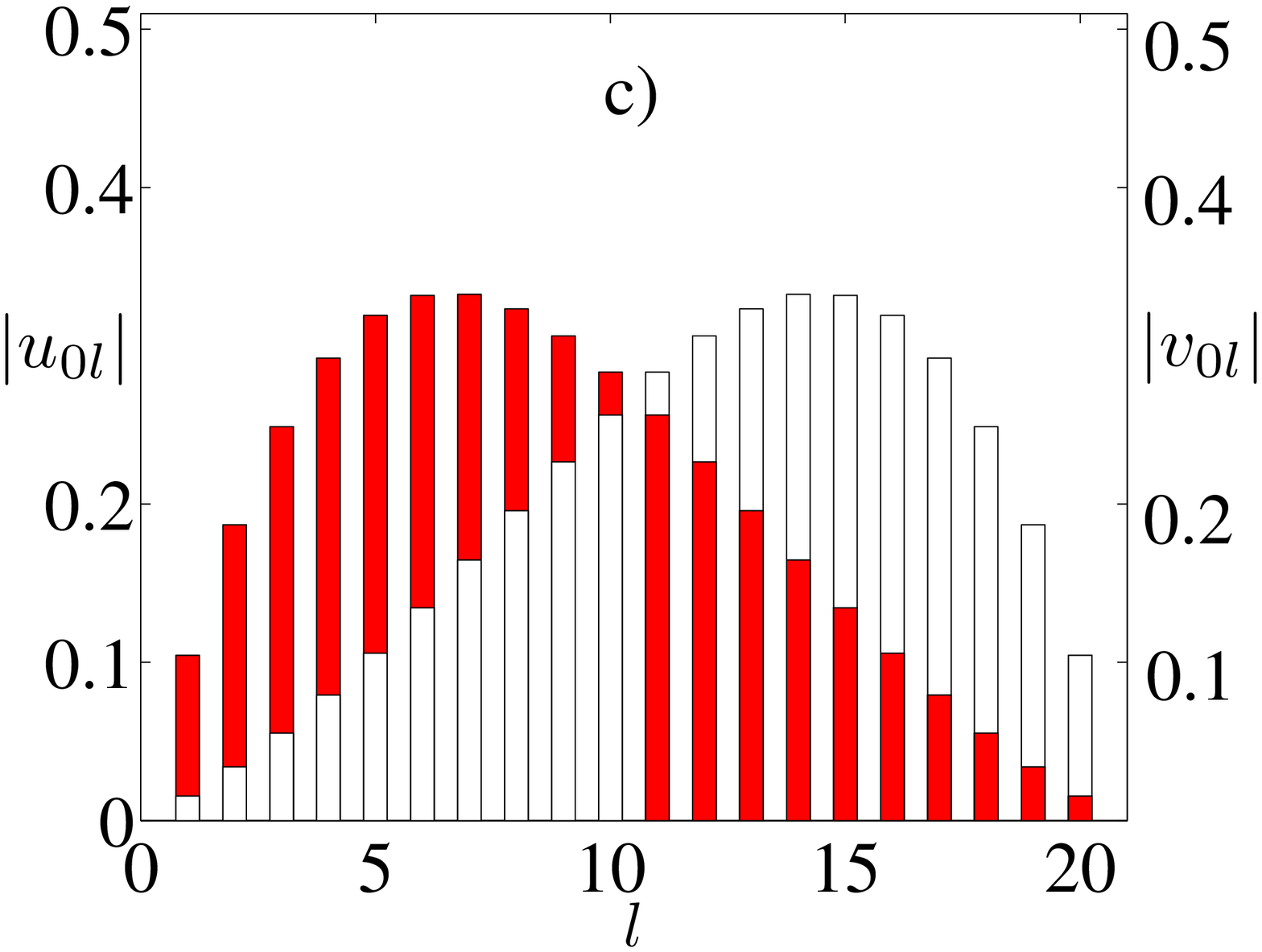}
		\caption{Spatial distribution of MM wave functions at $N=20$ for parameter values marked with stars in Fig. \ref{fig3}: (a) $|\Delta|=0.99t$, (b) $|\Delta|=0.5t$ and (c) $|\Delta|=0.1t$.}\label{fig5}
	\end{center}
\end{figure*}

When moving along the special lines in parameter space (51), the tendency of solutions to localize at the edges of the chain persists (Fig. \ref{fig5}a) because expression (52) is dominated by the exponential term $r^{l-1}$. But the smaller the ratio $|\Delta|/t$ becomes, the lower the degree of localization of the solution (Fig. \ref{fig5}b). For $|\Delta|/t<<1$, the solution delocalizes. This is most clearly seen on the extreme curve $m=1$, where the sum of the moduli of the coefficients ${w_{0l}^2+z_{0l}^2}$ has a maximum in the middle of the chain (Fig. \ref{fig5}c). We note that delocalized states with zero excitation energy also emerge due to the edge effects, because the excitation spectrum has a gap in a closed chain with $|\Delta|>0$.

Generally speaking, we used here the criterion for the realization of edge states in 1D systems proposed in [69]. A state described by a quasiparticle operator $\alpha_m$ is an edge state if, for the corresponding $u-v$ Bogoliubov coefficients (12)
\begin{eqnarray}
\label{Psi_surf}
u_{lm} = A_{m}e^{-\lambda_m\,l} + B_{m}e^{\lambda_m\,l} ;~~~
v_{lm} = C_{m}e^{-\chi_m\,l} + D_{m}e^{\chi_m\,l}
\end{eqnarray}
(where $A_{m}$, $B_{m}$, $C_{m}$, $D_{m}$ are constant coefficients), the real parts of the exponents are nonzero: $~Re(\lambda_m),~~Re(\chi_m)\neq0$. Otherwise, this state is nonedge (bulk). This criterion can be naturally generalized by taking the spin projection and the multiband nature of the system into account. It is applicable to 1D systems of any size; for semi-infinite chains, it coincides with the classical criterion for edge states (defined uniquely and rigorously):  $\lim_{l\rightarrow\infty}|u_{lm}|^2=0$;
$\lim_{l\rightarrow\infty}|v_{lm}|^2=0$. In addition, for 1D systems, the presented criterion for the realization of edge states unambiguously correlates with the need to find the energy of the state in the gap of the bulk spectrum.

\begin{figure}[h!]
	\begin{center}
		\includegraphics[width=0.9\textwidth]{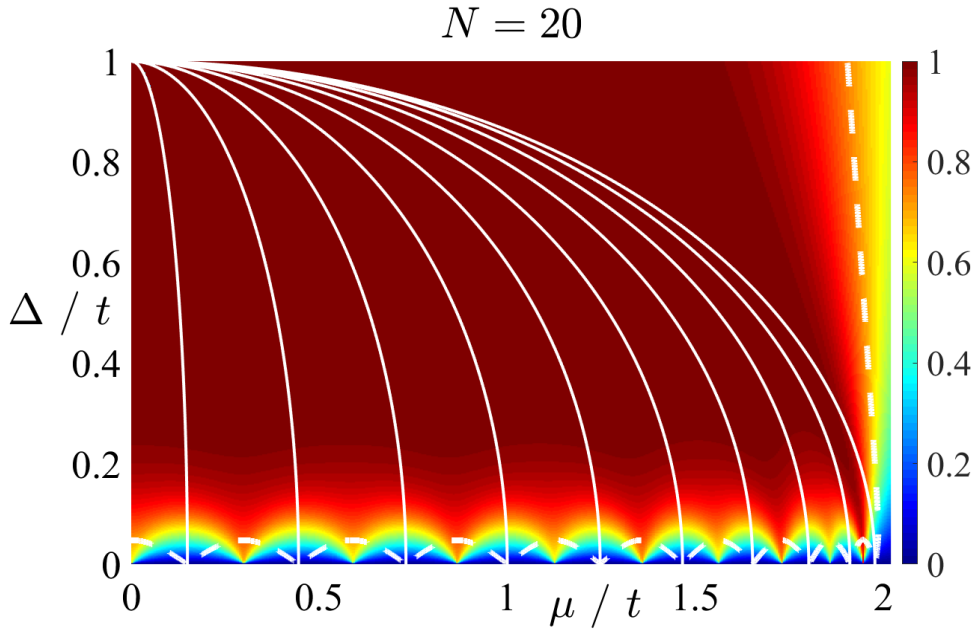}
		\caption{Map of the Majorana polarization $\mathscr{P}$ of a Kitaev chain with $N=20$ sites. Thin solid curves are lines of zero modes and changes in the FP $P$. Thick dashed lines bound the domain of the realization of edge states in the generalized sense proposed in [69].}\label{fig6}
	\end{center}
\end{figure}

For the Kitaev model, the boundaries in the parameter space that separate the conditions for the realization of edge and nonedge states are shown in Fig. \ref{fig6}. The finite size of the system leads, in addition to the previously noted realization of MMs only on special lines, to two effects. First, the boundaries of the realization of edge states, which in an infinitely long chain are given by $|\mu|=\pm2|t|$, become dependent on the superconducting gap $|\Delta|$. Second, at small values of $|\Delta|$, regions appear inside the domain where the edge state does not arise. These regions are located between the lines of Majorana hybridization modes. These regions in the parameter space exist because, as a result of a sufficiently strong overlap of edge excitations that tend to localize at opposite ends of the chain, the excitation energy is located in the bulk zone, and the nature of this excitation changes to the nonedge one. We can also see from Fig. \ref{fig5} that the domains where traditional $\mathscr{P} \cong 1$ MBSs are realized and the domains of Majorana hybridized states can differ (the cases $|\Delta|<<1$). Moreover, Majorana hybridized states can be realized in the case $\mathscr{P} << 1$. Therefore, in particular, Majorana hybridization states are not of interest for quantum computing. But their identification can be indicative of the conditions for the realization of a TNPD, which is discussed in Section \ref{sec2.5}.

\subsection{\label{sec2.5}Series of quantum transitions and caloric anomalies}

In addition to analyzing the fundamental features of MBSs in the Kitaev chain model, the options for detecting such states in practice have been actively studied. For example, in [70-72, 74, 75], various experimental possibilities of identifying topological phases and topological quantum transitions in the framework of Kitaev's chain model were considered. Two types of criteria were discussed. The first was based on the use of characteristics of the spectrum and eigenstates of a many-body system: the degeneration of the many-body ground state energy (for Hilbert space sectors with different FPs), the degeneration of the spectrum of the reduced many-body density matrix, and the asymptotic behavior of the one-particle density matrix at long distances were investigated [71-73]. Criteria of the second type were related to experimentally observed characteristics: the appearance of a peak in the differential conductance at zero bias [70], the anomalous behavior of compressibility and susceptibility [71, 72], magneto- and electrocaloric anomalies [74], and the appearance of Fano resonances in systems with nontrivial contact geometry [75]. The transport properties of quantum wires with MMs are discussed in detail in Sections \ref{sec3.3} and \ref{sec3.4} in the framework of a model that takes spin degrees of freedom into account.

Here, we briefly discuss the characteristic properties of the caloric effects in the TNPD of the Kitaev chain model, because the appearance of such effects is directly related to the realization of zero modes on special lines in the parameter space, which were discussed in Section \ref{sec2.4}.

The appearance of a zero mode on special parametric lines indicates the degeneracy of the many-body ground state. It is essential here that passing through such parametric points result in two states with different FPs replacing each other in their role of the ground state and that quantum transitions be realized in the system. Hence, the parametric lines of the realization of Majorana hybridization states in the system are parametric lines of quantum critical points. For an open Kitaev chain, the result of calculating the FP index and the Majorana polarization using formulas (23) and (49) allows constructing the phase diagram shown in Fig. \ref{fig6}. As can be seen from (51), as the number of sites $N$ increases, the number of FP-changing lines increases, and in the limit as $N\to \infty$ they form a quasicontinuum (a measure zero set) in the TNPD.

We thus see that the TNPD in the Kitaev chain model allows cascades of quantum transitions when the external parameters change. Such cascades can be detected using observable properties. In particular, the electron density
\begin{eqnarray}
\label{charge and spin}
\delta n_{l} = \langle 1| a^{+}_{l}a_{l}|1\rangle - \langle 0| a^{+}_{l}a_{l}|0\rangle = |u_{l0}|^2 - |v_{l0}|^2 = w_{l0}z_{l0}
\end{eqnarray}
is redistributed in the transition of a superconducting nanowire from the ground state $|0\rangle$ to the state $|1\rangle = \alpha_{0}^{+}|0\rangle$ with a filled MM. A similar argument can be adduced regarding the spin density when considering 1D systems with spin degrees of freedom. However, we believe that the TNPD can be detected most efficiently by measuring the caloric effects, because these are known to develop singularities when the system passes through quantum critical points.

Caloric effects (electric and magnetic) manifest themselves in a change in the temperature of the system with an adiabatic change in external parameters (in the case of a Kitaev chain model, this parameter is the electrochemical potential, and for wires with spin degrees of freedom, this can, for example, be the external magnetic field) and are defined as
\begin{eqnarray}
\label{MCE_ECE}
&~&{\left( {\frac{{\partial T}}{{\partial h}}} \right)_{S,\mu }} =  - T\Bigg(\frac {\partial \langle M \rangle/\partial T}{C(T)}\Bigg)_{\mu, h}; ~~~
{\left( {\frac{{\partial T}}{{\partial \mu}}} \right)_{S, h}} =  - T\Bigg(\frac {\partial \langle N \rangle/\partial T}{C(T)}\Bigg)_{\mu, h};
\end{eqnarray}
where $\langle N \rangle$, $\langle M \rangle$, and $C(T)$ are the electron concentration, specific magnetization, and specific heat capacity, and $S$, $\mu$, and $h$ are the entropy, chemical potential, and magnetic field strength.

\begin{figure}[h!]
	\begin{center}
		\includegraphics[width=0.7\textwidth]{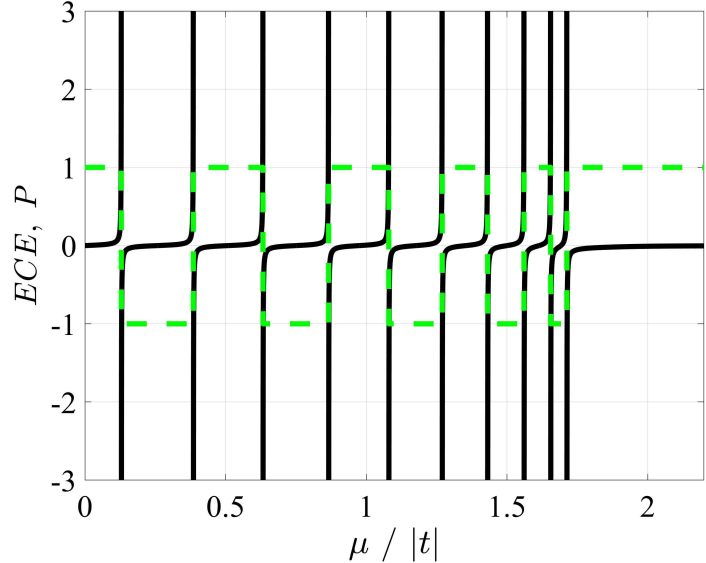}
		\caption{Dependence of the electrocaloric effect (ECE) (solid line) and FP index $P$ (dashed line drawn in accordance with formula (23)) of Kitaev chain at $\Delta/|t|=0.5$, $T/|t|=10^{-3}$, and $N=20$.}\label{fig7}
	\end{center}
\end{figure}

It was shown in [74] that for systems described by quadratic forms in Fermi operators, caloric effects must diverge at the quantum critical point and have practically zero magnitude far from it. This behavior is shown in Fig. \ref{fig7}. The dashed line shows the behavior of the FP index $P$ of the ground state of a nanowire. Solid lines show the dependences of the electrocaloric effect, exhibiting anomalous behavior at the points where P changes sign. Identifying these features can be a criterion for detecting the TNPD, in addition to the criteria already proposed in the literature (see, e.g., [75-78]).

\subsection{\label{sec2.6} Coulomb interaction effects}

\begin{figure}[h!]
	\begin{center}
		\includegraphics[width=0.7\textwidth]{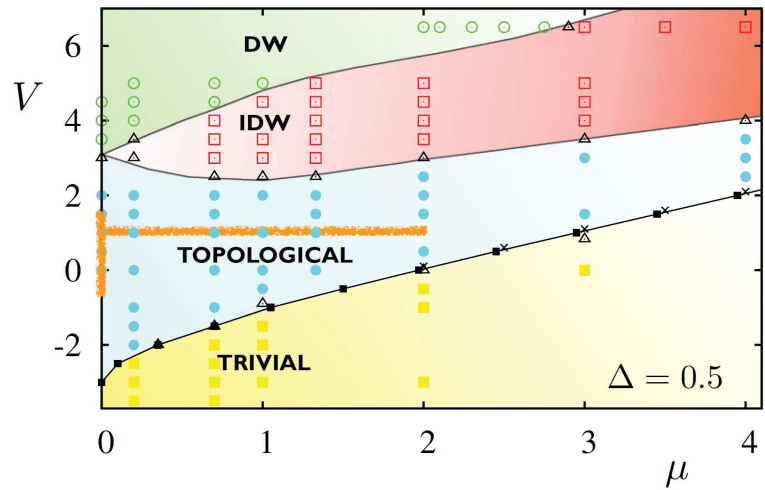}
		\caption{Phase diagram of a Kitaev chain with Coulomb interaction on neighboring sites in chemical potential-Coulomb interaction strength variables in accordance with formula (56) for $\Delta=0.5$ [70].}\label{fig8}
	\end{center}
\end{figure}

We briefly discuss the effect of the two-particle Coulomb interaction on the phase diagram of the Kitaev chain model. The Hamiltonian of this minimal model accounts for the intersite interaction of fermions (there is no single-site interaction, because the fermions are spinless) with amplitude $V$ and is given by
\begin{eqnarray}
\label{HamiltonianKitaev_V}
{H}_{K}  =  \sum_{l}\Big( (\epsilon - \mu)
a_{l}^{\dag}a_{l} -
t \left( a_{l}^{\dag}a_{l+1} + a_{l+1}^{\dag}a_{l} \right)+
\Delta a_{l}a_{l+1} + \Delta^* a_{l+1}^{\dag}a_{l}^{\dag} + V {n}_{l}{n}_{l+1} \Big),
\end{eqnarray}
where $n_{l} = a^{+}_{l}a_{l}$. It is now firmly established that the Coulomb interaction effect can lead to a violation of the topological classification or destruction of topological phases [79]. Moreover, the TIs introduced in studying a noninteracting fermionic system often become inapplicable to systems with interactions. But for 1D Fermi systems, a rigorous topological classification was given in [80-82], and properties were determined there that allow identifying topologically nontrivial phases. One such property is the multiple degeneration of the entanglement spectrum. In [70, 73], this approach was used together with the renormalization group method for the density matrix to construct the phase diagram of the Kitaev chain with the interaction shown in Fig. \ref{fig8}.

The criterion for the realization of a topologically nontrivial phase (`topological phase' in Fig. \ref{fig8}) was the degeneration of the energy of many-body states for Hilbert space sectors with different FPs. The trivial phase shown in Fig. \ref{fig8} is characterized by a nondegenerate ground state. Also, as a result of Coulomb repulsion, two different phases appear: commensurate and incommensurate charge density waves (CDWs) (see Fig. \ref{fig8}), whose topology is trivial. We see that electron correlations significantly change the range of parameters for the realization of topologically nontrivial phases, and a change in the strength of Coulomb interactions can lead to a sequence of topological phase transitions.

As noted in Section \ref{sec2.1}, taking hopping and superconducting pairing into account for next-to-nearest neighbors can lead to the realization of a topologically nontrivial phase with two or more pairs of MMs. The effect of Coulomb interactions on the structure of the phase diagram for such systems has not yet been studied. It should be expected, however, that these effects persist at a qualitative level for not too strong interactions.

In addition to studying the phase diagram, the effect of fermionic interactions on the structure of the ground state and spectral properties of the Kitaev chain were considered in [46-48, 83, 84]. In particular, analytic expressions for the energy and wave function of the ground state were obtained in selected regions of parameters: on a special line [83] and at a singular point [84], as was discussed in detail in Section \ref{sec2.1}. In the first case, the uniqueness of the ground state (up to a double degeneracy associated with the existence of MMs) and its adiabatic connection with the ground state of a noninteracting Kitaev chain in a topologically nontrivial phase were demonstrated. In this way, the ground-state wave function of the interacting system was constructed in a nontrivial phase.

In [84], for the singular point $\mu = 0$, $t = \Delta$, analytic expressions were obtained for many-body eigenfunctions and eigenvalues of the Hamiltonian, and a topological quantum transition was shown to occur when the coupling parameter changes. In addition, in [46-48, 83], the concept of MMs was generalized by taking many-body excitations into account. Many-body Majorana operators were then defined similarly to one-body operators, so as to be self-conjugate, commute with the Hamiltonian, and connect Hilbert space sectors with different FPs. It has been shown that in the regime of strong electron correlations, the contribution of many-body Majorana operators to the structure of a generalized MM can be significant [48].

We note that, in addition to the Kitaev chain model with four-fermion terms, other models of interacting 1D systems have been discussed that allow the existence of topological phases. Such models were studied in detail both analytically and numerically in [85, 86].

We also note paper [87], where the Kitaev chain model with strong intersite interaction was claimed to allow an effective realization in an array of Josephson junctions with varying electrical capacitances. This may open up additional possibilities for testing theoretical predictions regarding the effect of fermionic interactions on topological phases. In particular, in a recent experimental study [88], the Kitaev chain Hamiltonian was simulated by taking the nearest and next-to-nearest neighbors into account (under the condition $t =\Delta$ in both cases) using a qubit based on a superconducting ring. As the magnetic field was varied, the quasimomentum of the Hamiltonian under study effectively changed and the winding number $\nu$ (see (39)) of the Hamiltonian was measured. This treatment revealed the regimes of topologically trivial and nontrivial phases, in full agreement with theoretical predictions.

\subsection{\label{sec2.7} Effects of disorder and boundary conditions }

A contentious issue as regards the realization of MBSs is the role of defects and inhomogeneities. By defects and inhomogeneities, we mean random [72, 89] and regular [66, 90, 91] inhomogeneities, as well as different types of boundary conditions revealing finite-size effects [61-65]. For a Kitaev chain, it has been shown that a relatively weak disorder does not destroy the MM, keeping its excitation energy exponentially low. At that, the excitation energy of single-particle states which are not Majorana-like can decrease [72, 89]. This effect is of the first order of smallness in the magnitude of random disorder.

At the same time, it has been shown that the presence of disorder can significantly complicate the treatment of topological phases described in Sections \ref{sec2.2}-\ref{sec2.5}. This effect is especially significant when the disorder is accompanied by the fermion interaction in the system [72, 77]. The physical grounds for this complication are rooted in the realization of low-energy localized states that are similar to MBSs.

The most significant problem in identifying the MBSs in inhomogeneous 1D systems is the emergence of stable Andreev states, which have similarities in experimental manifestations with the MBSs. The occurrence of such excitations and the associated problems in experimental identification are discussed in more detail in Sections \ref{sec3.3} and \ref{sec3.4} when considering a superconducting nanowire with a strong spin-orbit Rashba coupling.

An important subject of study is the quantum dot-superconducting nanowire heterostructure, which is realized in experiments on ballistic transport with InAs or InSb semiconducting nanowires. The case of superconducting nanowire modeled by a Kitaev chain was treated analytically in recent study [66]. We also note that the properties of the Kitaev chain with disorder and with applied periodic and quasiperiodic potentials were discussed in [51]. The authors of [51] established a relation between the localization length of edge states in the system without a superconducting order parameter and the size of the superconducting gap required for the system to pass into a topologically nontrivial state.

The effect of boundary conditions on the formation of MMs was also studied in [65, 90-92], where a parameter was introduced that allows smoothly changing the type of boundary conditions in the system. In [65], one of the hopping and superconducting pairing amplitudes in the Kitaev model ($t$ and $\Delta$ in formula (1)) was normalized to a complex number $b$ whose modulus ranged from 0 to 1. The value $b=0$ ($b=1$) corresponds to open (periodic) boundary conditions, and the variation in phase also allowed considering antiperiodic boundary conditions. It was shown that a nonzero value of $b$ can be chosen such that MMs are realized. These effects are stable under the weak intersite Coulomb interaction of fermions. In [90, 91], infinite and closed Kitaev chains with a single impurity were considered. In the limit when the impurity produces an infinite-height potential barrier at a site, the model reduces to an open Kitaev chain.

In the studies mentioned above, the existence of Majorana states localized near an impurity was demonstrated, and the wave functions of these zero modes were found in an analytic form. It has been shown that excitations of this type can be realized in a topologically trivial phase of a homogeneous system and are stable under weak disorder. However, the numerical analysis carried out in [92] showed that the MMs localized near the impurity do not significantly change the transport features of the system. We also note that the concept of Majorana polarization was introduced in the study of finite-size 1D and quasi-1D topological superconductors [93, 94]. Majorana polarization allows identifying MBSs in open systems of the indicated type.

We also note that the Kitaev model can be reduced to an anisotropic Ising chain model by the Jordan-Wigner transformation [12, 50, 95]. The topological phases with a definite number of nonlocal Majorana states are then mapped into different magnetically ordered phases, with the trivial phase of the fermionic model corresponding to the phase of the spin system without short-range correlations. This isomorphism is often used in the study of model spin systems to obtain information on the properties of topological superconductors.

In [50], in particular, the Ising chain in a transverse magnetic field with three-spin interactions was investigated, which reduces to the Kitaev model with long-range hopping and superconducting pairing. The phase diagram for that system was constructed and the possibility of inducing several MMs was shown. This is consistent with the above results. It has been shown that, in the phase with an even number of MMs, there are lines in the parameter space such that, in crossing them, the oscillatory decay of the MM wave functions changes to a monotonic decay. A similar analysis of the extended 1D Ising model was carried out in [95], where the topological phase diagram of an isomorphic model and the structure of the spin wave function in various domains of the phase diagram were studied.

\section{\label{sec3}Majorana modes with spin degrees of freedom}

The simplicity of the Kitaev chain Hamiltonian allows transparent demonstrations of the fundamental features of elementary Majorana-type excitations in condensed media. One of the characteristic features of the Kitaev model is the absence of a spin variable. This is not a significant limitation when considering systems with $p$-type superconducting pairing. But from a practical standpoint, the more common condensed media are those with $s$ and $d$ symmetry types or with a chiral $d+id$ symmetry type of the order parameter. For such systems, several scenarios for the realization of MMs and methods for their experimental identification have been proposed.

In Sections \ref{sec3.1}-\ref{sec3.4} and \ref{sec4}, we present the results of studies of the conditions for the realization of MMs in models that take spin variables into account. Various techniques are considered for detecting these states, based on the study of the transport properties of systems including superconducting nanowires with spin-orbit coupling. The main idea of these methods is to detect the unique properties of MBSs: zero energy and spatial nonlocality.

\subsection{\label{sec3.1}Superconducting nanowire model}

A physical example of a topological superconductor with the $s$ symmetry type of order parameter is a semiconducting InSb or InAs nanowire, in which a strong spin-orbit coupling is realized. An aluminum layer 3-5 nm in thickness is epitaxially deposited on the wire. At low temperatures, the aluminum shell passes into the superconducting state, and a superconducting pairing potential is induced in the nanowire due to the proximity effect. Such a structure is referred to as a superconducting wire (SW). Important changes in the properties of SWs occur in a magnetic field [35, 36]. In the tight-binding approximation, the SW Hamiltonian can be written in the form
\begin{eqnarray}
\label{Ham_latt}
&~&{{H}_{W}} = \sum_{l,\sigma}\Big(\xi_{\sigma}a^{+}_{l\sigma}a^{}_{l\sigma} - \frac{t}{2} \left( a^{+}_{l\sigma}a^{ }_{l+1\sigma} + a^{+}_{l+1\sigma}a^{ }_{l\sigma}\right) \Big) + \nonumber\\
&+&\sum_{l}\Big(\Delta a_{l\uparrow}a_{l\downarrow} - \frac{\alpha}{2} \left( a^{+}_{l\uparrow}a^{ }_{l+1\downarrow} - a^{+}_{l\downarrow}a^{ }_{l+1\uparrow} \right) + h.c.\Big)+ \nonumber\\
&+&\sum_{l}\Big(Un_{l\uparrow}n_{l\downarrow} + V{n}_{l}{n}_{l+1}\Big).
\end{eqnarray}
where the terms in the first sum correspond to the 1D system of fermions with the hopping integral $t/2$, the on-site fermion energy depending on the spin projection measured from the level of the chemical potential $\mu$, $\xi_{\sigma}=\epsilon_{0}-\mu+\eta_{\sigma} h$, $h = \frac{1}{2}g \mu_B H$, $g$ is the Lande factor, $\mu_B$ is Bohr's magneton, $H$ is an external magnetic field, $a_{l\sigma} (a^{+}_{l\sigma})$ is the operator of annihilation (creation) of a fermion at the $l$th site with the spin projection $\sigma = \uparrow,~\downarrow$, $\eta_{\uparrow}=1$, and $\eta_{\downarrow}=-1$. The terms in the second sum are associated with the superconducting pairing potential $\Delta$ and the Rashba spin-orbit coupling with the parameter $\alpha$.

For generality, terms corresponding to the on-site ($U$) and intersite ($V$) Coulomb interaction of fermions have been added to the Hamiltonian. Next, ${ n}_l = { n}_{l\uparrow} + { n}_{l\downarrow}$ is the operator of the number of electrons on a site, with ${ n}_{l\sigma} = a^{+}_{l\sigma}a^{}_{l\sigma}$. We emphasize that the realization of MMs requires a large value of the $g$-factor ($|g|\simeq 50$). In magnetic fields up to $H \sim $ 1 T, superconductivity is then not destroyed and the SW can be in a topologically nontrivial phase.

We analyze the conditions under which the SW model with $U=V=0$, Eqn (57), can be reduced to the Kitaev model [35, 36]. For this, we write Hamiltonian (57) in the momentum representation,
\begin{eqnarray}
\label{Ham_latt_k}
&~&{{H}_{W}} = \sum_{k\sigma}\xi_{k\sigma}a^{+}_{k\sigma}a^{}_{k\sigma}
+\sum_{k}\left\{i\alpha_k a^{+}_{k\downarrow}a_{k\uparrow} +
\Delta a_{k\uparrow}a_{-k\downarrow}+ h.c. \right\},
\end{eqnarray}
where $\xi_{k\sigma}=\varepsilon_{k\sigma}-\mu,~~~
\varepsilon_{k\sigma}=\epsilon_{0}+\eta_{\sigma} h-t\cos k, ~~~\alpha_k=\alpha \sin k$.

We introduce new operators
\begin{eqnarray}
\label{d_p_Oprs}
&~&d_{k} = \cos\phi_k\cdot a_{k\downarrow}-i \sign(\alpha_k)\sin\phi_k\cdot a_{k\uparrow},\nonumber\\
&~&p_{k} = \sin\phi_k\cdot a_{k\downarrow}+i \sign(\alpha_k)\cos\phi_k\cdot a_{k\uparrow},\nonumber\\
&~& \cos\phi_k=\sqrt{(1+r_k)/2},~~\sin\phi_k=\sqrt{(1-r_k)/2},~~r_k=h/\sqrt{h^2 + \alpha_{k}^2},
\end{eqnarray}
and express $H_{W}$ in terms of them:
\begin{eqnarray}
\label{H_dp_repr}
{H}_{W} &=&  \sum_{k}\big[
(\varepsilon_{-}(k)-\mu)d^{+}_{k}d_{k} + \frac12\left(\Delta_{k} d^{+}_{k}d^{+}_{-k} + \Delta_{k}^{*}d_{-k}d_{k} \right)\big] + \nonumber\\
&+&\sum_{k}\big[
(\varepsilon_{+}(k)-\mu)p^{+}_{k}p_{k} - \frac12\left(\Delta_{k} p^{+}_{k}p^{+}_{-k} + \Delta_{k}^{*}p_{-k}p_{k} \right)\big] + \nonumber\\
&+&  \sum_{k} \left( A_{k} d^{}_{-k}p^{}_{k} +A_{k}^{*} p^{+}_{k}d^{+}_{-k} \right).
\end{eqnarray}
We here use notation\\
$\varepsilon_{\mp}(k)=\varepsilon(k)\mp\sqrt{(h^2+\alpha_k^2)},~~
\Delta_{k}= i\Delta\,\alpha_k/(\sqrt{h^2+\alpha_k^2}),$\\
$A_{k}~=~i\sign(\alpha_k)\Delta\, r_k$.\\
We see that, in terms of the new variables, the SW Hamiltonian describes two Fermi subsystems, and the parameter $A_{k}$ determines the coupling between them.

If $h > t$ and  $t >> |\alpha|$ then the $|\Delta|$ band splitting is such that the bottom of the upper band is above the ceiling of the lower band. In this case, for $\mu~<~t-h$, the properties of the SW are determined by the lower band. Therefore, in the leading approximation in the parameter $|\Delta|/t\ll 1$, the considered nanowire is described by the Kitaev chain Hamiltonian
\begin{eqnarray}
\label{H_d_tff}
{H}_{W} &=&  \sum_{k}\big[
(\varepsilon_{-}(k)-\mu)d^{+}_{k}d_{k} + \frac12\left(\Delta_{k} d^{+}_{k}d^{+}_{-k} + \Delta_{k}^{*}d_{-k}d_{k} \right)\big].
\end{eqnarray}
It is easy to verify that in the range $-t-h <\mu~<~t-h$, the ground state of the system has a negative FP and the SW is in a topologically nontrivial phase. In this case, MBSs are realized in the SW in an open geometry.

Qualitatively, the same behavior is also to be observed in a relatively weak magnetic field $h<t$, when the energies of the lower states of the $p$-subband and of the upper states of the $d$-subband overlap. In this case, for $\mu < h-t$, with the electron concentration $n < 1$ and the upper band practically unfilled, the SW is still effectively described by the lower band.

If the electron concentration is greater than unity and $\mu > t-h$, then the lower band is completely filled (up to corrections due to contributions proportional to $|\Delta_k|^2, |A_k|^2$). Then, the effective Hamiltonian of the SW is the Kitaev chain Hamiltonian for the upper band,
\begin{eqnarray}
\label{H_dp_repr}
{H}_{W} &=&  \sum_{k}\big[
(\varepsilon_{+}(k)-\mu)p^{+}_{k}p_{k} - \frac12\left(\Delta_{k} p^{+}_{k}p^{+}_{-k} + \Delta_{k}^{*}p_{-k}p_{k} \right)\big],
\end{eqnarray}
describing an electron ensemble with the electron concentration $n_p=n-1$.

Thus, the properties of SWs in sufficiently strong magnetic fields can be described by the Kitaev model in a wide range of parameters. At the same time, we emphasize that, for $h < t$ and $h-t<\mu < t-h$, taking the two-band structure into account is essential. In what follows, the transport properties of an SW are discussed in the framework of both the original model (57) and the Kitaev model.

\subsection{\label{sec3.2} Topological phases and Majorana modes in a superconducting nanowire}

An SW has the electron-hole symmetry and is characterized by broken rotation invariance in spin space. According to the classification of topological insulators and superconductors [9], this corresponds to the symmetry class BDI, characterized by the $\mathbb{Z}$ invariant, whose parity is expressed in terms of the FP of the ground state of a closed SW with an even number of sites and periodic boundary conditions.

When solving for the FP, it suffices to consider the operator structure of the Hamiltonian at symmetric points of the Brillouin zone [96, 97]. Such points (denoted by $K$ in what follows) satisfy the condition $-K+G=K$, where $G$ is a vector of the reciprocal lattice. As usual, points of the Brillouin zone are assumed coincident if they differ by the vector $G$. For the SW, there are two symmetric points, $K=0$ and $K=\pi$.

To substantiate the foregoing, we segregate the terms in (58) that are related to the symmetric points:
\begin{eqnarray}
\label{H_latt_k_repr}
{\mathscr{H}}=\sum_K{h}(K)+\sum\limits_{k \neq 0, \pi}{h}(k),
\end{eqnarray}
where
\begin{eqnarray}
\label{h_form}
{h}(K)&=&\sum_{\sigma}\xi_{K\sigma}{n}_{K\sigma}+\left(\Delta a_{K\uparrow}a_{K\downarrow}+h.c.\right),~~~ K=0,~\pi,\nonumber \\
{h}(k)&=&\sum_{\sigma}\xi_{k\sigma}\hat{n}_{k\sigma}+
\left(i\alpha_k a^{+}_{k\downarrow}a_{k\uparrow}+
\Delta a_{k\uparrow}a_{-k\downarrow}+h.c.\right).
\end{eqnarray}
The term corresponding to the spin-orbit coupling is absent in ${h}(K)$ because $\alpha_K$ vanishes at the symmetric points.

It follows from the structure of quadratic forms (64) that, for different magnetic fields, the ground state function can be represented in the form (with $\epsilon_0 = 0$ here and hereafter)
\begin{eqnarray}
\label{fun_gr_st}
|\Psi_0\rangle = \left\{
\begin{array}{rcl}
|\Psi^{(I)}_0\rangle& =&L_0L_{\pi}|\Phi\rangle,~~h<H^{-},~~h<H^{+};\\
|\Psi^{(II)}_0\rangle& =&L_0a^+_{\pi\downarrow}|\Phi\rangle,~~~~H^{-}<h<H^{+};\\
|\Psi^{(III)}_0\rangle& =&a^+_{0\downarrow}L_{\pi}|\Phi\rangle,~~~~H^{+}<h<H^{-};\\
|\Psi^{(IV)}_0\rangle& =&a^+_{0\downarrow}a^+_{\pi\downarrow}|\Phi\rangle,~~h>H^{-},~~h>H^{+},
\end{array}
\right.
\end{eqnarray}
where the characteristic magnetic field magnitudes are
\begin{eqnarray}
\label{H_char}
H^{-}=\sqrt{(\mu-t)^2+|\Delta|^2},~~~H^{+}=\sqrt{(\mu+t)^2+|\Delta|^2},
\end{eqnarray}
and $L_K$ is given in the operator form well known from the Bardeen-Cooper-Schrieffer theory:
\begin{eqnarray}
\label{def_L_K}
L_K = u_K-v_Ka^+_{K\downarrow}a^+_{K\uparrow},~~u_K=\sqrt{(1+x_K)/2},\nonumber\\
v_K=\frac{|\Delta|}{\Delta}\sqrt{(1-x_K)/2},~~x_K=\frac{\xi_K}{\sqrt{\xi_K^2+|\Delta|^2.}}.
\end{eqnarray}
The vector $|\Phi\rangle$ in $(\ref{fun_gr_st})$ can be represented as
\begin{eqnarray}
|\Phi\rangle=U^+|0\rangle,~~~~~~ U^+=\prod_{0 < k < \pi}U^+_{k},~~~~~~~~~~~~~~~~~~~~~~~~\\
U^+_{k} = A_{k} + B_{k}a^{+}_{-k\downarrow}a^{+}_{k\uparrow} +
C_{k}a^{+}_{k\downarrow}a^{+}_{-k\uparrow} +
D_{k}a^{+}_{-k\uparrow}a^{+}_{k\uparrow} + F_{k}a^{+}_{-k\downarrow}a^{+}_{k\downarrow} +
G_{k}a^{+}_{-k\downarrow}a^{+}_{k\uparrow}a^{+}_{k\downarrow}a^{+}_{-k\uparrow}\nonumber.
\end{eqnarray}
The vacuum state of the SW is denoted by $|0\rangle$. The coefficients entering the definition of the operator $U^+_k$ are easy to find from the solution of the Schroedinger equation. However, this is not needed to find the FP of the ground state, because, for any parameters of the system, $|\Phi\rangle$ has the form of a superposition of states with an even number of fermions. Therefore, the FP of $|\Phi\rangle$ is positive and does not change when the parameters change.

\begin{figure}[h!]
	\begin{center}
		\includegraphics[width=0.7\textwidth]{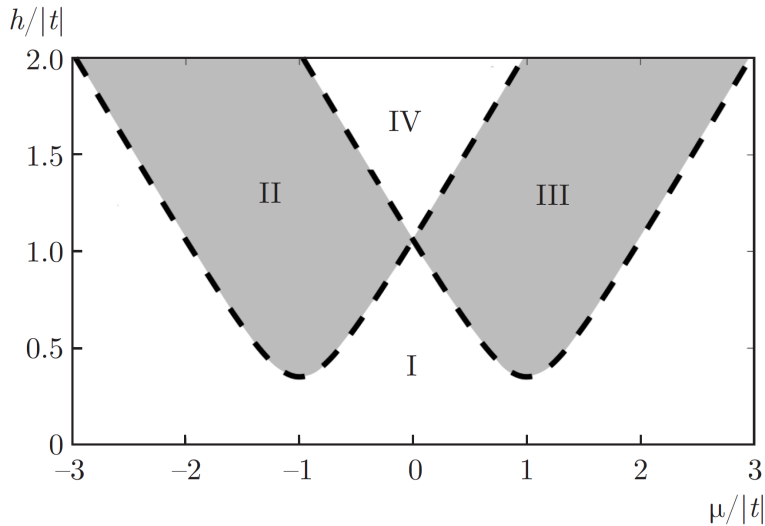}
		\caption{Diagram of a topologically nontrivial phase of a closed nanowire in chemical potential-external magnetic field variables. Parameter domain with $\mathscr{M}=\pm1$ corresponds to the trivial/nontrivial topological phase. Majorana number $\mathscr{M}=(-1)^{N_-}$, where $N_-$ is the parity of the number of fermionic modes with momenta $k=0,\pi$ and negative energy. The chosen parameter values are $\Delta=0.35t$ and $\alpha=0.22t$.}\label{fig9}
	\end{center}
\end{figure}

For a similar reason, the quadratic forms $L_K$ acting on $|\Phi\rangle$ and changing the state of the system preserve the positive FP. Therefore, the FP of $|\Psi^{(I)}_0\rangle$ is positive. It is also obvious that $|\Psi^{(IV)}_0\rangle$ has a positive FP. The states $|\Psi^{(II)}_0\rangle$ and $|\Psi^{(III)}_0\rangle$ have a negative FP, because the fermion creation operator entering the definition of these functions generates a superposition of states with an odd number of electrons when acting on $|\Phi\rangle$ and $L_K|\Phi\rangle$. It then follows that the conditions for the realization of states with a negative FP can be represented by the inequalities
\begin{eqnarray}
\label{eq_top}
\sqrt{(\mu-t)^2+|\Delta|^2}<h<\sqrt{(\mu+t)^2+|\Delta|^2},~~\mu > 0,\nonumber\\
\sqrt{(\mu+t)^2+|\Delta|^2}<h<\sqrt{(\mu-t)^2+|\Delta|^2},~~\mu < 0,
\end{eqnarray}
which determine the domain of the existence of a topologically nontrivial phase of the SW. In Fig. \ref{fig9}, this domain is shown in grey.

\begin{figure*}[h!]
	\begin{center}
		\includegraphics[width=0.48\textwidth]{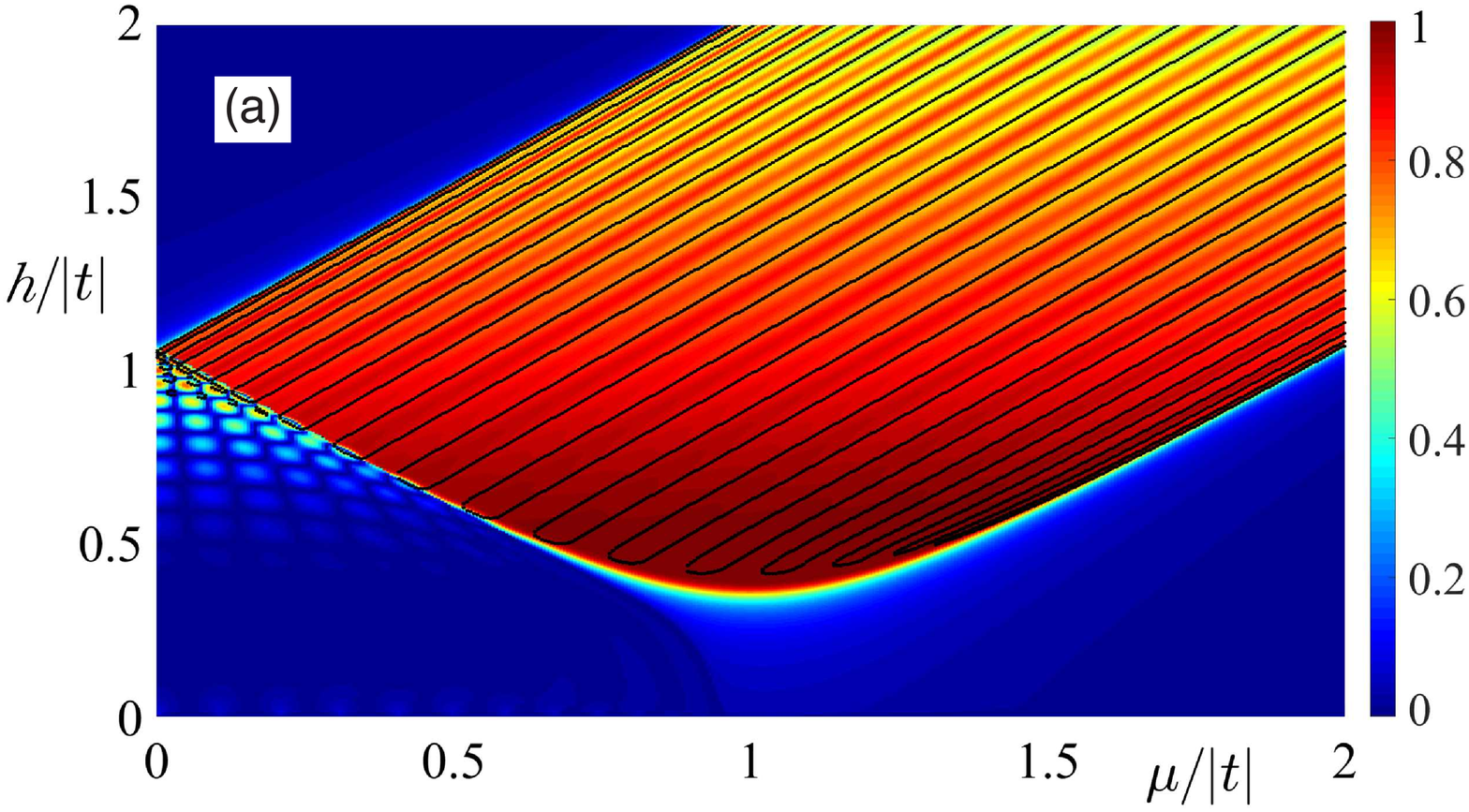}
		\includegraphics[width=0.48\textwidth]{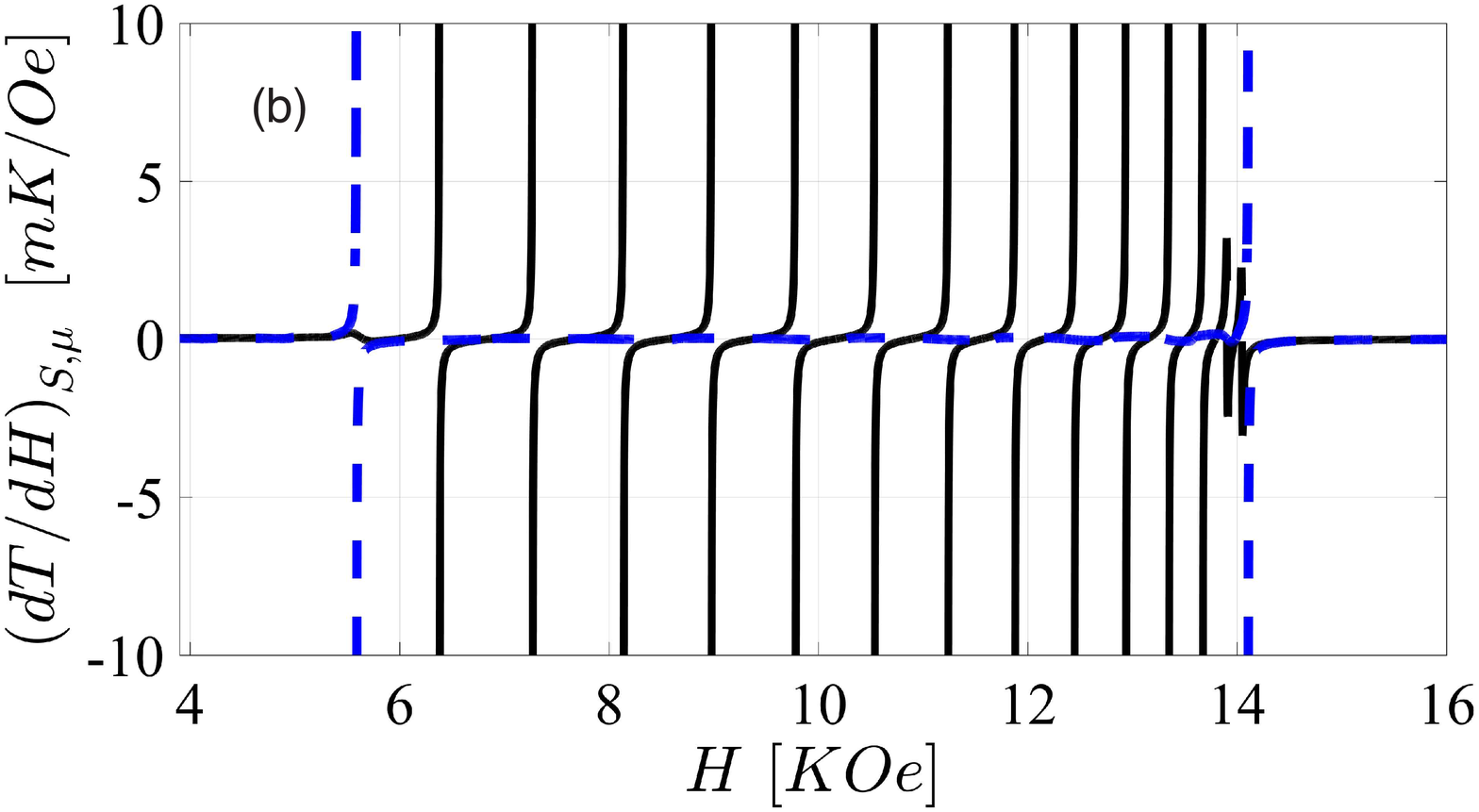}
		\caption{(a) Dependence of Majorana polarization ${\mathscr{P}}$ on magnetic field $h$ and chemical potential $\mu$ for an open chain at $N=30$, $\Delta=0.35|t|$, and $\alpha=0.22|t|$. Black lines show zero modes. (b) Dependence of the magnetocaloric effect on the external magnetic field at $T=10^{-3}|t|$ and $\mu/|t|=0.5$.}\label{fig10}
	\end{center}
\end{figure*}

In the diagonal representation with respect to the quasiparticle operators, the Hamiltonian becomes
\begin{eqnarray}
\label{H_k_alpha_repr}
{{H}}_{W} = E_{G} + \sum\limits_{K} \left\{|E_{K} - h|\alpha^{+}_{K}\alpha_{K}+
(E_{K} + h)\beta^{+}_{K}\beta_{K}\right\} \nonumber\\
+ \sum\limits_{k \neq K} \left [E_{-}(k)\alpha^{+}_{k}\alpha_{k} + E_{+}(k)\beta^{+}_{k}\beta_{k} \right],
\end{eqnarray}
where $E_G$ is the ground state energy of the SW, $E_{K} = \sqrt{\xi_{K}^2 + |\Delta|^2}$, and the quasiparticle energies $E_{\pm}(k)$ at nonsymmetric points are given by
\begin{eqnarray}
\label{Wire_spectrum}
&~&E_{\pm}(k) = \sqrt{\xi_{k}^2+{\alpha}_{k}^{2} + h^{2} + |\Delta|^2 \pm 2B_{k}},\nonumber\\
&~&B_{k} = \sqrt{\xi_{k}^{2}({\alpha}_{k}^{2} + h^{2}) + h^{2}|\Delta|^2}.
\end{eqnarray}

In long SWs with the parameters satisfying conditions (69) in the open geometry, MBSs with exponentially low MM energies and overlapping wave functions of these modes are realized. In particular, for such MBSs, the Majorana polarization is close to its maximum value ${\mathscr{P}} \cong 1$. As the length of the chain decreases, the domain of realization of MBSs with ${\mathscr{P}} \cong 1$ diminishes due to hybridization of the MM wave functions. For the same reason, the Majorana hybridization states realized on special lines in the parameter space exist in the nanowire. In passing through the points corresponding to the realization of such Majorana hybridization states in the phase space, the system undergoes quantum transitions with a change in the FP of the ground state, to be revealed in the study of caloric anomalies. For a short chain, this situation is shown in Fig. \ref{fig10}. The calculations were performed using the Bogoliubov-de Gennes matrix (see (16) and Sections \ref{sec2.2} and \ref{sec2.3}) with
\begin{eqnarray}
\label{Ham_kK}
{\hat{A}}  =  \left( {\begin{array}{*{20}{c}}
	{\hat{A}}_{\uparrow\uparrow}&{\hat{A}}_{\uparrow\downarrow}\\
	{\hat{A}}_{\uparrow\downarrow}^{+}&{\hat{A}}_{\downarrow\downarrow}\\
	\end{array}} \right);\hphantom{aa}
{\hat{B}}  =  \left( {\begin{array}{*{20}{c}}
	{\hat{B}}_{\uparrow\uparrow}&{\hat{B}}_{\uparrow\downarrow}\\
	-{\hat{B}}^{T}_{\uparrow\downarrow}&{\hat{B}}_{\downarrow\downarrow}\\
	\end{array}} \right),
\end{eqnarray}
where
\begin{eqnarray}
\label{A_B}
\left( {\hat{A}}_{\uparrow \downarrow} \right)_{l,l+1} &=& - \left( {\hat{A}}_{\uparrow \downarrow} \right)_{l+1,l} =
-\alpha/2,
~~\left( {\hat{B}}_{\uparrow \downarrow} \right)_{l,l} = - \Delta, \nonumber\\
\left( {\hat{A}}_{\sigma \sigma} \right)_{l,l} &=& -\mu + \eta_{\sigma} h,~
\left( {\hat{A}}_{\sigma \sigma} \right)_{l,l+1} = -t/2.
\end{eqnarray}
Expression (49) for the Majorana polarization can then be naturally generalized to the case where the $u-v$ and $w-z$ coefficients depend on spin variables. The caloric characteristics of the SW were considered for $\alpha/|t| \simeq 0.2$ and $\Delta/|t| \simeq 0.3$.

\subsection{\label{sec3.3} Principles for the identification
	of Majorana bound states and experimental studies}

One of the central tasks of Majorana studies is a reliable experimental confirmation of the realization of MBSs in condensed media. The identification problem is solved by using the characteristic features of states containing elementary Majorana-type excitations. In this section, we discuss modern approaches to the detection of such states. Most of the techniques are based on considering the features of quantum transport in structures with SWs.

\subsubsection{\label{sec3.3.1} Peak conductance measurement at zero voltage}

\begin{figure}[ht]
	\begin{center}
		\includegraphics[width=1\textwidth]{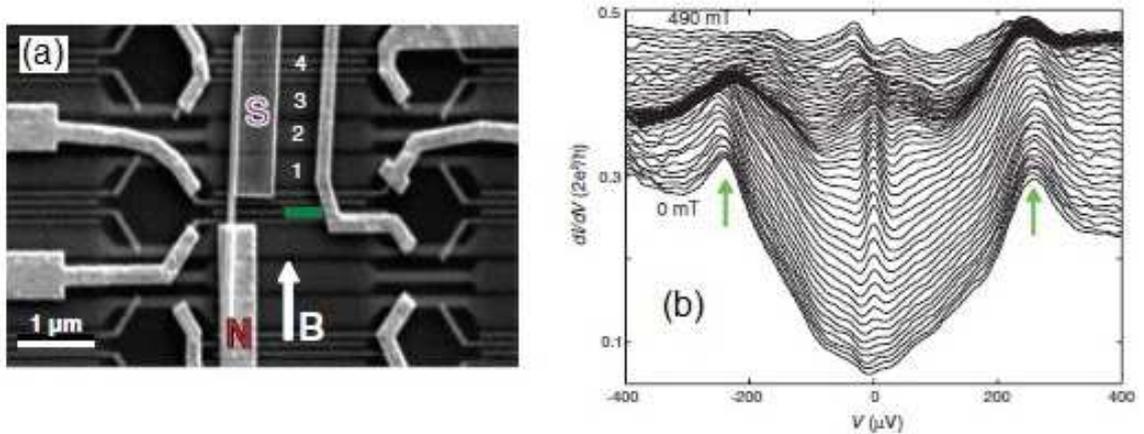}
		\caption{\label{mourik-12}  (a) Image of an InSb wire that is partly in the domain of normal, $N$, and partly superconducting, $S$, contacts in a magnetic field  $\textbf{B}$. 1-4 are the gate electrodes.
			(b) Conductance peaks occurring at zero voltage as the magnetic field increases  \cite{mourik-12}.}
	\end{center}
\end{figure}

After the discovery, under certain conditions, of an isomorphism between the 1D SW model considered in Section \ref{sec3.1}
and the Kitaev chain \cite{lutchyn-10,oreg-10},
experimental results on tunneling spectroscopy were obtained whose interpretation was related to the realization of MBSs in InAs and InSb semiconducting wires \cite{mourik-12,deng-12,das-12}.
In those studies, the differential conductance
$G=dI/dV$ of the contact between a paramagnetic metal and a hybrid semiconducting wire with a strong spin-orbit coupling was investigated (Fig. \ref{mourik-12}a).
The term 'hybrid' means a composite structure with a superconducting pairing potential in a semiconducting wire induced by the proximity effect due to contact with a massive superconductor. This structure is what we call the SW in what follows.

To clarify the features of the conductance that were discovered in the cited experimental studies, we discuss the metal-superconductor contact in more detail. At low voltage and low temperatures, the current is determined only by electron states with energies $\varepsilon$ near the Fermi level and $\varepsilon\ll\Delta$ (where $\Delta$ is the superconducting pairing strength). In this case, the ground state of the superconductor is described by the wave function given by a superposition of states with an even or an odd number of particles, and therefore the transported electron can be present in the superconductor with a nonzero probability only as part of a Cooper pair. This is called Andreev reflection \cite{andreev-64}. In this case, the electron is reflected from the interface into the normal metal in the form of a hole with an opposite spin projection. In the presence of a potential barrier in the contact region, there is also a nonzero probability of conventional backscattering of the electron. The probability of Andreev reflection is inversely proportional to the height of the potential barrier between the subsystems \cite{blonder-82}.

In the case of a topologically nontrivial phase, the ground state of the superconductor becomes doubly degenerate: there are two many-body wave functions with the same energy but different parities. In other words, an electron with the Fermi energy can also be present in the superconductor only as a result of Andreev reflection. However, this process now goes via a zero-energy Majorana excitation. As a result, the Andreev reflection acquires a resonant character: the corresponding probability is equal to unity, regardless of the value of the hopping integral between the subsystems. This effect is explained by the fact that, due to the Hermitian self-conjugation of the Majorana operator, the tunneling interaction of the MM localized in the interface region is the same for the electron and hole reservoirs of a normal metal \cite{law-09}.

As a consequence, the conductance at zero bias has a maximum with the height $2G_{0}$, where $G_{0}=e^2/h$ is the conductance quantum. The factor 2 here indicates the charge $2e$ transferred in the process of Andreev reflection \cite{law-09,flensberg-10}. In addition, the peak must be stable under fluctuations of the chemical potential and the magnetic field \cite{wu-12,dassarma-12,rainis-13}.

In the first experiments on SW tunneling spectroscopy \cite{mourik-12,deng-12,das-12}, a conductance peak was also recorded at zero bias; it appeared with an increase in the magnetic field, was preserved only in the case of the simultaneous presence of spin-orbit coupling and superconducting s-pairing, and was explained by resonant transfer through an MBS (Fig. \ref{mourik-12}b). However, the data obtained showed the absence of stable quantization of the conductance, whose values were often much less than $2G_{0}$ \cite{mourik-12,deng-12,das-12,deng-16,nichele-17}.

This discrepancy has resulted in a wide discussion about the possible causes of weak conductance and various alternative mechanisms leading to a resonance singularity without involving topologically nontrivial states. In particular, a serious problem in observing a $2G_{0}$ stable peak was the insufficiently strong superconducting pairing induced in a semiconducting wire, which resulted in a nonzero conductance at source-drain electric field energies that were less than the width of the induced superconducting gap. This effect can be explained by several causes: disorder in the wire, thermal fluctuations, roughness of the electrode-wire interface, tunneling barrier fluctuations, and electron-electron and electron-phonon interactions \cite{takei-13,liu-17a}.
It was shown in addition that intragap quasiparticle states not only complicate the interpretation of experimental data but also can negatively affect the topological protection of MBSs, because they can participate in the so-called braiding of the quasiparticle worldlines and thereby introduce errors into the final quantum state  \cite{cheng-12,rainis-12}.

We note that, after the appearance of the first experiments on tunneling spectroscopy of semiconducting wires, it was demonstrated that the conductance resonance at zero bias can result from the appearance of low-energy states in wires with several subbands, both with and without disorder. Importantly, these low-energy states are in no way related to the topological phase transition in the system \cite{liu-12,bagrets-12,rieder-12,pan-20}.
Such states can arise if the Zeeman energy exceeds the splitting between the subbands \cite{kells-12a}.
Other scenarios for the onset of resonance are the Kondo effect, which can coexist with superconducting pairing in nonzero magnetic fields \cite{lee-12},
and weak antilocalization \cite{pikulin-12}.

Further advances in the technology of epitaxial growth of hybrid SWs allowed implementing induced pairing of a higher quality, eliminating the factors of disorder and interface roughness \cite{krogstrup-15,chang-15,gul-17}, and attaining the ballistic transport mode \cite{gazibegovic-17,zhang-17}.
This made it possible to observe a stable
$2G_{0}$ quantization of the conductance at zero bias in the TNPD \cite{zhang-18a,franz-18}.

\begin{figure}[ht]
	\begin{center}
		\includegraphics[width=1\textwidth]{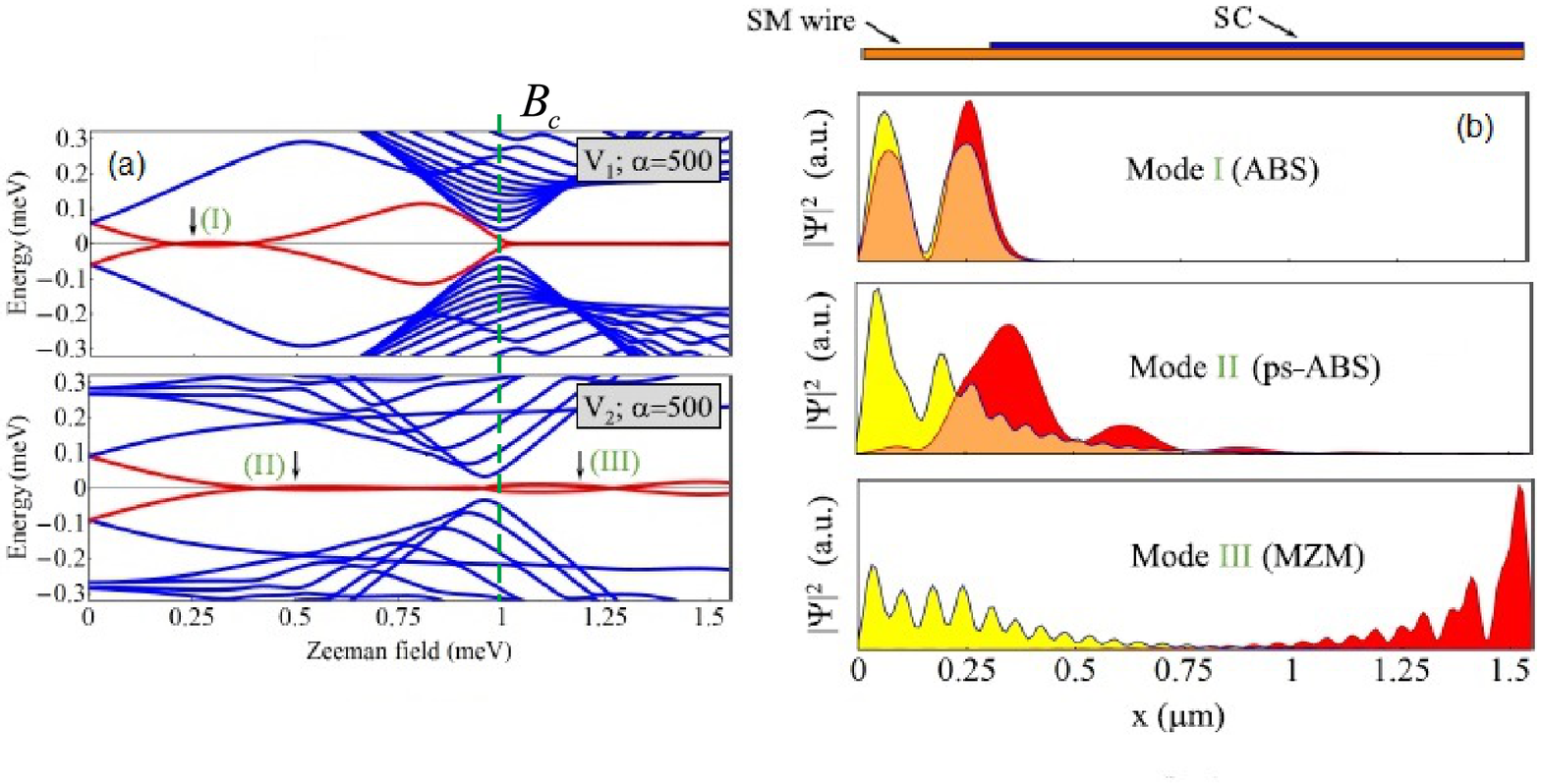}
		\caption{\label{moore-18b} Andreev and Majorana bound states in the QD/SW system. (a) Spectra for potential profiles $V_{1}$ (top) and $V_{2}$ (bottom). (b) Spatial distribution of the probability density of Majorana wave functions for lowest-energy excitation (see red lines in panels (a)) for various magnetic fields, shown with the arrows in panels (a) \cite{moore-18b}.}
	\end{center}
\end{figure}

But even the progress achieved does not completely eliminate the ambiguity in interpreting the latest results \cite{zhang-20}. In particular, there are alternative scenarios according to which local conductance measurements give a $2G_{0}$ resonance at zero voltage. In an SW, low-energy Andreev bound states (ABSs) can form (1) due to a smooth electrostatic potential at the end of the wire \cite{kells-12b}; (2) due to quasi-one-dimensionality \cite{stanescu-13};  (3) when MM wave functions overlap significantly \cite{haim-15a}; (4) in the case of a strongly inhomogeneous potential along the entire length of the multiband hybrid structure \cite{chen-19,woods-19}.

In many experiments devoted to the detection of MBSs, there is a semiconducting domain between the metal electrode and the SW region, where the induced superconducting pairing is either substantially suppressed or absent altogether, and the potential profile can be significantly different from that in the wire due to the effect of the gate electrodes. As a result, a quantum dot (QD) is realized in that domain. In Fig. \ref{mourik-12}a this domain is located between the $N$ and $S$ rectangular regions.

It was found in a number of studies that, as the chemical potential or the magnetic field increases, the two ABSs that arise in such a QD can merge and form a zero mode. In Fig. \ref{moore-18b}a,
the appearance and behavior of such states in magnetic fields lower than the critical field $B_{c}$ are shown with red lines. As we can see, such ABSs are preserved in a certain range of parameters in the topologically trivial phase. Under a subsequent topological phase transition, these ABSs turn into MBSs (see the evolution of the probability density of Majorana wave functions in Fig. \ref{moore-18b}b) \cite{liu-17b,moore-18a,moore-18b}.

Thus, a quantized conductance resonance can arise in both topologically trivial and topologically nontrivial phases. Its identical properties in both cases, such as pinning at zero bias and oscillations in a changing magnetic field, additionally complicate the detection of MBSs by local tunneling spectroscopy, because it is difficult to determine the exact values of the chemical potential and of the induced superconducting gap under experimental conditions \cite{lee-14,haim-15a,liu-17b,liu-18,reeg-18,moore-18a,moore-18b}.

It follows from the foregoing that the conductance peak at zero bias is a necessary but not sufficient condition for the unambiguous detection of MBSs in an SW. Therefore, it is urgent to search for alternative ways to detect these excitations.

\subsubsection{\label{sec3.3.2} Detecting the nonlocality of a Majorana bound state}

\begin{figure}[ht]
	\begin{center}
		\includegraphics[width=1\textwidth]{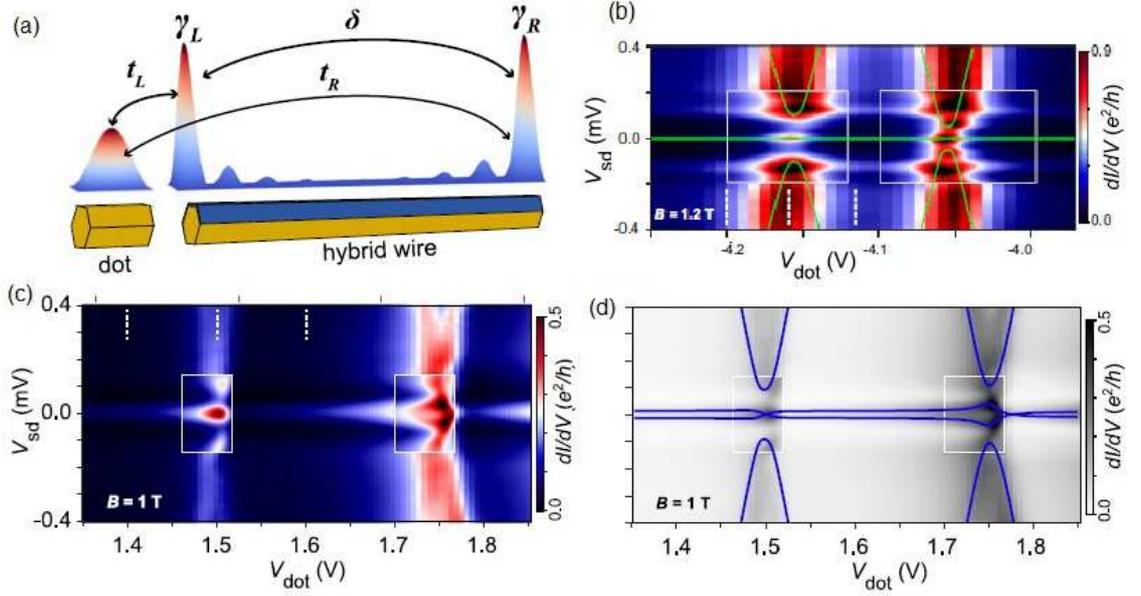}
		\caption{\label{deng-18} Investigating the nonlocality of MBSs by local tunneling spectroscopy of the QD/SW structure. (a) Interaction of QD energy levels with each of the MMs, $\gamma_{L}$ and $\gamma_{R}$. (b)
			Dependence of conductance on source-drain voltage $V_{sd}$ and gate voltage $V_{dot}$
			(conductance map) in the case where Majorana fermions do not interact
			($\delta\approx0$). Green lines show results of model calculations.
			(c) Map of conductance in the case of nonvanishing hybridization of wave functions of Majorana fermions
			($\delta\neq0$). (d) Results of numerical computations corresponding to the case in Fig. (c) \cite{deng-18}.}
	\end{center}
\end{figure}

One of the new methods for identifying an MBS is with the use of the above-mentioned nonlocality of this state. To detect two spatially separated MMs, it was proposed to investigate the conductance correlations at opposite ends of the wire in a two-contact circuit \cite{moore-18b}. In a recently performed experiment, however, no consistent behavior of these quantities was found \cite{yu-20}. In addition, analyzing quantum entanglement and dissonance of the state of two QDs interacting with different MMs can be useful \cite{li-15}.

Despite the noted disadvantages of local transport measurements in the QD/SW system, the spatial nonlocality of MBSs can also be studied in this case \cite{prada-17,clarke-17,deng-18}. As noted in Section \ref{sec2.1}, Majorana-type Bogoliubov excitation can be represented as a superposition of MM operators whose wave functions are localized at opposite ends of the SW. The QD then interacts with each MM separately, as can be seen from Fig. \ref{deng-18}a. 

In turn, the magnitude of hybridization with the right-hand Majorana wave function should actually be much less than with the left-hand one, i.e., $t_{R} \ll t_{L}$.
If there is a true MBS with well-separated zero modes, then its energy level remains unchanged when coinciding with the spin-dependent QD energy levels (in Fig. \ref{deng-18}b, see the green straight line $V_{sd}=0$, which coincides with the experimental data).
This property is a direct manifestation of the topological protection of MBSs.

But if the MM wave functions overlap, then the hybridization of an electron or hole state in the QD and the MBS leads to significant splittings (anticrossings) of the energy levels of the MBS at the points of intersection with the QD energy levels (the anticrossing at $V_{dot}\approx1.75~V$ in Fig.  \ref{deng-18}c, d).
The magnitude of this splitting depends on the spin polarization of the outer MM located farther from the QD. In addition, it has been noted that, as a result of a decrease in the nonlocality of the MBS, the peak of the QD density of states at zero frequency acquires a dependence on the spin projection \cite{ricco-19}.
The nonlocality of the MBS can also be detected if the QD is in the Kondo regime \cite{lee-13}.

\subsubsection{\label{sec3.3.3} Analysis of the fluctuation characteristics of transport current}

Another tool that can yield additional proof of the existence of MBSs is the analysis of current fluctuations. The study of autocorrelations in a system where two MMs are connected to opposite contacts shows that the ratio of the shot noise of the contact to the current in it (the Fano factor of the contact) is equal to unity in the weak conductance regime \cite{nilsson-08,law-09,wu-12,liu-13}.
The main contribution to the current is then made by the processes of Andreev cross reflection on the MBS or direct charge transfer through the MBS, which is a direct consequence of the nonlocality of this quasiparticle excitation. In addition, in the weak conductance regime, the cross-correlation of currents in opposite contacts is positive and reaches a maximum at $\varepsilon_{0}\gg eV$, where $\varepsilon_{0}$ is the MBS energy \cite{nilsson-08}.
In the opposite limit, $\varepsilon_{0}\ll eV$,
cross-correlations tend to zero, which indicates the dominance of the processes of resonant local Andreev reflection on the MBS \cite{bolech-07,liu-13}.
A similar behavior of cross-correlations is observed in the topologically trivial phase. When only one of the two MMs is coupled to a contact in a single-contact geometry, its Fano factor is equal to $2$ in the low-conductance regime, because the processes of local Andreev reflection on the MBS again prevail in that case \cite{wu-12}.

In a number of studies, the features of current fluctuations were studied in a system where both contacts were coupled to a QD that, in turn, interacted with one of the two MMs via tunneling \cite{liu-15a,haim-15a,haim-15b,valentini-16,smirnov-17,smirnov-18}.
In such a situation, the universal behavior of noise in the contact was demonstrated both at a low voltage and in an essentially nonequilibrium regime \cite{liu-15a,smirnov-17,smirnov-18}.
In contrast to the cross-correlation of currents in contacts associated with different MMs, this characteristic turns out to be negative and tends to zero in the high-voltage limit \cite{haim-15a,haim-15b,valentini-16}.
This behavior also allows distinguishing the existence of an MM from other scenarios of the occurrence of a conductance peak at zero bias (Kondo effect, ABSs).

\subsubsection{\label{sec3.3.4} Study of the spin polarization of the Majorana bound state}

Another characteristic of the MBS is its spin polarization 
\cite{leijnse-11,sticlet-12,nagai-14}.
It was shown in \cite{sticlet-12} that, if a magnetic field $\mathbf{B}$ is directed along the $z$-axis and the effective Rashba field vector $\mathbf{B}_{SO}$ is collinear to the $y$-axis, then the spin polarization at the ends of the SW changes in the $xz$ plane. This characteristic can be used as a local (coordinate-dependent) order parameter in describing a topological phase transition \cite{sticlet-12,guigou-16,serina-18}.The magnitude and direction of the spin polarization vector can depend on both the nature of the spin-orbit coupling \cite{sticlet-12} and the magnitude and direction of the external magnetic field \cite{valkov-17b,valkov-17c,prada-17}.
In high magnetic fields and high values of the spin-orbit coupling corresponding to the experimentally observed ones, the $z$ component is dominant, while the $x$ component has the order $O\left(\Delta/B,~\alpha/B\right)$
and measures the spin polarization  tilt. We note that this spin polarization component is proportional to a similar component of the local Majorana polarization if $\mathbf{B} \parallel z$ and
$\mathbf{B}_{SO} \parallel y$ \cite{sticlet-12,guigou-16}.
In the general case, the relation between the x and z components can be determined, for example, based on the anticrossing values of the QD and MBS energy levels under the resonance conditions for these subsystems (Fig. \ref{deng-18}a, d) \cite{prada-17,deng-18}.

In transport processes, a nonzero spin polarization of the MBS gives rise to Andreev scattering without changing the spin projection \cite{he-14a,he-14b} and noncollinear Andreev reflection \cite{wu-14}. Thus, the existence of an MBS can be verified by spin-polarized tunneling spectroscopy. In \cite{valkov-18},
the symmetry breaking of spin-polarized currents in the SW was shown to occur due to the spin polarization of the MBS. Based on this effect, it is possible to implement a current switch with the direction of the current controlled by the gate magnetic field.

\subsubsection{\label{sec3.3.5} Method based on the Coulomb blockade effect}

\begin{figure}[ht]
	\begin{center}
		\includegraphics[width=0.65\textwidth]{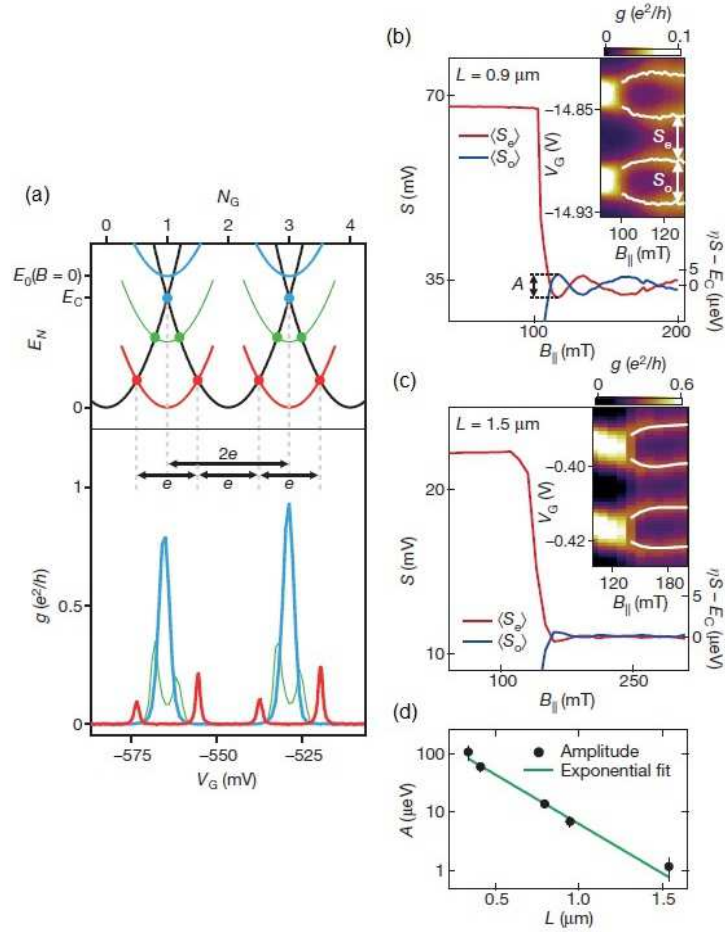}
		\caption{\label{albrecht-16} Emergence of topological protection of an MBS in electron transport through an SW in the Coulomb blockade regime. (a) Above: spectrum of the device as a function of normalized gate voltage $N_{G}$. Below: conductance at the zero source-drain voltage as a function of gate voltage $V_{G}$ ,for different magnetic fields. (b, c) Dependence of averaged distances between conductance peaks $\left\langle S_{e,o}\right\rangle$ on the magnetic field $B_{||}$ for wire lengths $L=0.9~\mu m$ and $L=1.5~\mu m$, respectivelly. Insets: map of the conductance in coordinates $\left(B_{||},V_{G}\right)$. (d) Dependence of oscillation amplitude $\left\langle S_{e,o}\right\rangle$ on the length of the wire \cite{albrecht-16}.}
	\end{center}
\end{figure}

The results discussed in Sections \ref{sec3.3.1}-\ref{sec3.3.4} pertain to the situation where the SW is grounded directly. However, a number of effects that are promising from the standpoint of MBS detection and applications also occur when the charging energy $E_{c}$ is taken into account in the processes of tunneling between the wire and the contacts, or, in other words, in the Coulomb blockade regime. Such a dependence can arise when the hybrid wire under study is sufficiently short. Then, if the ABS energy $E_{0}$ in Fig. \ref{albrecht-16}a is greater than $E_{c}$, the conductance peaks at zero bias associated with the degeneration of states with $2N$ and $2N+2$ electrons are separated by the field on the gate, which is proportional to $2e$,
and hence transport is implemented by Cooper pairs (the blue parabolas and the blue curve at the top and bottom of Fig. \ref{albrecht-16}a)
\cite{hekking-93,hergenrother-94,albrecht-16}.

As the magnetic field increases, the quasiparticle energy decreases due to the Zeeman effect. For $E_{0}<E_{c}$, when the gate voltage changes, a state with an odd number of particles can become the ground state, and the conductance peaks then split (the green parabolas and the green curve at the top and bottom of Fig. \ref{albrecht-16}a). If an MBS is realized ($E_{0}=0$),
the distance between the conductance maxima becomes proportional to $1e$ (the red parabolas and the red curve at the top and bottom of Fig. \ref{albrecht-16}a).
Because adding the charge $2e$ to the wire in a topologically nontrivial phase becomes less energetically advantageous than adding the charge $e$, the processes of local Andreev reflection on the MBS are suppressed. As a result, the leading contribution to the current starts being made by the processes of one-electron tunneling; therefore, in the regime of a strong Coulomb blockade, the conductance maximum is $G_{0}$ rather than $2G_{0}$ \cite{averin-92,fu-10,zazunov-11}. Moreover, a twofold decrease in the conductance peak is also observed at nonzero temperatures \cite{hutzen-12}. Because coherent transport between opposite contacts then occurs via two spatially separated noninteracting MMs (which is equivalent to passing through a single QD), such processes can be interpreted as quantum teleportation \cite{fu-10}. If the MM wave functions overlap, the transport time between them becomes nonzero and inversely proportional to the magnitude of this hybridization \cite{khaymovich-17}.

In the experiment in \cite{albrecht-16}, an SW in the Coulomb blockade mode was used to demonstrate the topological protection of the MBS \cite{kitaev-01,dassarma-12}. For this, the average values of the distance were measured between the conductance peaks resulting from the degeneration of states with an even and odd number of particles, $\left\langle S_{e,o}\right\rangle$.
It can be seen from \ref{albrecht-16}b,c that these quantities oscillate with an amplitude $A$  when the Zeeman field changes. Moreover, as can be seen from the insets in Fig. \ref{albrecht-16}, $S_{e,o}$ for $B_{||}\approx120~mT$
is half of $S_{e}$ for $B_{||}\approx90~mT$,
which indicates the $1e$-periodicity of conductance resonances and single-electron transport in stronger fields. Most important, however, is that the amplitude $A$ decays exponentially as the wire length increases, as can be seen from Fig. \ref{albrecht-16}d. 

We note that another effect characteristic of MBSs, an increase in the amplitude of oscillations with an increase in the magnetic field, was not found in \cite{albrecht-16}. This can be explained by the combined effect of the temperature factor, presence of a few subbands or ABSs localized in QD domains between the contacts and the hybrid structure \cite{chiu-17}.

\subsubsection{\label{sec3.3.6} Detection of a Majorana bound state in interference structures}

Starting with the classic work by Aharonov and Bohm \cite{aharonov-59},
the effect of the electromagnetic potential on the motion of a quantum particle has been used in many studies to analyze the features of coherent transport in mesoscopic structures (see, e.g.,
\cite{yacoby-94,kobayashi-03,cabosart-14}).
This effect was also investigated in systems containing MBSs. We can distinguish several versions of Aharonov-Bohm interferometers that include a topological superconductor.

In standard geometry, the SW is located in one of the arms of the ring. The topological phase transition in such a structure can then be inferred from the doubling of the period of the Aharonov-Bohm oscillations in conductance, irrespective of the degree of disorder; this is explained by the possibility of individual Fermi quasiparticles, rather than Cooper pairs, being transferred \cite{akhmerov-11,whiticar-20} (for the same reason, the $4\pi$-periodic Josephson effect exists in a junction of two topological superconductors
\cite{kitaev-01,ioselevich-11,snelder-13}). When considering the persistent current in such a ring, the nontrivial phase is characterized by the appearance of an $h/e$ harmonic, which remains finite in a zero magnetic flux and has the sign determined by the ground-state FP of the SW \cite{jacquod-13}. The parity is then fixed by transferring the wire into the Coulomb blockade regime. The same system was considered when the differences between interference patterns for MBSs and ABSs were determined in structures with zero
\cite{tripathi-16} and nonzero charge energies \cite{sau-15,hell-18}.

\begin{figure}[ht]
	\begin{center}
		\includegraphics[width=0.88\textwidth]{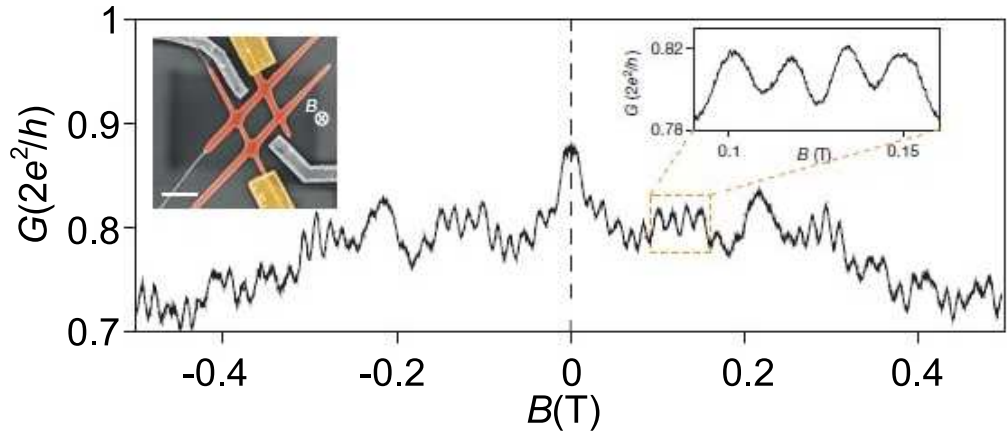}
		\caption{\label{gazibegovic-17} Aharonov-Bohm oscillations in the conductance of a ring made of InSb wires (shown in orange in the inset in the left-hand part of the figure) \cite{gazibegovic-17}.}
	\end{center}
\end{figure}

A variant of nonstandard geometry is provided by the MM/QD/MM scheme, where a topological superconductor has a ring shape and a QD is placed between two MMs located at the opposite ends of the wire. This system was proposed as a topological qubit in which the operation of non-abelian rotation in the space of degenerate ground states can be implemented by means of one-electron tunneling into the QD in the Coulomb blockade regime. The magnetic flux through such a ring allows it to be transferred to the desired energy degeneration point  \cite{flensberg-11,liu-11}.
The described MM/QD/MM configuration has been used to analyze the differences between the topological $h/e$-periodicity of conductance and the nontopological one observed in the case of a normal rather than a superconducting ring \cite{chiu-18}.
In \cite{sau-15}, the QD placed between the MMs was replaced with an extended 1D electrode connected to a contact. It has been shown that conductance oscillations with a period of $4\pi$ (if the flux quantum is defined as $\phi_{0}=h/2e$) are observed in the topologically nontrivial phase due to the processes of nonlocal tunneling through the MBS, while no oscillations are observed in the trivial phase.

The destructive interference accompanying coherent transport in the Aharonov-Bohm ring gives rise to Fano resonance \cite{fano-61}. Several methods have been proposed for detecting MBSs based on this effect. In \cite{dessotti-14}, a modification of the Fano resonance was studied for the conductance of a ring consisting of two QDs connected in parallel with the contacts. In that case, one of the QDs interacts with a Kitaev chain. Such a chain placed between QDs was considered in \cite{shang-14}. In  \cite{ueda-14,jiang-15}, an SW serves as one of the contacts in a two-contact circuit. In addition, the Fano resonance and its properties were studied in the local transport regime \cite{schuray-17}.
Additionally, the contributions of the local and crossed Andreev reflection processes to the Fano effect were analyzed for various geometries of an interferometer with an MBS \cite{zhang-18b}.

In most theoretical studies of the transport characteristics of an Aharonov-Bohm ring made of two arms, the contacts in one of the arms are assumed to be directly connected via tunneling. The interaction of the SW located in the other arm is also described by the processes of direct hopping to the left and right contacts. As a result, the presence of lead wires in the normal phase (hereafter referred to as normal wires, or NWs), which connect the SW to contacts or the contacts with each other, is not taken into account. On the other hand, it was shown recently that quantum devices based on semiconducting wires, including the Aharonov-Bohm ring, can be constructed such that these objects themselves, both in the normal and superconducting phases, serve as the arms (see the inset on the left side in Fig. \ref{gazibegovic-17})
\cite{gazibegovic-17}.

\begin{figure}[ht]
	\begin{center}
		\includegraphics[width=0.88\textwidth]{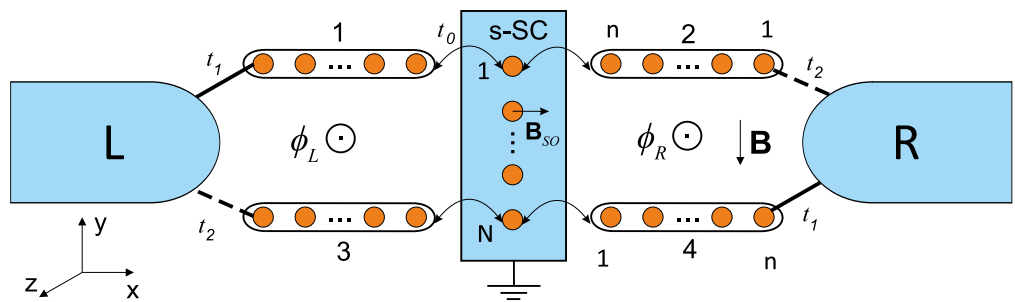}
		\caption{\label{ABring} Aharonov-Bohm ring made of a superconducting nanowire (s-type superconductor, s-SC) whose ends are connected in parallel with metallic contacts via lead normal-phase wires
			\cite{valkov-19,aksenov-20}.}
	\end{center}
\end{figure}

In view of the foregoing, we now consider an interferometer (or a ring) consisting of four NWs: two of them make up the upper arm, and the other two, the lower arm (Fig. \ref{ABring}).
The SW separates the upper left and the lower left NWs from the upper right and lower right NWs and at the same time  connects the upper and lower arms, playing the role of a bridge. The ring is placed in an external magnetic field
$B$, applied parallel to the SW (the magnetic field component perpendicular to the plane of the ring is responsible for the appearance of the Aharonov-Bohm phase). The magnetic field induces a topological phase transition in the bridge and allows topologically trivial low-energy excitations to exist in the NW. As a result, the interference interaction of carriers propagating in the NW and bridge transport channels leads to the Fano effect. Using the method of nonequilibrium Green's functions \cite{keldysh-64,arseev-15,arseev-17}, we study the influence of the magnetic field, the strength of superconducting pairing, and other factors on these Fano resonances.

We note that a similar geometry was also investigated in \cite{shang-14}.
However, the authors of \cite{shang-14} did not use the microscopic description of the SW, and considered QDs instead of NWs. A ring with a bridge containing fractional fermions was discussed in \cite{rainis-14} in the tight-binding framework, but the coupling of these states to low-energy modes of the NW was not analyzed there.

\subsection{\label{sec3.4} Features of coherent transport in the Aharonov-Bohm ring with a bridge in the topological superconducting phase}

\subsubsection{\label{sec3.4.1} Current in an interference device in terms of nonequilibrium Green's functions}

The features of coherent quantum transport that are discussed below are associated with the presence of an SW. To use the SW Hamiltonian \eqref{Ham_latt}, ${H}_{W}$, with $U=V=0$, introduced in Section \ref{sec3.2}, we choose the axes as shown in Fig.  \ref{ABring}.
The NWs that make up the arms of the ring (see Fig. \ref{ABring}) are assumed to be identical. The NW Hamiltonians ${H}_{1-4}$ are obtained from \eqref{Ham_latt} with $\Delta=\alpha=0$.
The tunnel interaction between the SW and an NW is described by the Hamiltonian
\begin{eqnarray} \label{HWl}
{H}_{T} =-t_{0}\sum\limits_{\sigma}\left[\left(b_{Ln\sigma}^{+}+b_{Rn\sigma}^{+}\right)a_{1\sigma}
+\left(d_{L1\sigma}^{+}+d_{R1\sigma}^{+}\right)a_{N\sigma}\right]+ h.c.,
\end{eqnarray}
where $t_{0}$ is the parameter of hopping between edge sites of the SW and the NW, the operator $b_{L\left(R\right)n\sigma}^{+}$ creates an electron with spin $\sigma$ on the $n$th site of the upper left (right) NW, and the operator $d_{L\left(R\right)1\sigma}^{+}$ creates an electron with spin $\sigma$ on the first site of the lower left (right) NW. In turn, the (SW + NW) system is also coupled to the contacts, as described by the Hamiltonian
\begin{eqnarray} \label{V}
&&H_{V} =-\sum \limits_{k\sigma}\left[c_{Lk\sigma}^{+}\left(t_{1}e^{-i\frac{\Phi_L}{2}}b_{L1\sigma}+
t_{2}e^{i\frac{\Phi_L}{2}}d_{Ln\sigma}\right)\right. + \\
&&~~~~~~~~~~~~~~~~~+\left. c_{Rk\sigma}^{+}\left(t_{2}e^{i\frac{\Phi_R}{2}}b_{R1\sigma} +
t_{1}e^{-i\frac{\Phi_R}{2}}d_{Rn\sigma}\right)\right]+ h.c.,\nonumber
\end{eqnarray}
which simultaneously serves as the interaction operator
when using the diagram technique for nonequilibrium
Green's functions. Here, the operator $c_{L\left(R\right)k\sigma}^{+}$ creates an
electron with the wave vector $k$ and spin $\sigma$ in the left
(right) contact, $t_{1,2}$ are the hopping parameters between
contacts and the device,
$\Phi_{L\left(R\right)}=\frac{\phi_{L\left(R\right)}}{\phi_0}$; $\phi_{L\left(R\right)}$ is the
magnetic flux through the left (right) semiring, and $\phi_0=\hbar/e$ is the flux quantum (we set $\hbar=1$ in what follows).
The Hamiltonian of the $i$th contact ($i=L,~R$) has the form
$\hat{H}_{i}=\sum_{k}\left(\varepsilon_{k}-\mu_{i}\right)c_{ik\sigma}^{+}c_{ik\sigma}$,
where $\mu_{L,R}=\mu \pm eV/2$ is the electrochemical potential of the contacts, taking the applied bias voltage into account.

We analyze the transport properties of the interference device using the method of nonequilibrium Green's functions, which are determined from equations written in the site representation. To calculate the stationary current within this approach, we diagonalize the ring Hamiltonian ${H}_{D}={H}_{W}+\sum\limits_{i=1}^{4}{H}_{i}+{H}_{T}$,
using the Nambu operators
$\hat{f}_{l}=\left(f_{l\uparrow}~f_{l\downarrow}^{+}~ f_{l\downarrow}~f_{l\uparrow}^{+}\right)^T$,
where $f_{l\sigma}$ is the annihilation operator of an electron with spin $\sigma$ on the $l$th site of the NW or SW \cite{valkov-19}.
The matrix nonequilibrium Green's function of the device is then defined as
\begin{equation} \label{GF}
\hat{G}^{ab}\left(\tau,\tau'\right)=-i\left\langle {\rm{T}}_{C}\hat{\Psi}\left(\tau_{a}\right)\otimes
\hat{\Psi}^{+}\left(\tau'_{b}\right)\right\rangle,~
\end{equation}
where ${\rm{T}}_{C}$ is the ordering operator on the Keldysh contour that includes the lower (superscrip $+$) and upper (superscript $-$)
branches \cite{keldysh-64}; $a,b=+,-$; $\hat{\Psi}$ contains the Nambu operators of the SW and all the NWs:
\begin{equation} \label{Psi}
\hat{\Psi}=\left(\hat{b}_{L1}...\hat{b}_{Ln}\hat{d}_{L1}...\hat{d}_{Ln}\hat{a}_{1}...\hat{a}_{N}\hat{b}_{R1}...\hat{b}_{Rn}\hat{d}_{R1}...\hat{d}_{Rn}\right)^{T}.
\end{equation}

The electron current in the left contact $I=e\left\langle\dot{N}_{L}\right\rangle$ (where $N_{L}=\sum_{k\sigma}c_{Lk\sigma}^{+}c_{Lk\sigma}$ is the particle number operator in the left contact), can be expressed in terms of the above Green's functions as
\begin{eqnarray} \label{IL1}
I=
2e\sum\limits_{k}\Tr\Biggl[\hat{\sigma}{\rm{Re}}\Biggl\{\hat{t}_{L1}^{+}\left(t\right)\hat{G}_{k,L1}^{+-}\left(t,t\right)\textbf{\textcolor{blue}{+}}\hat{t}_{Ln}^{+}\left(t\right)\hat{G}_{k,Ln}^{+-}\left(t,t\right) \Biggr\} \Biggr],
\end{eqnarray}
where $\hat{\sigma}=\diag\left(1,-1,1,-1\right)$. As a result of unitary transformation \cite{rogovin-74,zeng-03},
the dependence on the source-drain voltage is carried over to the operator $H_{V}$, such that the matrices
$\hat{t}_{L1\left(n\right)}\left(t\right)$ become time dependent,
\begin{eqnarray} \label{tL12}
&&\hat{t}_{L1}=\frac{t_{1}}{2}\hat{\sigma}\hat{T}\hat{\Phi}_{L},~\hat{t}_{Ln}=\frac{t_{2}}{2}\hat{\sigma}\hat{T}\hat{\Phi}_{L}^{+},\\
&&\hat{T}=\diag\left(e^{-i\frac{eV}{2}t},e^{i\frac{eV}{2}t},e^{-i\frac{eV}{2}t},e^{i\frac{eV}{2}t}\right),~\nonumber\\
&&\hat{\Phi}_{L}=\diag\left(e^{- i\frac{\Phi_{L}}{2}},e^{ i\frac{\Phi_{L}}{2}},e^{- i\frac{\Phi_{L}}{2}},e^{ i\frac{\Phi_{L}}{2}}\right).~\nonumber
\end{eqnarray}

In \eqref{IL1}, the mixed Green's functions are
$\hat{G}_{k,L1}^{+-}=i\left\langle \hat{b}_{L1}^{+}\otimes\hat{c}_{Lk}\right\rangle$ and
$\hat{G}_{k,Ln}^{+-}=i\left\langle \hat{d}_{Ln}^{+}\otimes\hat{c}_{Lk}\right\rangle$.
Because $\hat{H}_{D}$ has the
form of a free-particle Hamiltonian in the space of Nambu
operators, the averages involved in $\hat{G}_{k,L1}^{+-}$ and $\hat{G}_{k,Ln}^{+-}$ are
evaluated using the same principle as those used for the averages of ${\rm{T}}_{C}$-ordered products of secondary-quantization operators
\cite{vonsovsky-77,arseev-15}. Hence, as $t\rightarrow0$ expression \eqref{IL1} becomes
\begin{eqnarray} \label{IL2}
I&=&
2e\int\limits_{C}d\tau_{1}\Tr\Biggl[\hat{\sigma}{\rm{Re}}\Biggl\{\hat{\Sigma}_{L1,L1}^{+a}\left(-\tau_{1}\right)\hat{G}_{L1,L1}^{a-}\left(\tau_{1}\right)+\hat{\Sigma}_{Ln,Ln}^{+a}\left(-\tau_{1}\right)\hat{G}_{Ln,Ln}^{a-}\left(\tau_{1}\right)+\nonumber\\
&+&\hat{\Sigma}_{L1,Ln}^{+a}\left(-\tau_{1}\right)\hat{G}_{Ln,L1}^{a-}\left(\tau_{1}\right)+\hat{\Sigma}_{Ln,L1}^{+a}\left(-\tau_{1}\right)\hat{G}_{L1,Ln}^{a-}\left(\tau_{1}\right) \Biggr\} \Biggr],
\end{eqnarray}
where $\hat{\Sigma}_{Li,Lj}^{+a}\left(-\tau_{1}\right)=\hat{t}_{Li}^{+}\left(0\right)\hat{g}_{Lk}^{+a}\left(-\tau_{1}\right)\hat{t}_{Lj}\left(\tau_{1}\right)$ is the $ij$ block of
the self-energy matrix function describing the effect of the left contact on the ring ($i,j=1,n$), and $\hat{g}_{Lk}^{+a}\left(-\tau_{1}\right)=-i\left\langle {\rm{T}}_{C}\hat{c}_{Lk}\left(0\right)\otimes\hat{c}_{Lk}^{+}\left(\tau_{1}\right)\right\rangle_{0}$ is the bare matrix Green's function of the left contact.

Integrating over the time $\tau_{1}$ and using the Fourier transform, we arrive at
\begin{eqnarray} \label{IL3}
I=
e\sum_{i,j=1,n}\int\limits_{-\infty}^{+\infty}\frac{d\omega}{\pi}
\Tr\Biggl[\hat{\sigma}{\rm{Re}}\Biggl\{
\hat{\Sigma}_{Li,Lj}^{r}\left(\omega\right)\hat{G}_{Lj,Li}^{+-}\left(\omega\right)+ \Biggr.\Biggr.\Biggl.\Biggl.\hat{\Sigma}_{Li,Lj}^{+-}\left(\omega\right)
\hat{G}_{Lj,Li}^{a}\left(\omega\right) \Biggr\}\Biggr].
\end{eqnarray}
Because there are no many-body interactions in the system, the Green's functions in the integrand in
\eqref{IL3} are defined with all the tunneling processes between the device and the contacts taken into account \cite{arseev-15}.
In particular, which is a block of the advanced matrix Green's function $\hat{G}_{Lj,Li}^{a}$ of the device, is to be found from the Dyson equation
\begin{equation}\label{Ga}
\hat{G}^{a}=\left[\left(\omega-\hat{H}_{D}-\hat{\Sigma}^{r}\left(\omega\right)\right)^{-1}\right]^{+},
\end{equation}
where $\hat{\Sigma}^{r}\left(\omega\right)=\hat{\Sigma}^{r}_{L}\left(\omega\right)+\hat{\Sigma}^{r}_{R}\left(\omega\right)$ is the matrix of the retarded self-energy function, reflecting the effect of both contacts on the interferometer. In what follows, we use the approximation of wide-band contacts, in accordance with which the real parts of the self-energy functions can be neglected and the imaginary parts can be considered constant. Then, the nonzero blocks of $\hat{\Sigma}^{r}$ are
\begin{eqnarray}\label{Sr}
&&\hat{\Sigma}^{r}_{L1,L1}=\hat{\Sigma}^{r}_{Rn,Rn}=-\frac{i}{2}\hat{\Gamma}_{11}, \hat{\Sigma}^{r}_{R1,R1}=\hat{\Sigma}^{r}_{Ln,Ln}=-\frac{i}{2}\hat{\Gamma}_{22},\nonumber\\
&&\hat{\Sigma}^{r}_{L1,Ln}=-\frac{i}{2}\hat{\Gamma}_{12}\left(\hat{\Phi}_{L}^{2}\right)^{+},\hat{\Sigma}^{r}_{Ln,L1}=-\frac{i}{2}\hat{\Gamma}_{12}\hat{\Phi}_{L}^{2},\\
&&\hat{\Sigma}^{r}_{R1,Rn}=-\frac{i}{2}\hat{\Gamma}_{12}\left(\hat{\Phi}_{R}^{2}\right)^{+},\hat{\Sigma}^{r}_{Rn,R1}=-\frac{i}{2}\hat{\Gamma}_{12}\hat{\Phi}_{R}^{2},\nonumber
\end{eqnarray}
where $\hat{\Gamma}_{ii}=\Gamma_{ii}\hat{I}_{4}$, $\Gamma_{ii}=2\pi t_{i}^{2}\rho$ is the level broadening function due to the coupling to the contact ($i=1,2$), $\rho$ is the density of states of the contact, $\Gamma_{12}=\sqrt{\Gamma_{11}\Gamma_{22}}$, $\hat{I}_{4}$ is the $4\times4$ identity matrix, and
$$\hat{\Phi}_{R}=\diag\left(e^{ i\frac{\Phi_{R}}{2}},e^{- i\frac{\Phi_{R}}{2}},e^{ i\frac{\Phi_{R}}{2}},e^{- i\frac{\Phi_{R}}{2}}\right)$$.
When dealing with an asymmetric (symmetric) ring, we assume that $\Gamma_{22}=\Gamma_{11}/2=0.01$ ($\Gamma_{22}=\Gamma_{11}=0.01$). The values of $\Gamma_{11,22}$
are given in units of $t$.

The $\hat{G}_{Li,Lj}^{+-}$ blocks in \eqref{IL3} can be found by solving the Keldysh equation
$\hat{G}^{+-}=\hat{G}^{r}\hat{\Sigma}^{+-}\hat{G}^{a}$. The nonzero blocks of $\hat{\Sigma}^{+-}$ are given by
\begin{eqnarray}
&&\hat{\Sigma}_{\alpha i,\alpha j}^{+-}=-2\hat{\Sigma}_{\alpha i,\alpha j}^{r}
\hat{F}_{\alpha},~~~~\alpha=L,R,~~~~i,j=1,n,\\
&&\hat{F}_{L\left(R\right)}=\diag\biggl(f\left(\omega \pm eV/2\right),~f\left(\omega \mp eV/2\right),\biggr.
\biggl.f\left(\omega \pm eV/2\right),~f\left(\omega \mp eV/2\right)\biggr),\nonumber
\end{eqnarray}
where $f\left(\omega \pm eV/2\right)$ is the Fermi-Dirac function.

\subsubsection{\label{sec3.4.2}Breit-Wigner and Fano resonances as a topological phase marker}

\begin{figure*}[ht]
	\begin{center}
		\includegraphics[width=0.48\textwidth]{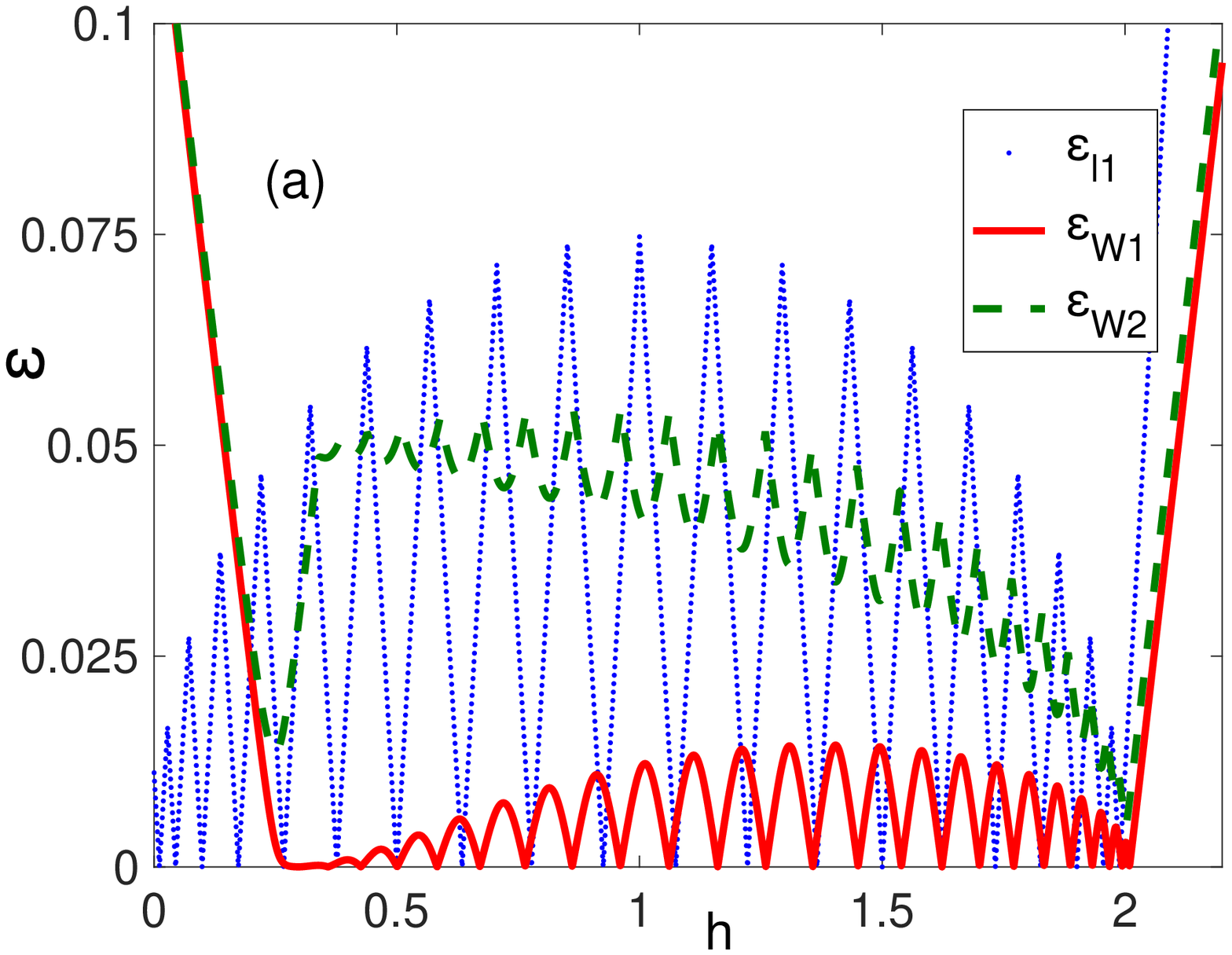}
		\includegraphics[width=0.48\textwidth]{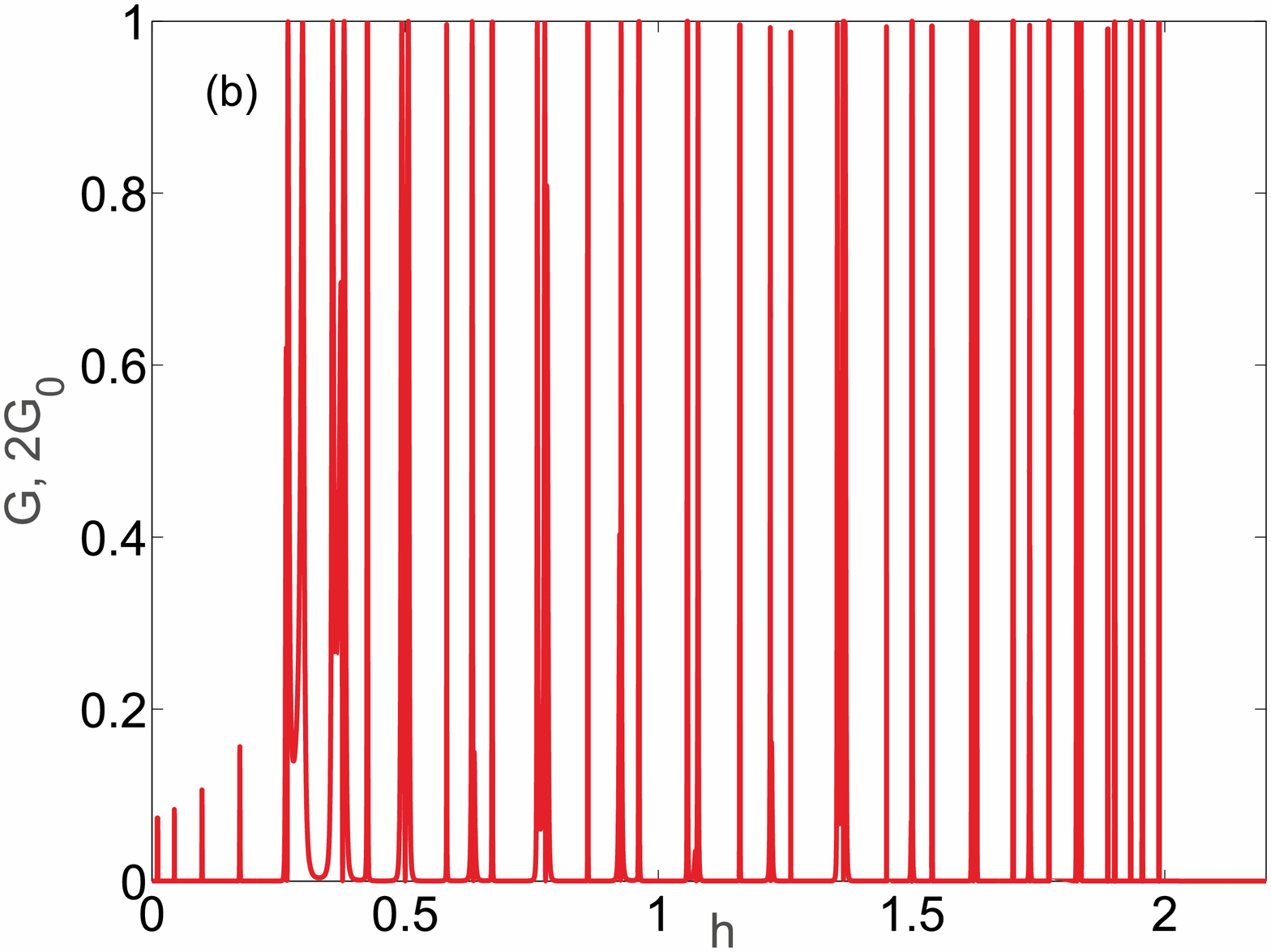}
		\caption{\label{EG_Vz} Dependence (a) of the energy of one-electron excitations of arm wire, $\varepsilon_{l1}$, and the bridge, $\varepsilon_{W1,W2}$, and (b) of the conductance of the ring on the energy of a magnetic field applied in the plane of the device. The parameters are $U=V=0$ and $\Phi=0$.}
	\end{center}
\end{figure*}

As follows from Section \ref{sec3.3}, the main MBS features whose detection is attempted in modern experiments are their zero energy and nonlocality. Hence, in considering the transport properties of the interference device shown in Fig. \ref{ABring}, we are interested only in those properties that are determined by the presence of low-energy excitations in both the NW and the SW, and by the nature of the spatial distribution of excitations with zero or nearly zero energy in the SW. Direct numerical calculations are therefore carried out in the linear response regime, when $eV\to0$. 

It is assumed in what follows that the SW and NW are relatively short, $N=30$ and $n=20$, which significantly reduces the amount of computation. As a result of this assumption, however, an unnaturally large value of the lattice constant
$a=50$ nm \cite{rainis-13} has to be used in order to consider the topologically nontrivial phase in the framework of the parameters corresponding to experiments ($\Delta=250~\mu eV$,
$\alpha_R=0.2~eV\cdot\mathring{A}$, $g=50$, $B\sim 0.1-1~T$) \cite{mourik-12}. With the effective mass of carriers in semiconducting wires $m^{*}=0.015m_{0}$ \cite{mourik-12}, we have the hopping parameter
$t=\frac{\hbar^2}{m^{*}a^2}$, which is used as the unit of measurement in what follows.
The main parameters of the SW used to construct most of the plots in Section \ref{sec3.4} are then $\alpha=0.195$ and $\Delta=0.243$. The other quantities take the values $k_{B}T\approx0$, $\mu=0$, and
$t_{0}=0.1$.

We start with the case without interparticle interactions in the SW, $U=V=0$ \cite{valkov-19}. We turn to the dependence of the conductance on the energy of the magnetic field $h$ in the plane of the device. Transport in the system is determined by single-particle excitations in the vicinity of the Fermi level. The corresponding energies $\varepsilon_{l1}$
(blue dashed curve) of an individual NW and the energies $\varepsilon_{W1}$ and $\varepsilon_{W2}$(red solid and green dashed curves) of the SW as functions of $h$ are shown in Fig. \ref{EG_Vz}a. We see that the energy $\varepsilon_{l1}$ oscillates, vanishing periodically for $h\lesssim2$. In turn, in the TNPD defined by inequality \eqref{eq_top} with $\varepsilon_{0}=t$ \cite{lutchyn-10,oreg-10}, the SW energy
$\varepsilon_{W1}$ splits off from $\varepsilon_{W2}$ and also vanishes periodically. The MBS zeros of $\varepsilon_{W1}$ therefore coexist with the zeros of $\varepsilon_{l1}$, which have a topologically trivial nature.

The dependence of conductance on the magnetic field energy for a symmetric ring is shown in Fig. \ref{EG_Vz}b. In weak fields $h\lesssim0.25$, the peaks of $G\left(h=\varepsilon_{l1}\right)$ are significantly suppressed because $\varepsilon_{W1}\gg0$ (Fig. \ref{EG_Vz}a). In this domain, resonances at which
$G\left(h=\varepsilon_{l1}\right)\rightarrow1$ can arise, for example, if $\Phi_{i}\neq0$.
The bridge in the topological superconducting phase corresponds to a collection of resonances of two types (the range $0.25 \lesssim h \lesssim 2$). The first are symmetric Breit-Wigner resonances and the second are asymmetric resonances or Fano resonances. In strong magnetic fields, $h\gtrsim2$, the conductance is close to zero because $\varepsilon_{l1},~\varepsilon_{W1}\gg0$.

To better understand the causes of the appearance of resonance features at $0.25 \lesssim h \lesssim 2$,
we turn to the simplest situation where each NW is made of only one site, and the SW, of two. As a result, there is a structure made of six QDs, with the upper and lower arms connected in parallel with the contacts. Because the magnetic field approximately gives rise to spin-polarized transport in the ring, we analyze the six-QD system in the spinless case. The energy spectrum of such a structure with $\alpha=\Delta=0$ is given by
\begin{eqnarray}\label{spec6QD}
\varepsilon_{1,2}=\xi_{\downarrow}\equiv\varepsilon,~\varepsilon_{3,4}=\varepsilon+\frac{1}{4}\sqrt{t^2+8t_{0}^2}\pm \frac{t}{4},~~
\varepsilon_{5,6}=\varepsilon-\frac{1}{4}\sqrt{t^2+8t_{0}^2}\pm \frac{t}{4}.
\end{eqnarray}
In a closed system, as we see from \eqref{spec6QD}, there are a pair of degenerate states  ($\varepsilon_{1,2}$) and two pairs of bound and antibound states ($\varepsilon_{3,4}$ and $\varepsilon_{5,6}$). The presence of such solutions is conducive to the realization of bound states in continuum (BICs) \cite{neumann-29}, i.e., discrete-spectrum states that are not coupled to the contacts in the case of an open system, when $\Gamma\neq0$ \cite{volya-03,sadreev-06}.
The appearance of states of this type is a natural consequence of the non-one-dimensionality of the structure under consideration
\cite{lee-99,sadreev-03}.

\begin{figure*}[ht]
	\begin{center}
		\includegraphics[width=0.48\textwidth]{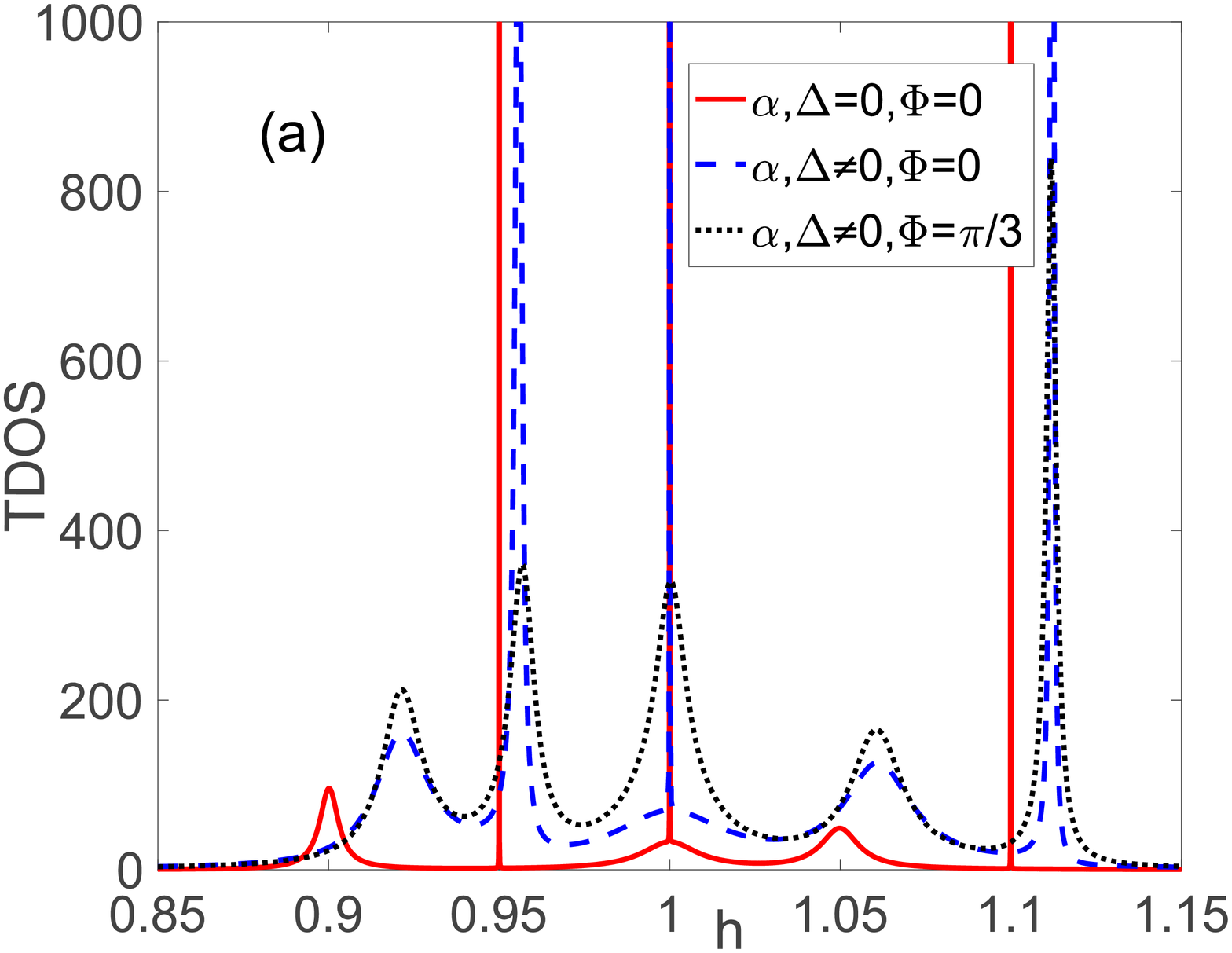}
		\includegraphics[width=0.48\textwidth]{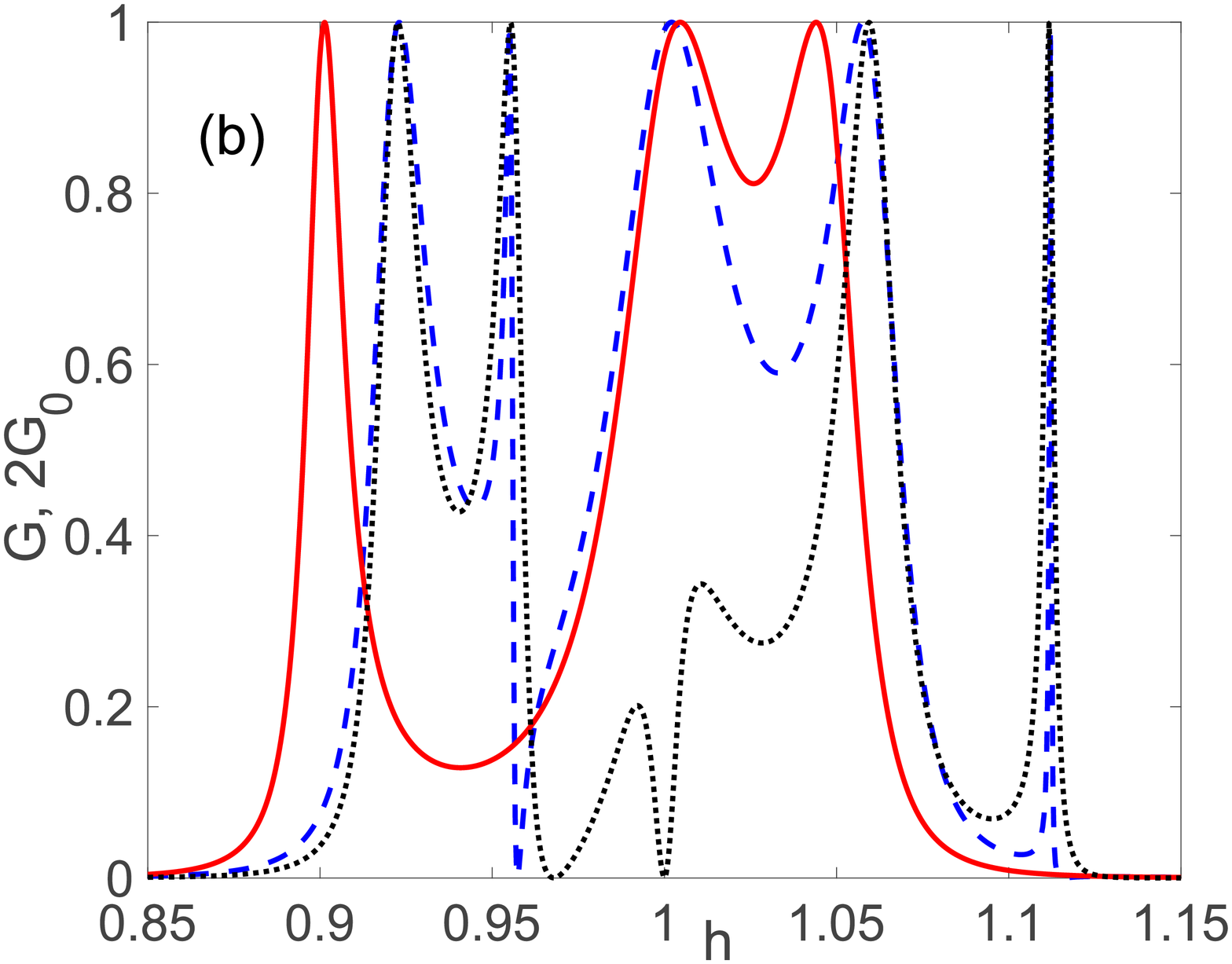}
		\caption{\label{SQD} Dependence of (a) density of states and (b) conductance of a structure made of six QDs on the energy of the magnetic field at  $t=0.1$.}
	\end{center}
\end{figure*}

The existence of BICs in a structure of six QDs is illustrated in Fig. \ref{SQD}a, which shows the dependence of the total density of states ($\rm{TDOS}$) on the Zeeman energy,
\begin{equation}\label{TDOS_6QD}
\rm{TDOS}\left(\omega=0;~h\right)=-\Tr\left[{\rm{Im}}\left\{\hat{G}^{r}\left(\omega=0;~h\right)\right\}\right]/\pi,
\end{equation}
where $\omega$ is the particle energy. The case $\alpha=\Delta=0$ corresponds to the red curve. In the case of degenerate states, the wide maximum of the $\rm{TDOS}$ at $h=1$, which belongs to the state with energy $\varepsilon_{1}$, coincides with a narrow peak due to the BIC with energy $\varepsilon_{2}$ (the width of this peak is proportional to $\delta^2$, where $\delta$ is the infinitesimal parameter responsible for the analytic continuation of $\hat{G}^{r}\left(\omega\right)$). In addition, for two pairs of bound and antibound states with energies $\varepsilon_{3,4}$ and $\varepsilon_{5,6}$ the wide maximum and the narrow peak of the BIC are spatially separated. In Fig. \ref{SQD}a in accordance with \eqref{spec6QD}, they are located to the left and to the right of $h=1$.  All three broad maxima in the density of states manifest themselves as symmetric resonances in conductance (the red solid line in Fig. \ref{SQD}b). In turn, the presence of BICs leaves the transport properties unaffected.

For $\alpha,~\Delta\neq0$, a nonzero broadening of the BIC levels close to $\varepsilon_{4,6}$, occurs, because the spin-orbit coupling leads to the breaking of the spatial symmetry of the ring eigenstates (blue dashed line in Fig. \ref{SQD}a) \cite{nowak-11}. As a consequence, Fano resonances appear in conductance at the same values of $h$ as broadened peaks in the density of states do. Further, the lifetime of a BIC associated with a pair of degenerate states can be made finite by introducing the Aharonov-Bohm phase \cite{lu-05,guevara-06}. All three BIC peaks in the density of states then acquire a nonzero width, and the conductance has three asymmetric Fano peaks (the black dotted curves in Fig. \ref{SQD}a,b). It follows from \eqref{spec6QD} that the energy splitting in a pair of bound and antibound states is determined by the parameter $t$. In particular, if $t=1$ (and hence $t\gg t_{0}$),
then two broad maxima and two BIC peaks in the density of states occur in the vicinity of $h=1$. In addition, one wide maximum (BIC peak) occurs at a distance to the left (right) of this point.

\begin{figure*}[ht]
	\begin{center}
		\includegraphics[width=0.48\textwidth]{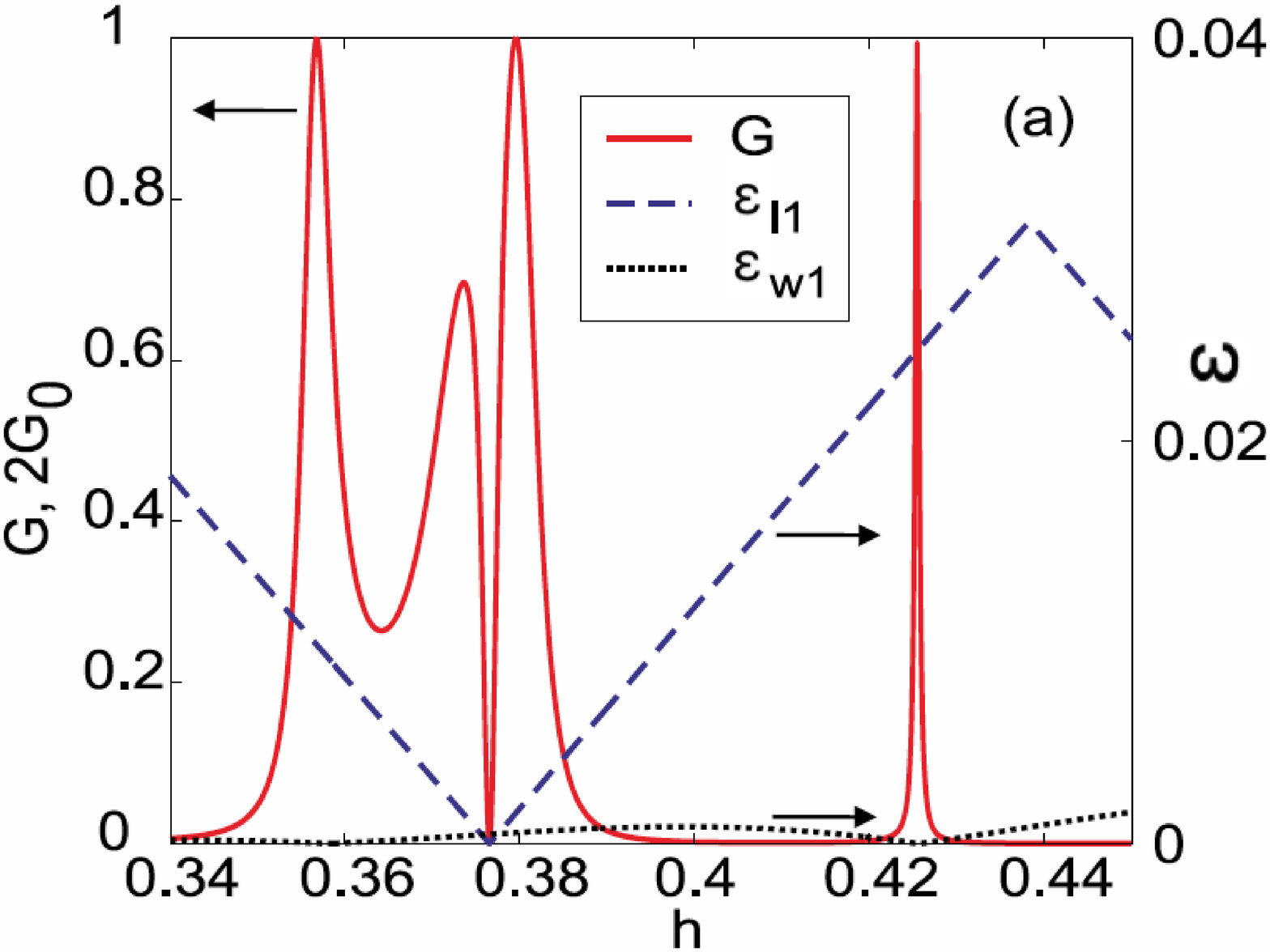}
		\includegraphics[width=0.48\textwidth]{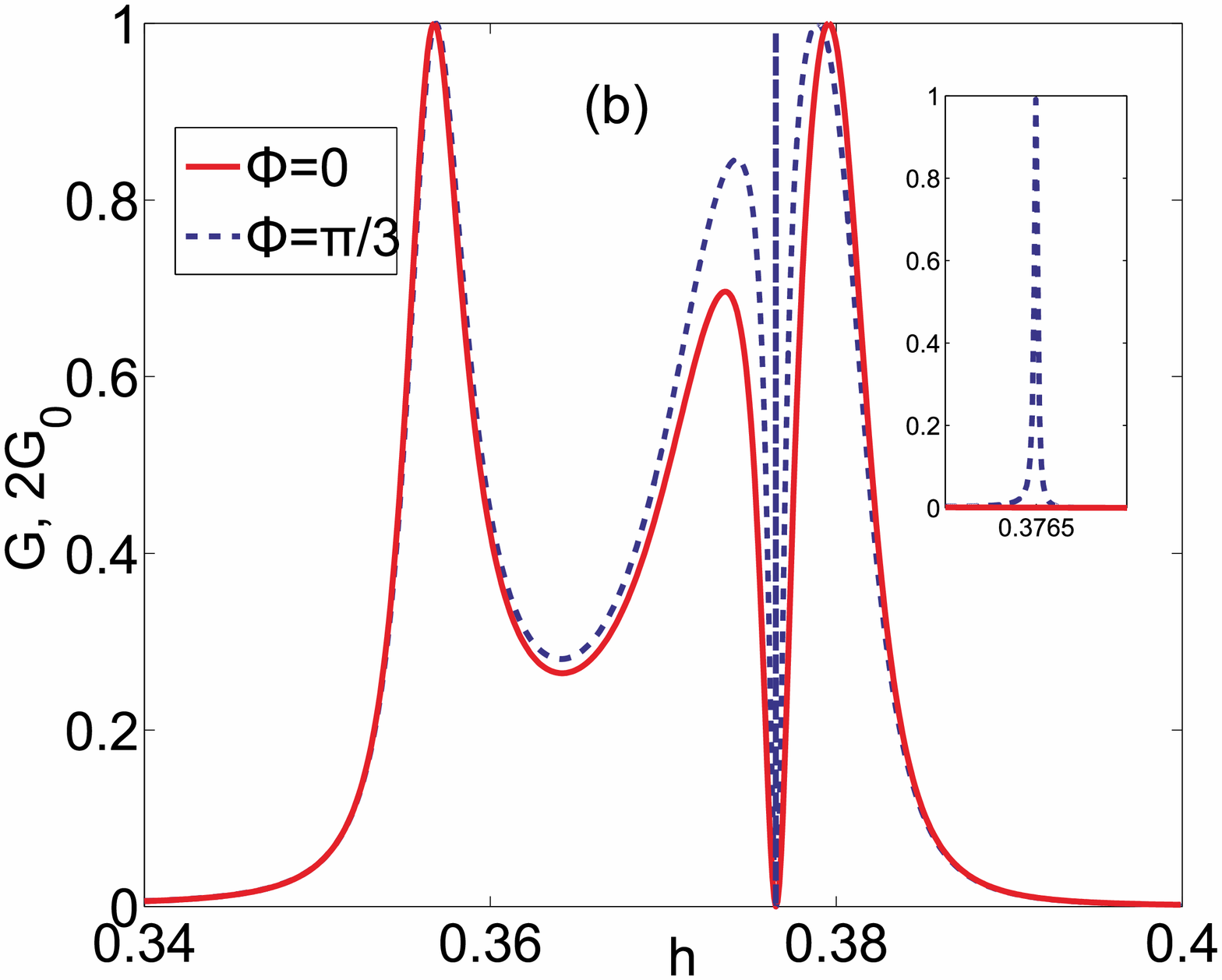}
		\caption{\label{BW&F} (a) Breit-Wigner and Fano resonances in the conductance of a ring in the case where the superconducting bridge is in a topologically nontrivial phase. Position of Breit-Wigner/Fano resonance in conductance (left vertical axis) coincides with the energy minimum $\varepsilon_{W1}$/$\varepsilon_{l1}$ (right vertical axis). (b) Effect of Aharonov-Bohm phase on conductance. Inset: inducing the Fano resonance at $\Phi_{L}=\Phi_{R}=\Phi\neq0$.}
	\end{center}
\end{figure*}

As the system size increases and the ring is restored, the number of maxima and BIC peaks increases. It is then essential that the BIC lifetime become finite only if the SW lowest excitation energy
$\varepsilon_{W1}$  is close to zero, which in the case under consideration implies the realization of a topologically nontrivial phase (Fig. \ref{EG_Vz}a). 

A typical structure of conductance resonances in the TNPD is shown in Fig.
\ref{BW&F}a. This dependence is modified compared with the above-described picture of transport through the structure of six QDs. In particular, one of the Fano resonances in Fig. \ref{SQD}b transforms into a set of Breit-Wigner peaks for the ring. As a result, only the Fano resonances near the minima of $\varepsilon_{l1}\left(h\right)$
remain (one of them is shown in Fig. \ref{BW&F}a). At the same time, the positions of the Breit-Wigner resonances are determined by the zeros of $\varepsilon_{W1}\left(h\right)$. At fixed $h$, $\Delta$ and $\alpha$ the widths of respective symmetric and asymmetric resonances depend on $\varepsilon_{l1}$ and $\varepsilon_{W1}$: the higher
$\varepsilon_{l1,W1}$, the narrower the resonances. 

As noted above, a nonzero Aharonov-Bohm phase makes the corresponding BICs amenable to observation in transport characteristics. In Fig. \ref{BW&F}b we show a new very narrow Fano resonance occurring at $\Phi_{L}=\Phi_{R}=\Phi\neq0$, whose position coincides with a zero of
$\varepsilon_{l1}\left(h\right)$.

Thus, the presence of two types of resonances in conductance can be qualitatively explained by the presence of a few interacting channels for the transport of carriers in the device. Indeed, if the leading contribution is made by the channel related to the SW
($\mu=\varepsilon_{W1}$), then the Breit-Wigner resonance is realized. If, on the contrary, the transport channel mainly involves the NW ($\mu=\varepsilon_{l1}$), then interference occurs in accordance with the Fano scenario. Here, we can trace a similarity with double QDs, whose conducting properties were studied in detail in \cite{guevara-03,orellana-04,guevara-06}.In particular, a similar realization of a wide Breit-Wigner resonance and a narrow Fano resonance was noted, which was interpreted as a manifestation of the Dicke effect known from optics
\cite{shahbazyan-94,vorrath-03,orellana-04}. It amounts to the appearance of wide and narrow peaks in the luminescence spectrum of a pair of atoms \cite{dicke-53}. he first peak is associated with a short-lived collective excitation (super-radiant state) and the second, with a long-lived excitation (subradiant state). The realization of symmetric and asymmetric resonances in transport in other low-dimensional structures has also been noted \cite{myoung-19}. In our case, for a certain choice of parameters, a similar behavior of conductance can be interpreted as the topological Dicke effect, because two types of resonance occur precisely in the TNPD.

\subsubsection{\label{sec3.4.3} Effect of Coulomb interactions and disorder}

To describe the effect of Coulomb repulsion in an SW on the obtained resonance features of the conductance of an interferometer, we use the generalized mean-field approximation, whose applicability domain is determined by the inequalities $U,~V \ll t$. In the framework of this approach, the four-operator terms in SW Hamiltonian \eqref{Ham_latt} can be written as \cite{valkov-17}
\begin{eqnarray}\label{GMFA}
&&Un_{l\uparrow}n_{l\downarrow} = U\left[\langle n_{l\uparrow}\rangle n_{l\downarrow}+\langle n_{l\downarrow}\rangle n_{l\uparrow}-\langle a^{+}_{l\downarrow} a_{l\uparrow}\rangle a^{+}_{l\uparrow} a_{l\downarrow}-\right.\nonumber\\
&&
\left.-\langle a^{+}_{l\uparrow} a_{l\downarrow}\rangle a^{+}_{l\downarrow} a_{l\uparrow}-\langle a^{+}_{l\uparrow} a^{+}_{l\downarrow}\rangle a_{l\uparrow} a_{l\downarrow}-\langle a_{l\uparrow} a_{l\downarrow}\rangle a^{+}_{l\uparrow} a^{+}_{l\downarrow}\right],\nonumber\\
&&V\sum\limits_{\sigma\sigma'}n_{l\sigma}n_{l+1,\sigma'}=V\sum\limits_{\sigma\sigma'}
\left[\langle n_{l\sigma}\rangle n_{l+1,\sigma'}+\langle n_{l+1,\sigma}\rangle n_{l\sigma}-\right.\nonumber\\
&&\left.-\langle a^{+}_{l+1,\sigma} a_{l\sigma'}\rangle a^{+}_{l\sigma'} a_{l+1,\sigma}-\langle a^{+}_{l\sigma} a_{l+1,\sigma'}\rangle a^{+}_{l+1,\sigma'} a_{l\sigma}-\right.\nonumber\\
&&\left.-\langle a^{+}_{l\sigma} a^{+}_{l+1,\sigma'}\rangle a_{l\sigma} a_{l+1,\sigma'}-\langle a_{l\sigma} a_{l+1,\sigma'}\rangle a^{+}_{l\sigma} a^{+}_{l+1,\sigma'}\right].
\end{eqnarray}
It is important that, when even weak Hubbard repulsion is switched on, an increase in the effective Zeeman splitting and a decrease in the superconducting pairing potential result in shifting the boundaries of the regions of the existence of topological phases of a wire with a spin-orbit coupling \cite{stoudenmire-11}. These effects, respectively proportional to $\langle a^{+}_{l\sigma} a_{l\sigma}\rangle$ and $\langle a_{l\sigma} a_{l\bar{\sigma}}\rangle$, must be taken into account when considering transport, because they have the same order of smallness as $h$ and $\Delta$.

The normal and anomalous averages in \eqref{GMFA} are to be found self-consistently using the spin-dependent coefficients, $u_{ml\sigma}$ and $v_{mj\sigma}$ of the Bogoliubov transformation,
\begin{eqnarray}\label{avs}
&&\langle a^{+}_{l\sigma} a_{j\sigma'}\rangle = \sum\limits_{m=1}^{2N}\left[f_{m}u_{ml\sigma}u^{*}_{mj\sigma'}+
\left(1-f_{m}\right)v^{*}_{ml\sigma}v_{mj\sigma'}\right],\nonumber\\
&&\langle a_{l\sigma} a_{j\sigma'}\rangle = \sum\limits_{m=1}^{2N}\left[f_{m}v_{ml\sigma}u^{*}_{mj\sigma'}+
\left(1-f_{m}\right)u^{*}_{ml\sigma}v_{mj\sigma'}\right],
\end{eqnarray}
where $f_{m}$ is the Fermi function at an energy equal to the $m$th quasiparticle excitation energy. When Coulomb interactions are taken into account in approximation \eqref{GMFA}, renormalization occurs and new matrix elements appear in the matrix of the Hamiltonian $\hat{H}_{W}$.

\begin{figure}[ht]
	\begin{center}
		\includegraphics[width=0.78\textwidth]{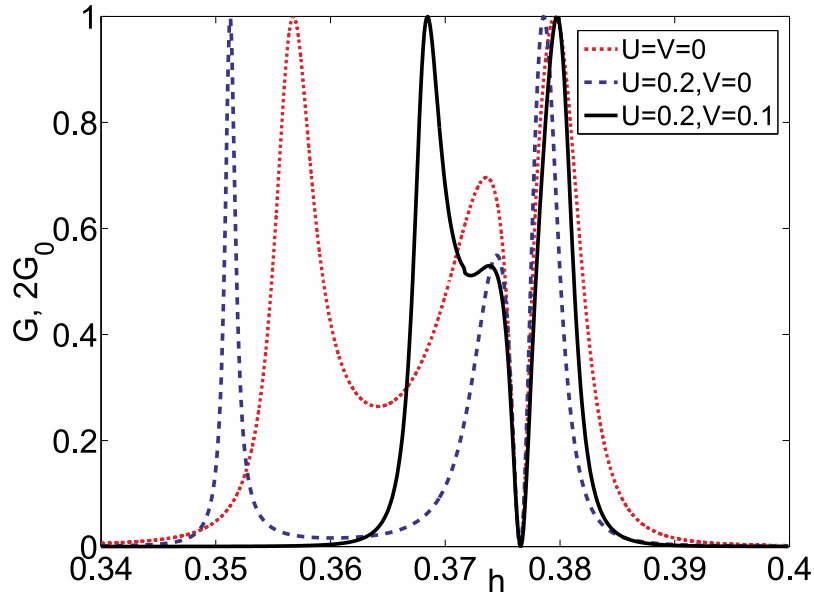}
		\caption{\label{GVz_UV} Effect of Coulomb interactions on symmetric and asymmetric resonances of conductance.}
	\end{center}
\end{figure}

The effect of Coulomb interactions on the properties of the conductance of the ring in the TNPD is shown in Fig. \ref{GVz_UV}. One-site correlations can be seen to slightly shift the maximum of the asymmetric Fano peak, while its minimum (or antiresonance) preserves its position (cf. the red dashed and blue dashed curves). At the same time, the Breit-Wigner peak shifts to the left, because a corresponding shift is also acquired by the zero energy of MBS for  $U\neq0$. For $V\neq0$, the Breit-Wigner resonance shifts in the opposite direction (black solid curve). The Fano resonance, in turn, remains in the same position, and its width is practically unchanged. We note that similar effects are observed when diagonal disorder is taken into account in the SW,  $\xi_{\sigma}+\delta\xi_{l},~l=1,...,N$ (where $\delta\xi_{l}$ is a random addition to the one-particle energy at site $l$, taking values in the interval $\left[-1/2,~1/2\right]$).

\subsubsection{\label{sec3.4.4} Dependence of Fano resonance properties on the type of low-energy excitations of the bridge}

\begin{figure*}[ht]
	\begin{center}
		\includegraphics[width=0.48\textwidth]{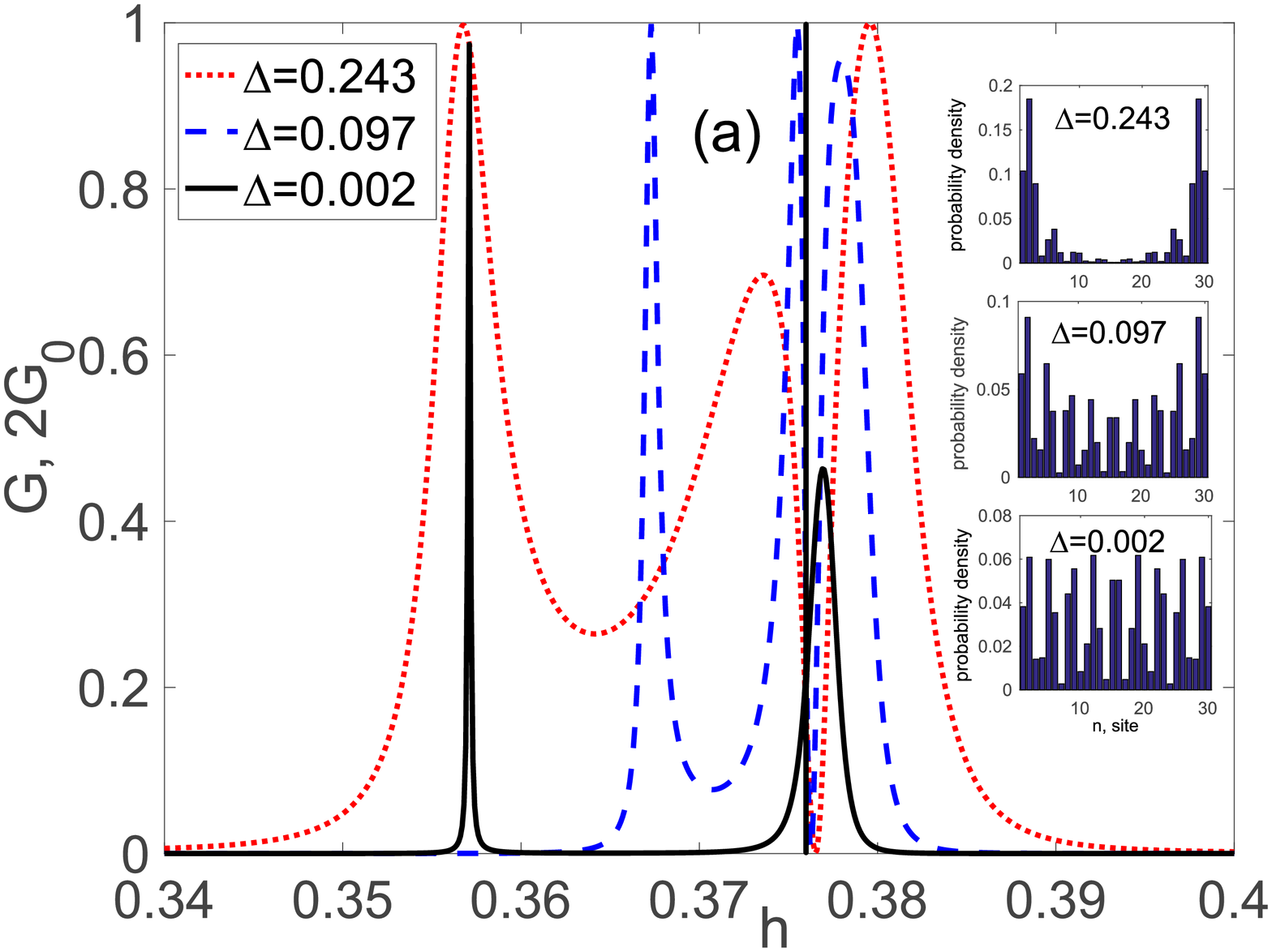}
		\includegraphics[width=0.48\textwidth]{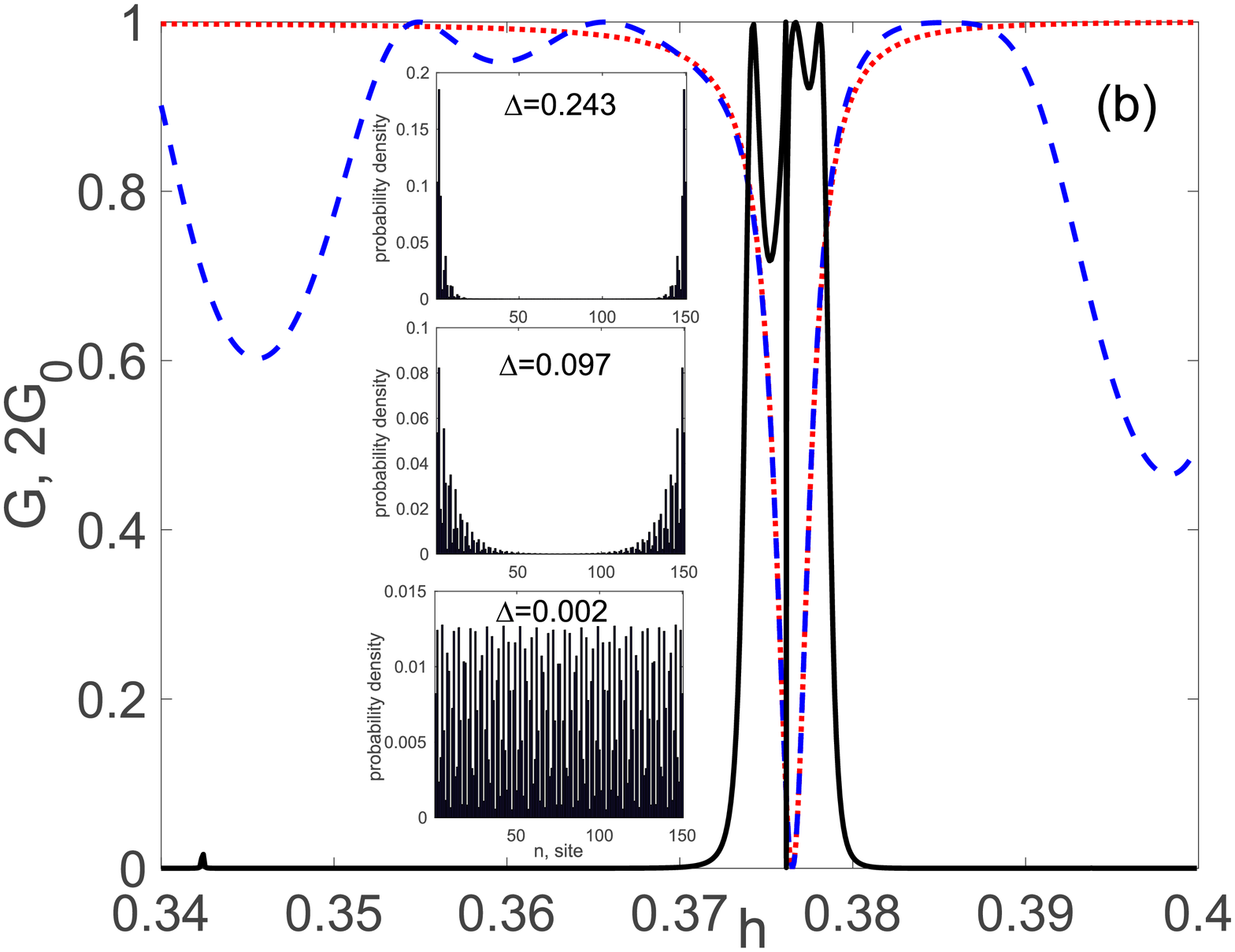}
		\caption{\label{GVz_Del} Dependence of Fano resonance on the superconducting gap of a wire with (a) $N=30$ and (b) $N=150$ (b). Insets: spatial distribution of probability density (PD) of the state with energy $\varepsilon_{W1}$ at different $\Delta$ and $h=0.375$.}
	\end{center}
\end{figure*}

It follows from the results in Section \ref{sec3.4.2}, that, regardless of the length of the superconducting bridge, Fano resonances appear for any nonzero $\alpha$, $\Delta$ and $\mu$ if the magnetic field satisfies condition
\eqref{eq_top}. In other words, this resonance is present regardless of the spatial distribution of the probability density of the state with energy $\varepsilon_{W1}$, which changes from edge (Majorana) to bulk (Andreev) as the gap width in the SW excitation spectrum decreases \cite{haim-15a}.
But the properties of the Fano resonance are substantially dependent on the type of this state, which can be controlled using a magnetic field or by varying the magnitude of the superconducting gap. The latter case is shown in Fig.  \ref{GVz_Del},
where the red dotted and blue dashed curves show the $G\left(h\right)$ dependences corresponding to  $\Delta=0.243$
and $\Delta=0.097$. As we can see, the width of the asymmetric peak decreases significantly if an ABS is realized at
$\Delta=0.097$ (cf. the upper and central insets in Fig. \ref{GVz_Del}a).

The state with energy $\varepsilon_{W1}$ at $\Delta=0.097$ can be transformed into the Majorana-type one by taking a longer wire (see the central inset in Fig. \ref{GVz_Del}b). As a result, the width of the Fano resonance is restored and stays unchanged until the overlap of the wave functions of the two MMs tends to zero (red dotted and blue dashed curves in Fig. \ref{GVz_Del}b). If, further, $\Delta\rightarrow0$, then the Fano resonance collapses (its width becomes infinitesimal) even though $\varepsilon_{W1}\approx0$, i.e., a BIC arises \cite{kim-99a,kim-99b}. This situation corresponds to the black solid curves in Fig. \ref{GVz_Del}a,b.

\begin{figure}[ht]
	\begin{center}
		\includegraphics[width=0.78\textwidth]{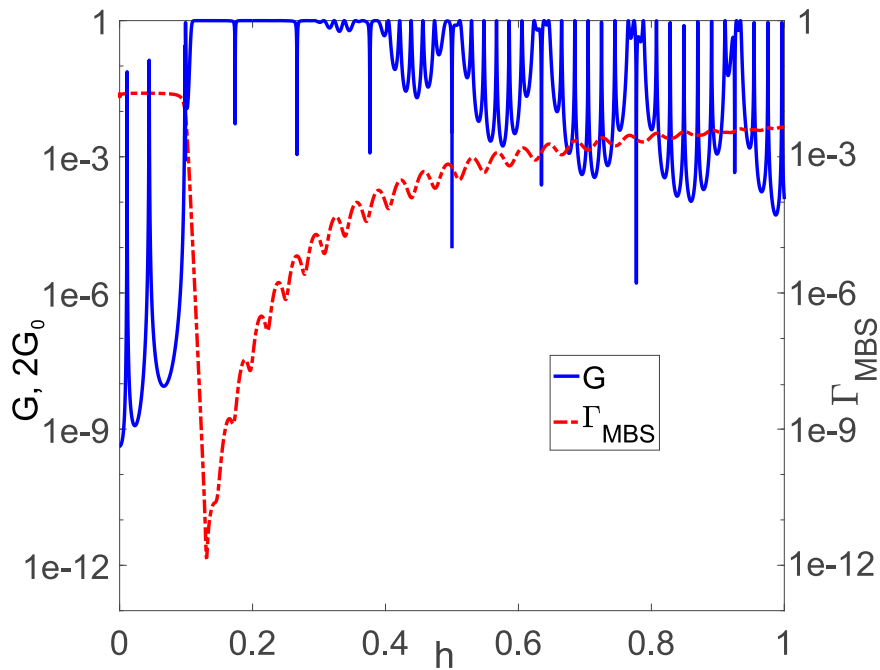}
		\caption{\label{GVz_MSAS} Dependence of conductance (left vertical axis) and overlap $\Gamma_{MBS}$ of Majorana wave functions (right vertical axis) on the Zeeman energy at $\Delta=0.097$ and $N=150$.}
	\end{center}
\end{figure}

As we have discussed, the transition from an MBS to an ABS can be observed as the magnetic field increases. This evolution can be easily traced, for example, using the Majorana polarization introduced above. For clarity, we restrict ourself to a simpler and more intuitive quantity characterizing the overlap of the MM wave functions as the sum of the probability densities of the first Bogoliubov excitation at two central sites of the SW,
\begin{equation}\label{olp}
\Gamma_{MBS}=\sum\limits_{\sigma;l=N/2}^{N/2+1}\left(\left| u_{1l\sigma}\right|^2 +\left| v_{1l\sigma}\right|^2\right),
\end{equation}
where $u_{1l\sigma},~v_{1l\sigma}$ are spin-dependent coefficients of the Bogoliubov transformation for an excitation with energy
$\varepsilon_{W1}$ at the $l$th site. The behavior of this quantity as a function of the magnetic field energy is shown by the red dashed-dotted curve in Fig. \ref{GVz_MSAS} on a logarithmic scale. In the topologically trivial phase
($h<\Delta$) $\Gamma_{MBS}$ remains practically unchanged, and the conductance exhibits small-amplitude peaks (blue solid curve in Fig. \ref{GVz_MSAS}). In the case of a topological quantum phase transition, $\Gamma_{MBS}$
decreases sharply (by a factor of $10^{10}$) due to the change in the type of excitation from bulk to edge. In the topologically nontrivial phase ($h>\Delta$), two domains can be distinguished. For $h\approx 0.1$ --- $0.3$, the MM overlap increases, but is still insignificant. In this range of Zeeman energies, the high transmission regime is dominant, $G=1$, with periodically arising Fano antiresonances (with $G=0$). On the contrary, for $h> 0.3$ a low transmission regime sets in, $G\ll1$,  with periodically arising Breit-Wigner resonances with $G=1$.
In this domain, the hybridization of the Majorana wave functions becomes substantial in strong magnetic fields.

It follows from \eqref{spec6QD} that the positions of the maxima in the density of states and the corresponding conductance resonances are directly dependent on the hopping parameters. Numerical calculations show that, as the size of the system increases, the effect of the parameter $t_{0}$ on the Fano resonances is largely determined by the type of spatial distribution of the state with energy $\varepsilon_{W1}$. Namely, as $t_{0}$ increases, the asymmetric peak preserves its position in the case of an MBS. If the state transforms into an ABS, an increase in  $t_{0}$ is accompanied by a shift in the Fano resonance.

\subsubsection{\label{sec3.4.5} Asymmetric ring}

\begin{figure}[ht]
	\begin{center}
		\includegraphics[width=0.78\textwidth]{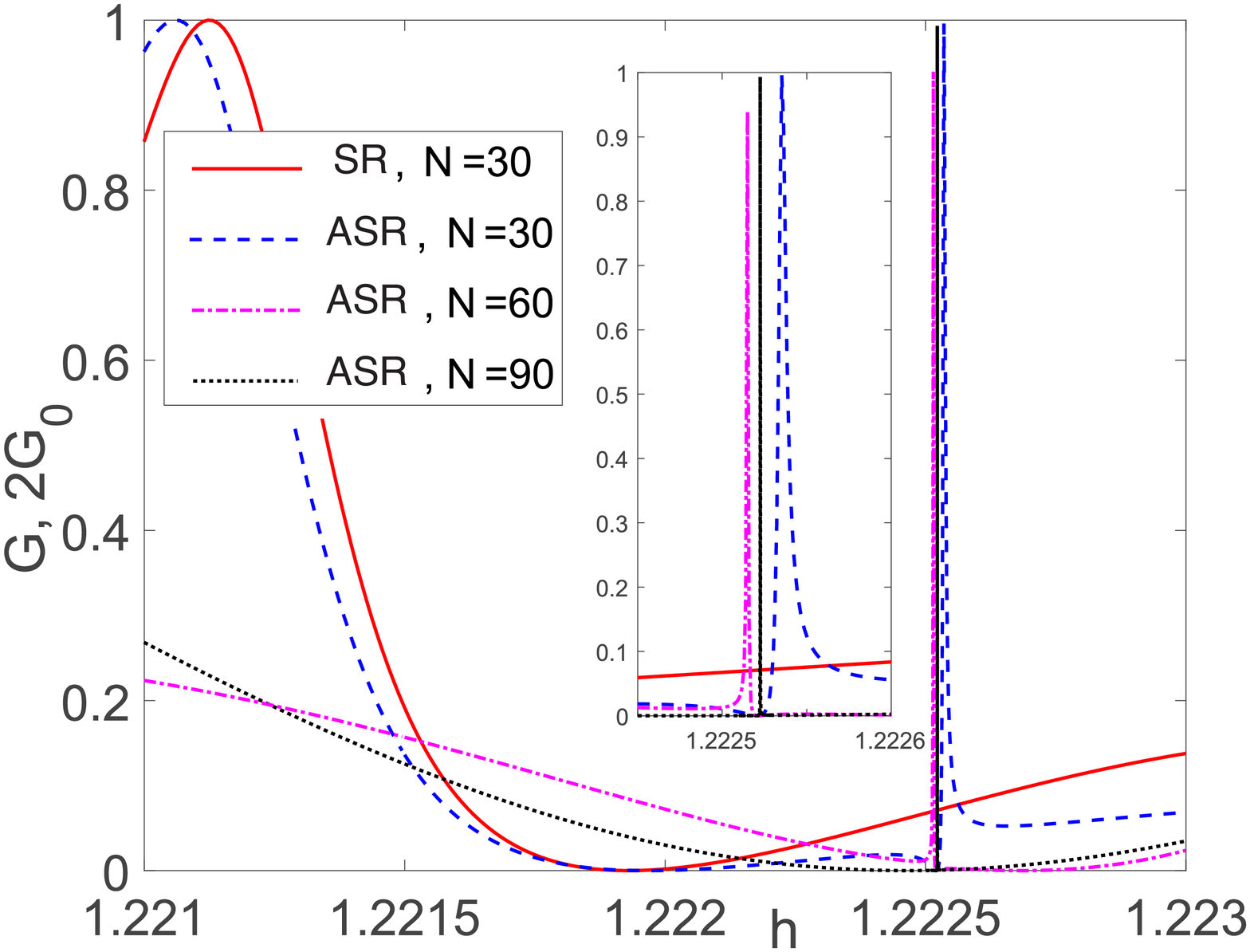}
		\caption{\label{GVz_asym} Collapse of Fano resonance induced by asymmetry of tunneling processes in a ring with an increase in the nonlocality of the MBS. SR: symmetric ring, ASR: asymmetric ring. Inset: collapse at a higher resolution in $h$.}
	\end{center}
\end{figure}

Additional transport features associated with the nonlocality of MBSs are manifested in asymmetric rings \cite{aksenov-20}. We recall that asymmetry is understood as a difference in the diagonal broadening parameters,
$\Gamma_{11}=2\Gamma_{22}$, while preserving the symmetry of the tunneling processes between the NW and the SW. In this case, an additional narrow Fano resonance arises (cf. the solid and dashed curves in Fig. \ref{GVz_asym}), which is identical to the resonance induced by a nonzero Aharonov-Bohm phase in Fig.  \ref{BW&F}b. In Fig. \ref{GVz_asym},this resonance is realized at $h\approx1.2225$. As the length of the bridge increases, the wide Fano antiresonance located around $h\approx1.222$ at $N=30$  (dashed line) shifts toward the narrow one. At the same time, the asymmetry-induced Fano resonance collapses as $N$ increases, as can be clearly seen in the inset in Fig. \ref{GVz_asym}, and a BIC is formed. In other words, we can speak of a kind of topological blockade of the Fano effect, because the Fano resonance disappears only if a true MBS with two decoupled MMs is realized.

To explain the mechanism leading to the collapse of the Fano resonance, it is important to recall that this asymmetric peak corresponds to a BIC that arises due to the degeneracy of the zero-energy eigenstates of a closed system. The disappearance of a Fano resonance can therefore be indicative of an increase in the degeneracy multiplicity of this state if the overlap of the Majorana wave functions becomes negligible. To test this hypothesis, we turn to the spinless model of a ring with $n=1$. In this situation, we regard a Kitaev chain with an even number of sites as a bridge connecting four QDs with energies $\xi_{j}$
($j=1,...,4$) \cite{kitaev-01}. Diagonalizing the Hamiltonian of the device with $\xi_{j}=\xi=\varepsilon-\mu=0$, we obtain the equation for the spectrum
\begin{eqnarray}\label{CMs}
\varepsilon^{4}\left(\varepsilon\cdot P_{1}-2t_{0}^{2}\delta_{1}^{N/2-1}\right)\left(\varepsilon\cdot P_{2}+2t_{0}^{2}\delta_{1}^{N/2-1}\right)
\left(\varepsilon\cdot P_{3}-2t_{0}^{2}\delta_{2}^{N/2-1}\right)\left(\varepsilon\cdot P_{4}+2t_{0}^{2}\delta_{2}^{N/2-1}\right)=0,\nonumber
\end{eqnarray}
where $\delta_{1,2}=t \mp \Delta$; $P_{i}$ is the $i$th polynomial of degree $N/2$, and
$P_{2,4}=P_{1,3}$ due to the electron-hole symmetry $\varepsilon\to-\varepsilon$.
It follows from this equation that, at singular points of the Kitaev model, $\Delta=\pm t$, where the MM wave functions do not overlap, the degeneracy multiplicity of the zero-energy state does indeed increase for $N>2$, leading to a suppression of the narrow Fano resonance in Fig. \ref{GVz_asym}.

\begin{figure}[ht]
	\begin{center}
		\includegraphics[width=0.78\textwidth]{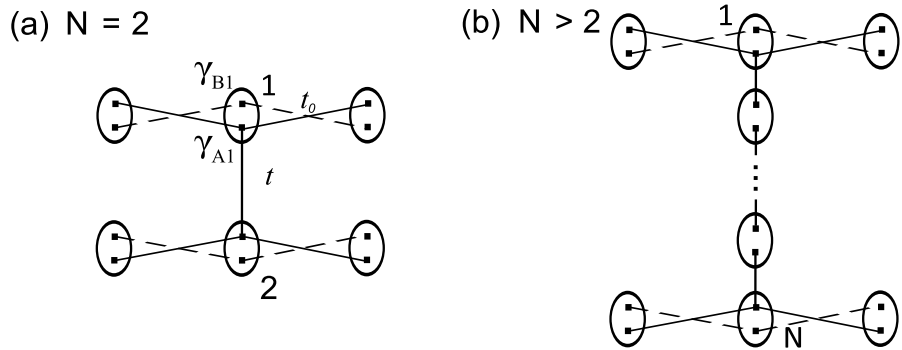}
		\caption{\label{ringMaj} Ring with $n=1$ in terms of Majorana operators for (a) $N=2$ and (b) $N>2$ at $\Delta=t$ and $\xi_{j}=\xi=0$.}
	\end{center}
\end{figure}

For clarity, we discuss the system in the representation
of self-conjugate Majorana operators
$\gamma_{jl}=\gamma_{jl}^{+}$ ($j=A,B$): $a_{l}=\left(\gamma_{Al}+i\gamma_{Bl}\right)/2$. In Fig. \ref{ringMaj} we schematically depict devices in
the framework of such a description at the singular point of
the Kitaev model $\Delta=t$ with $N=2$ and
$N>2$ (straight lines show the coupling between different kinds of Majorana operators). As we can see, the upper and lower arms remain connected in the first case. The eigenergies of the chains
shown with dashed straight lines, which have only two
links in the horizontal direction, are $\varepsilon_{1}=0$ and
$\varepsilon_{2,3}=\pm t_{0}/\sqrt{2}$. In turn, the energies of the chain with hopping between the arms taken into account (shown with solid straight lines) are $\varepsilon_{4,5}=0$,
$\varepsilon_{6,7}=\left(\sqrt{t^2+2t_{0}^2}\pm t\right)/2$ and
$\varepsilon_{8,9}=-\left(\sqrt{t^2+2t_{0}^2}\pm t\right)/2$. Asaresult,thezero-energystate is fourfold degenerate.

In the second case, $N>2$, the device splits into upper and lower identical subsystems, which are not coupled to each other, because the Majorana operators belonging to one site do not interact for $\xi=0$. Each subsystem includes two chains. In this case, the energies of the second chain shown by solid straight lines containing a structural element in the vertical direction (similar to the Fano-Anderson model  \cite{miroshnichenko-10}) are
$\varepsilon_{1,2}=0$ and $\varepsilon_{3,4}=\pm \sqrt{t^2+t_{0}^2/2}$. As a result, the zero-energy level is sixfold degenerate in this case.
Thus, it is the appearance of T-shaped structures of Majorana operators that leads to the suppression of Fano resonance in the asymmetric ring. We emphasize that this effect, being related solely to the nonlocality of the MBS, has a universal character and arises in the most general situation characteristic of an experiment, when all the parameters of tunneling between different subsystems are different. In addition, it becomes obvious from Fig. \ref{ringMaj} that, in the simplest case of  $t=0$ (two decoupled arms), the Fano resonance is not suppressed.

\subsubsection{\label{sec3.4.6} T-shaped geometry}

\begin{figure*}[ht]
	\begin{center}
		\includegraphics[width=0.5\textwidth]{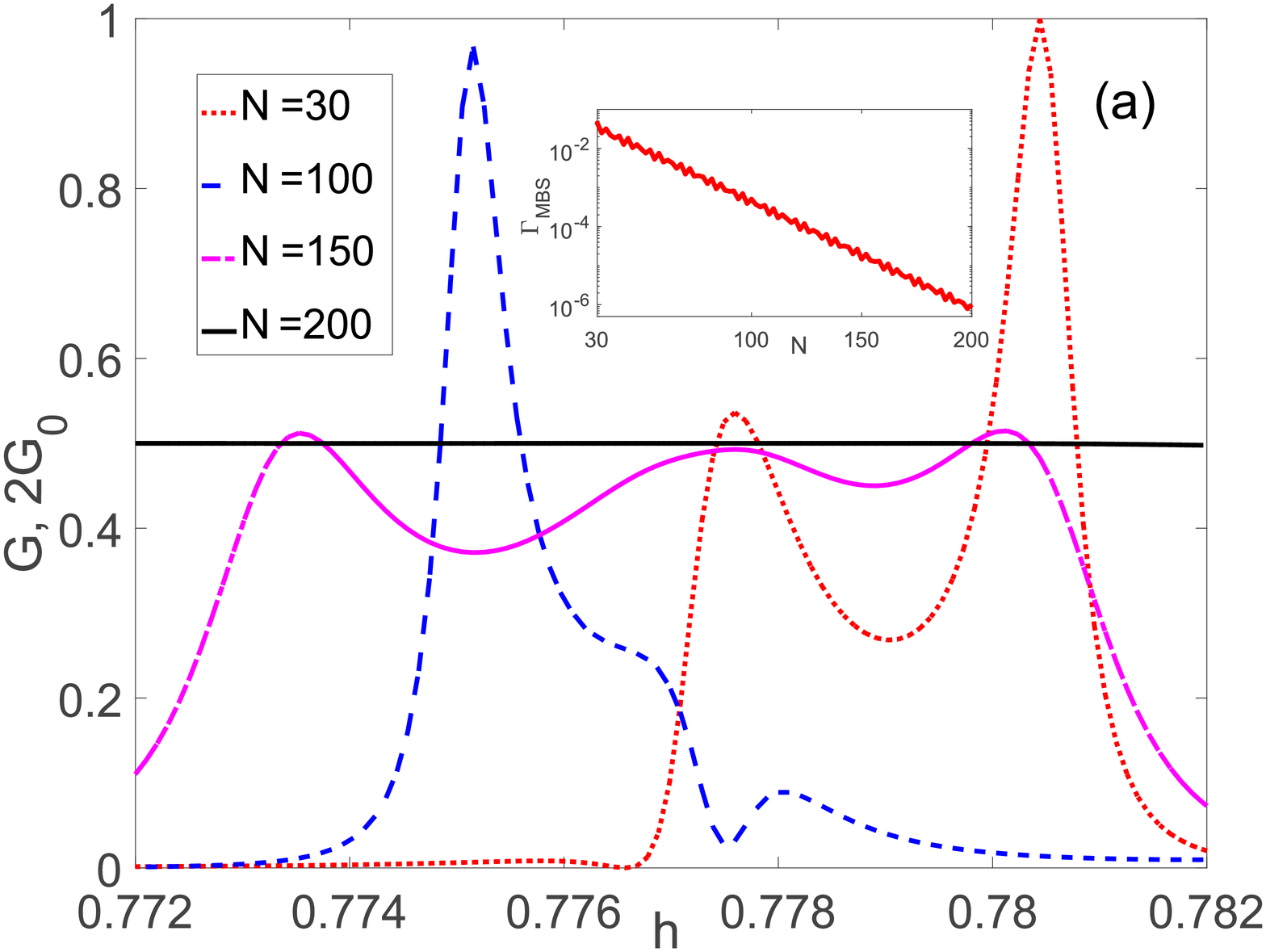}
		\includegraphics[width=0.48\textwidth]{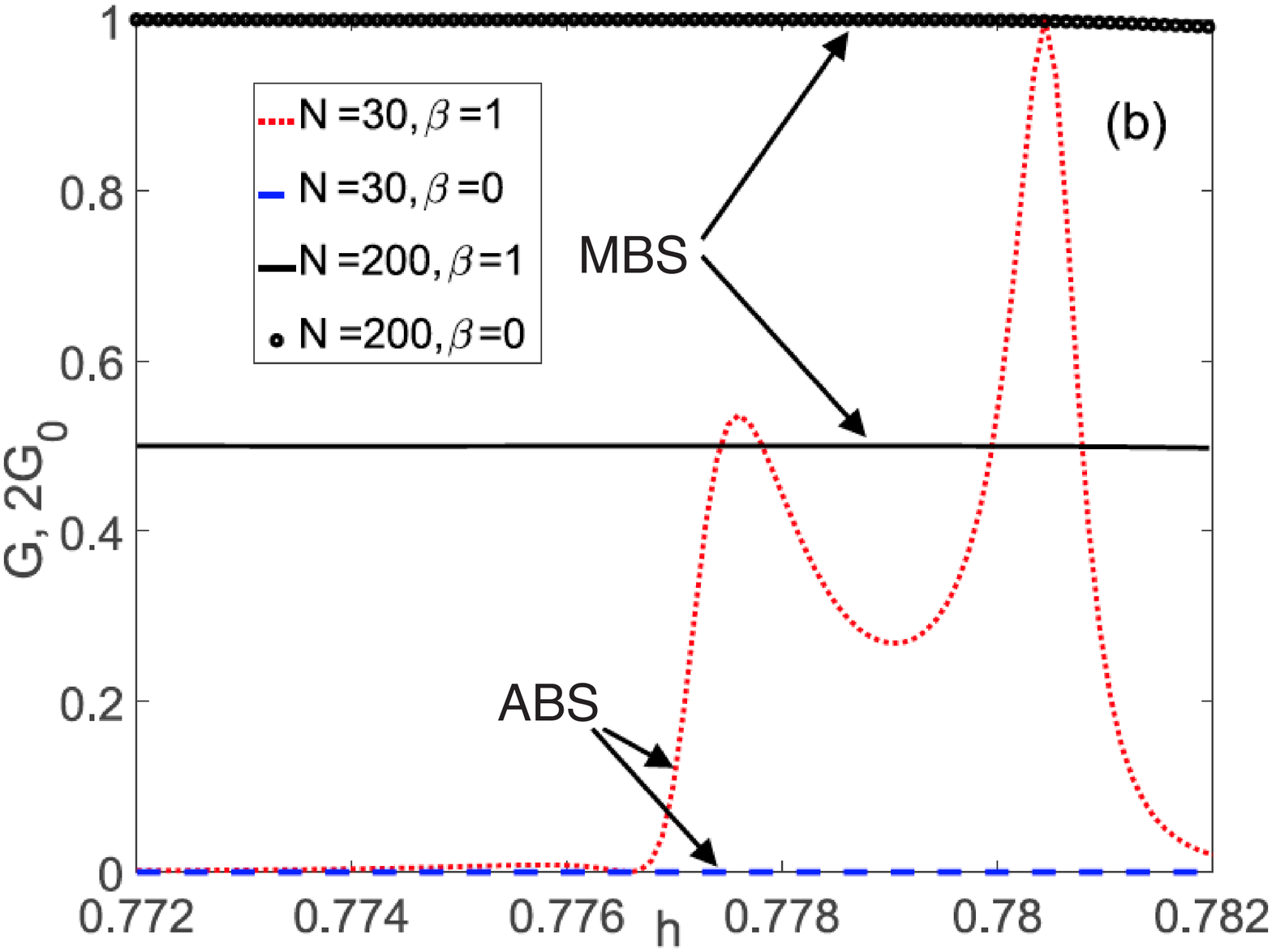}
		\caption{\label{Tgeom} Dependence of conductance of the device on Zeeman energy in the case of T-shaped geometry. (a) Effect of spatial distribution of the state with energy $\varepsilon_{W1}$ on Fano resonance in the nonlocal transport regime. Inset: overlap of Majorana wave functions \eqref{olp} depending on the length of the superconducting nanowire at $h=0.777$. (b) Comparison of conductance in the nonlocal ($\beta=1$) and local ($\beta=0$) transport regimes.}
	\end{center}
\end{figure*}

The fundamental difference between quantum transport through an MBS and an ABS can be clearly demonstrated by considering one of the limit cases of an asymmetric ring, the T-shaped geometry, where there is no coupling between the lower NWs and the contacts. In Fig. \ref{Tgeom}a, we show the dependences of the conductance of the ring on the Zeeman energy when the SW is in a nontrivial phase for several values of $N$ (which is the number of bridge sites). An exponential decrease in the overlap of the MM wave functions with an increase in the wire length is shown in the inset in Fig. \ref{Tgeom}. We can see that  $\Gamma_{MBS}$ decreases by two orders of magnitude if the number of sites in the wire increases from $N=30$ to $N=100$. The state with the energy $\varepsilon_{W1}$ then transforms from an Andreev state into a Majorana state and the antiresonance in the vicinity of $h=0.777$ elevates; hence, $G\neq0$ in this case (cf. the red dotted and blue dashed lines in Fig. \ref{Tgeom}a). With a further increase in the length of the SW, $\Gamma_{MBS}$ decreases by several orders of magnitude. As a result, a conductance plateau of the height $G_0$ arises (purple dashed-dotted and black solid lines).

This behavior is due to the fact that, in the T-shaped transport scheme, the coupling of the lower left and right NWs to the conducting channel (or contacts) is dependent on the type of low-energy state in the bridge. In the case of an MBS  ($N\gtrsim100$), the coupling of these structural elements to the contacts is absent, and the Fano resonance disappears. If an ABS is realized ($N<100$), then the above NWs are lateral, as in the case of the Fano-Anderson model  \cite{miroshnichenko-10}. In that case, a Fano resonance is realized.

More significant differences can be seen when we turn to the features of local transport via MBSs and ABSs. In the case of MBSs, the height of the conductance plateau doubles upon passing from the nonlocal, $\beta=1$, to the local, $\beta=0$, regime  (where $\beta$ is the asymmetry parameter; see Hamiltonian \eqref{H_MBS}). The corresponding dependences
$G\left(h\right)$ are shown by the solid line and black circles in Fig. \ref{Tgeom}b. In the case of ABSs, the conductance of the local configuration is suppressed (dashed line in Fig.  \ref{Tgeom}b).

The properties of the conductance of the T-shaped structure found in numerical calculations can be described analytically. For this, we have to clarify that transport through an MBS essentially implies the coupling of the contacts to only one Majorana operator. But tunneling through an ABS, which corresponds to the Bogoliubov operator
$\alpha^{+}$, actually implies the coupling to both MMs,  $\gamma_{A}$ and $\gamma_{B}$, because
$\alpha^{+}=\left(\gamma_{A}+i\gamma_{B}\right)/2$.

We systematically consider the transport properties of a T-shaped structure in which either an MBS or ABS is realized. To analyze transport in the case of an MBS, we use the effective low-energy SW Hamiltonian that describes the coupling of a pair of MMs with strength $\varepsilon_{0}$, i.e.,
${H}_{W}=i\varepsilon_0\gamma_{A}\gamma_{B}/2$ \cite{kitaev-01}. For simplicity, we replace the NW with a single-level QD with energies $\xi_{j}$ ($j=1,...,4$) and discuss the spinless case. Then,
\begin{eqnarray}\label{H_MBS}
&&{H}_{T}=-t_{0}\left[b_{1}^{+}-b_{1}+\beta\left(b_{2}^{+}-b_{2}\right)\right]\gamma_{A}
-t_{0}\left[d_{3}^{+}-d_{3}+d_{4}^{+}-d_{4}\right]\gamma_{B},\\
&&H_{V}=-\sum\limits_{k}\left(t_{1}c_{Lk}^{+}b_{1}+t_{2}c_{Rk}^{+}b_{2}+h.c.\right).\nonumber
\end{eqnarray}

Appling formula \eqref{IL3} to a symmetric T-shaped system ($t_{1}=t_{2}$) leads to the expression for the conductance of the left contact in the linear response regime,
\begin{equation}\label{GLTsh}
\hat{G}_{L}=\hat{G}_{0}\frac{\Gamma^2}{2}\left[2\left| \hat{F}_{11e}^{r}\left(0\right) \right|^2+\left|
\hat{G}_{12e}^{r}\left(0\right) \right|^2+\left| \hat{G}_{12h}^{r}\left(0\right) \right|^2  \right].
\end{equation}
where $\hat{F}^{r}_{11e}\left(\omega\right)$, $\hat{G}^{r}_{12e}\left(\omega\right)$ and 
$\hat{G}^{r}_{12h}\left(\omega\right)$ are Fourier transforms of the anomalous and normal Green's functions:
\begin{eqnarray}
&&\hat{F}^{r}_{11e}\left(t-t'\right)=-i\Theta\left(t-t'\right)\left\langle \left\{ b^{+}_{1}\left(t\right), b^{+}_{1}\left(t'\right) \right\} \right\rangle, \nonumber\\
&&\hat{G}^{r}_{12e}\left(t-t'\right)=-i\Theta\left(t-t'\right)\left\langle \left\{ b_{1}\left(t\right), b^{+}_{2}\left(t'\right) \right\} \right\rangle, \nonumber\\
&&\hat{G}^{r}_{12h}\left(t-t'\right)=-i\Theta\left(t-t'\right)\left\langle \left\{ b^{+}_{1}\left(t\right), b_{2}\left(t'\right) \right\} \right\rangle. \nonumber
\end{eqnarray}
Here, $\Theta$ is the Heaviside function.  We note that the term proportional to $\hat{F}^{r}_{11e}$ in \eqref{GLTsh} describes the contribution of the processes of local Andreev reflection. The next two terms are responsible for the direct charge transfer processes. The contribution of the processes of crossed Andreev reflection is absent because $\mu_{L\left(R\right)}=\mu\pm eV/2$
\cite{wu-14}.

Using the method of the equations of motion \cite{zubarev-60}, we obtain the required Green's functions:
\begin{eqnarray}\label{FG1_MBS}
&&\hat{F}^{r}_{11e}=-\frac{2t_{0}^2zZ_{B}C_{2}}{zZ_{T}Z_{B}-\varepsilon_{0}^2C_{T}C_{B}}, \\
&&\hat{G}^{r}_{12e\left(h\right)}=\frac{2\beta t_{0}^2zZ_{B}C_{1h\left(e\right)}C_{2h\left(e\right)}}{z Z_{T}Z_{B}-\varepsilon_{0}^2C_{T}C_{B}},\nonumber
\end{eqnarray}
where $z=\omega+i\delta$, $C_{je\left(h\right)}=z\mp\xi_{j}+i\delta_{jj_T}\Gamma/2$ $\left(j_T=1,2\right)$, $C_{j}=C_{je}C_{jh}$, $C_{T\left(B\right)}=C_{1\left(3\right)}C_{2\left(4\right)}$, $Z_{T}=z C_{T}-2t_{0}^{2}\left(C_{1e}+C_{1h}\right)\left(\beta^2C_{1}+C_{2}\right)$,
$Z_{B}=C_{B}-4t_{0}^2\left(C_{3}+C_{4}\right)$.
If $\xi_{1}=\xi_{2}=\xi_{3}=\xi_{4}=0$, then
\begin{equation}\label{FG2_MBS}
\hat{G}^{r}_{12e}=\hat{G}^{r}_{12h}=-\beta \hat{F}^{r}_{11e}=\frac{2\beta t_{0}^{2}\left(z^2-8t_{1}^2\right)}{C_{1e}Z_{1}},
\end{equation}
where $Z_{1}=\left(z^2-8t_{0}^2\right)\left(zC_{1e}-4t_{0}^{2}\left(1+\beta^2\right)\right)-\varepsilon_{0}^2zC_{1e}$.

Hence, at $\beta=1$, the conductance becomes equal to $G_{0}/2$. The resulting conductance value can be explained qualitatively if we consider the entire system in the representation of Majorana operators. From the form of the operator representing the coupling between the QDs and the MBS \eqref{H_MBS}, t follows that the left and right contacts are coupled only by a chain containing a vertical connection, shown by solid
straight lines in Fig. \ref{ringMaj}a. 

In the general case, as with ordinary fermions and
electromagnetic fields \cite{miroshnichenko-05}, the presence of a conductance resonance or antiresonance at $\omega=0$ depends both on the number of coupled Majorana operators in the lower branch and on the particular operator that directly couples to  $\gamma_{A}$. In the situation under consideration, the lower chain includes three operators: $\gamma_{3B}$, $\gamma_{B}$ and $\gamma_{4B}$. Moreover, because its eigenvector corresponding to a zero-energy state is proportional to $\sin \left(\frac{\pi l}{2}\right)$ (where $l=1,2,3$) and ${H}_{W}=i\varepsilon_0\gamma_{A}\gamma_{B}/2$, the upper and lower chains effectively decouple. In other words, the Fano effect is absent and resonance transmission is observed.

We note that in considering the T-shaped geometry, the case with only one Majorana operator in the lower chain is also possible. It occurs if the SW is represented by the Kitaev chain model at a singular point and $\xi_{1}=\xi_{2}=0$. In terms of Majorana operators, such a system is presented in Fig. \ref{ringMaj}b. Similarly to what was shown in \cite{liu-11}, the total conductance is equal to the sum of contributions from chains with two and three links. In the first case, it is $G_{0}/2$, and in the second, $0$ due to the Fano effect.

In the local transport regime $\beta=0$, we have $G_{L}=G_{0}$. The same results can be obtained for $\xi_{1}=\xi_{2}=\xi_{3}=\xi_{4}\neq0$ and $\varepsilon_{0}=0$ (see Fig. \ref{Tgeom}). In the general case of $\xi_{j},~\varepsilon_{0}\neq0$, conductance \eqref{GLTsh} tends to zero.

We turn to transport in a T-shaped structure with an ABS of energy $\xi$. The corresponding low-energy SW Hamiltonian is ${H}_{W}=\xi\alpha^{+}\alpha$. Because the ABS and the original secondary quantization operators are related by a Bogoliubov transformation, $a_{1\left(N\right)}\approx u\alpha\pm v\alpha^{+}$,
we can represent the Hamiltonian of tunneling between the QD and the ABS in the form \cite{tripathi-16}
\begin{eqnarray} \label{HWl_ABS}
{H}_{T} =-t_{0e}\left(b_{1}^{+}+\beta b_{2}^{+}+d_{3}^{+}+d_{4}^{+}\right)\alpha-t_{0h}\left(b_{1}+\beta b_{2}+d_{3}+d_{4}\right)\alpha+h.c.,\nonumber
\end{eqnarray}
where $t_{0e\left(h\right)}$ is the electron (hole) tunneling amplitude. 

Solving the equations of motion for the Green's functions
with $\xi_{1}=\xi_{2}=\xi_{3}=\xi_{4}=0$ gives
\begin{eqnarray}\label{FG1_ABS}
&&\hat{F}^{r}_{11e}=-\frac{2t_{0e}t_{0h}z^3}{Z_{2}}, \\
&&\hat{G}^{r}_{12e\left(h\right)}=\frac{\beta z}{C_{1e}Z_{2}}\left[zC_{1e}\left(t_{0e}^2C_{h\left(e\right)}+t_{0h}^2C_{e\left(h\right)}\right)\textbf{\textcolor{blue}{-}}\left(t_{0e}^2-t_{0h}^2\right)^2\left(z\left(1+\beta^2\right)+2C_{1e}\right)\right],\nonumber
\end{eqnarray}
where $C_{e\left(h\right)}=z\mp\xi$ and
\begin{eqnarray}\label{Z2}
&&Z_{2}=z^2C_{1e}^2C_{e}C_{h}+\left(t_{0e}^2-t_{0h}^2\right)^2\left(z\left(1+\beta^2\right)+2C_{1e}\right)^2-\nonumber\\
&&~~~~~~~~~~~~~~~~~~-2z^2C_{1e}\left(t_{0e}^2+t_{0h}^2\right)\left(z\left(1+\beta^2\right)+2C_{1e}\right).\nonumber
\end{eqnarray}
It follows from \eqref{FG1_ABS} that, in the linear response approximation, the conductance vanishes in both local and nonlocal regimes. As in the case of an MBS, the obtained result can be explained qualitatively by moving to the representation in terms of Majorana operators. At $\xi=0$, the Hamiltonian of the system ABS $+$ four QDs then becomes
\begin{equation} \label{H_ABSQQD}
H=i\frac{t_{0e}+t_{0h}}{2}\left(\gamma_{1A}+\beta\gamma_{2A}+ \gamma_{3A}+\gamma_{4A}\right)\gamma_{B}-i\frac{t_{0e}-t_{0h}}{2}\left(\gamma_{1B}+\beta\gamma_{2B}+ \gamma_{3B}+\gamma_{4B}\right)\gamma_{A}
\end{equation}
Thus, there are two chains, in each of which the central operator $\gamma_{A\left(B\right)}$ interacts laterally with two others, $\gamma_{3B\left(3A\right)}$ and $\gamma_{4B\left(4A\right)}$ (which are not coupled to each other). At $\omega=0$, $G=0$ in both chains due to the Fano effect.

In the general case of $\xi_{j}\neq0$, the Green's functions have a rather cumbersome form and are not presented here.  However, we emphasize that,in this case, $F^{r}_{11e}\left(0\right)=0$ for $\beta=0,1$, and $G^{r}_{12e\left(h\right)}\left(0\right)\neq0$ for $\beta=1$. This behavior is in qualitative agreement with the results of the numerical calculations presented in Fig. \ref{Tgeom}.

Thus, the considered Aharonov-Bohm ring with a superconducting bridge allows a topological phase transition in the system to be detected due to the appearance of Fano resonances. In addition, an analysis of the properties of these asymmetric features of the conductance allows distinguishing a true MBS, which has an edge character, from a low-energy excitation of a bulk type, i.e., an ABS.

\section{\label{sec4}Majorana modes in systems
	with noncollinear magnetic order}

\subsection{\label{sec4.1}Introduction to the problem}

Recently, work has intensified on revealing topologically
nontrivial superconducting phases and MMs not only in
systems with spin-orbit coupling but also in a wide class of
structures and materials with magnetic ordering. Interest in
the topological properties of such systems increased significantly after a connection was established in~\cite{braunecker-10} between the
Hamiltonians describing materials with noncollinear or spiral
spin ordering and those of the systems considered in Section \ref{sec3}.

To demonstrate this connection, we subject Hamiltonian \eqref{Ham_latt} to a unitary transformation corresponding to the
successive rotation of the coordinate system in the spin space
through the angle $\pi/2$ around the $y$- and $x$-axes,
\begin{eqnarray}
H_W \to \tilde{H}_W & = & U H U^{\dag},
\end{eqnarray}
where the unitary transformation operator has the form
\begin{eqnarray}
U = \prod_l
\left[ \exp\left(-i\frac{\pi}{2}\sigma_l^x\right) \exp\left(-i \frac{\pi}{2}
\sigma_l^y \right) \right].
\end{eqnarray}
We use the notation $\sigma_l^{x,y,z}$ for spin operators of itinerant
electrons on the $l$th site. Given the transformation law for
Fermi operators
\begin{eqnarray}
\label{Trnsf}
a_{l \sigma} \to \tilde{a}_{l \sigma} & = U a_{l \sigma} U^{\dag}  =
&\frac{1+i\eta_{\sigma}}{2} a_{l \sigma} + i\frac{1-i\eta_{\sigma}}{2} a_{l \bar{\sigma}},
\end{eqnarray}
we obtain the transformed Hamiltonian \eqref{Ham_latt} (with $U = 0$ and $V = 0$)
\begin{eqnarray}
\label{Ham_latt_tr}
&~&\tilde{H}_W = \sum_{l,\sigma}\Big((\varepsilon_0 - \mu)a^{+}_{l\sigma}a^{}_{l\sigma} - \frac{t}{2} \left( a^{+}_{l\sigma}a^{ }_{l+1\sigma} + h.c. \right) \Big)
- \nonumber\\
&-& \frac{i\alpha}{2} \sum_{l} \Big( \left( a^{+}_{l\uparrow}a^{ }_{l+1\uparrow} - a^{+}_{l\downarrow}a^{ }_{l+1\downarrow} \right) - \left( h.c. \right) \Big)
- \nonumber\\
&-& h\sum_{l}\Big( a^+_{l\uparrow}a_{l\downarrow} + a^+_{l\downarrow}a_{l\uparrow} \Big)
+\sum_{l}\Big(\Delta a_{l\uparrow}a_{l\downarrow} + h.c.   \Big),
\end{eqnarray}
where $\alpha$ is the spin-orbit coupling constant. The structure of
the Hamiltonian corresponds to the case \cite{braunecker-10} where the
magnetic field is directed along the $x$-axis and the spin-orbit
coupling vector $\textbf{B}_{SO}$ is parallel to the quantization axis $z$.

Following \cite{braunecker-10}, we perform the transformation $U_2 = \prod_l
\exp\left(-i Q l \sigma_l^z\right)$, under which
\begin{eqnarray}
\label{Trnsf2}
a_{l \sigma} \to a'_{l \sigma} = U_2 a_{l \sigma} U_2^{\dag}  = a_{l \sigma} \exp\left( -i \eta_{\sigma} Q l/2 \right).
\end{eqnarray}
We relate the transformation parameter $Q$ to the hopping
parameters and the spin-orbit coupling as $Qa = 2 \arccos\left( \left( \sqrt{ 1 + \alpha^2/t^2 } \right)^{-1} \right)$, where $a$ is the lattice constant. In
the site representation, the Hamiltonian then becomes
\begin{eqnarray}
\label{Ham_latt_tr2}
H_W' & = & \sum_{l,\sigma}\Big((\varepsilon_0 - \mu)a^{+}_{l\sigma}a^{}_{l\sigma} - \frac{\tilde{t}}{2} \left( a^{+}_{l\sigma}a^{ }_{l+1\sigma} + h.c. \right) \Big)
- \nonumber \\
&-& \sum_{l,\sigma, \sigma'} { a^{+}_{l\sigma} \left( \vec{h}_l \vec{\sigma} \right)_{\sigma \sigma'} a^{}_{l\sigma'} }
+\sum_{l}\Big(\Delta a_{l\uparrow}a_{l\downarrow} + h.c.   \Big),
\end{eqnarray}
where $\tilde{t} = t\sqrt{ 1 + \alpha^2/t^2 }$ and $\vec{\sigma} $ is a vector composed of Pauli
matrices. The transformed Hamiltonian corresponds to an
ensemble of fermions with the superconducting gap $\Delta$ in the magnetic field of a helicoidal structure $\vec{h}_l = h \left( \cos(Ql), -\sin(Ql), 0 \right)$. We can see that the spin-orbit
coupling parameter $\alpha$ determines the renormalization of the
hopping integral and the structure of the helicoidal spin
ordering in terms of the wave number $Q$.

Based on this correspondence and using the results on the
prediction of MMs in superconducting systems with spin-orbit coupling in a uniform magnetic field~\cite{sato-09, sau-10, lutchyn-10},
a new class of topologically nontrivial systems without spin-orbit
coupling but with inhomogeneous magnetic ordering was
proposed.

\begin{figure}[ht]
	\begin{center}
		\includegraphics[width=0.78\textwidth]{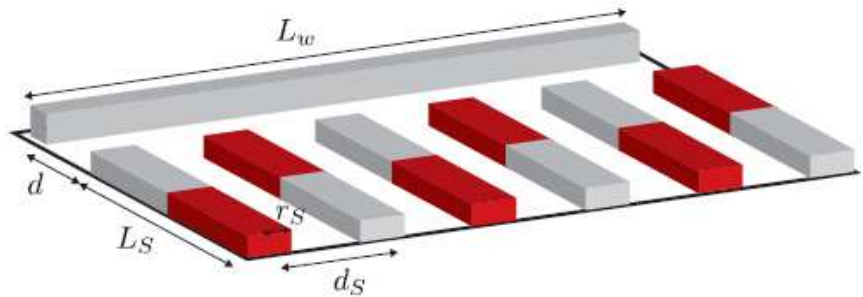}
		\caption{\label{fig_kjaergaard} Schematic of a quantum nanowire placed in an inhomogeneous magnetic field \cite{kjaergaard-12}.}
	\end{center}
\end{figure}

It was shown in \cite{choy-11} that topologically nontrivial phases
exist in chains of magnetic nanoparticles (or atoms) with
arbitrary magnetization directions placed on a superconducting substrate. It was assumed that the distance between the
nanoparticles is less than the coherence length of the superconductor. It was argued that an inhomogeneous external
magnetic field can induce MMs in a nanowire with induced
superconductivity without a pronounced spin-orbit coupling \cite{kjaergaard-12, egger-12, kornich-20}. Such a field can be produced by placing submicrometer magnets near the nanowire, as shown in Fig.~\ref{fig_kjaergaard}. Subsequently, it was proposed to use quasi-1D conductors,
in which the required magnetic structure of localized magnetic moments sets in due to the Ruderman-Kittel-Kasuya-Yosida (RKKI) interaction~\cite{braunecker-13, klinovaya-13, vazifeh-13}. In this case,
the pitch of the magnetic helicoid is determined by the Fermi
momentum of conduction electrons.

It is known that the introduction of an isolated magnetic
impurity into a superconductor gives rise to Yu-Shiba-Rusinov states \cite{yu-65, shiba-68, rusinov-69} whose energies lie inside the superconducting gap. In diluted chains of magnetic atoms,
different impurity states can overlap, giving rise to impurity
bands. Such regimes were considered in detail in a recent
review \cite{choi-19}. Here, we summarize the main results for chains
in the case where the magnetic atoms are so close that their
atomic orbitals overlap.

\begin{figure}[ht]
	\begin{center}
		\includegraphics[width=0.78\textwidth]{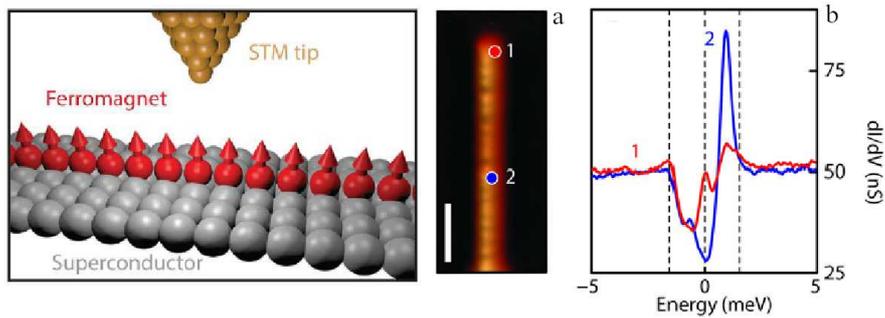}
		\caption{\label{fig_nadjperge} (a) Schematic of scanning tunneling spectroscopy experiment with a spin chain for observing MMs. (b) Dependence
			of the conductance on energy for different positions of the microscope tip
			($1$ and $2$). Zero-energy peak for dependence $1$ is associated with an MM.
			Vertical dashed lines intersecting the horizontal axis at nonvanishing
			energy values determine magnitude of the superconducting gap on a lead
			substrate \cite{nadj-perge-14}.}
	\end{center}
\end{figure}

The first experimental studies of a chain of Fe atoms on
superconducting lead by scanning tunneling microscopy were
carried out in \cite{nadj-perge-13, nadj-perge-14}. The length of the chains reached
several hundred angstroms. The presence of a conductance
peak at zero voltage was found when the microscope tip was
located near the edge of the magnetic chain, which could be
associated with the presence of an MM (Fig.~\ref{fig_nadjperge}). In the
same experiment, at nonzero energies not exceeding the
superconducting gap, the appearance of Yu-Shiba-Rusinov
impurity states was observed.

A similar system was investigated using a tunnel microscope with a superconducting tip \cite{ruby-15} and using atomic force microscopy \cite{pawlak-16}. In both studies, a conductance peak was
found, but it was not observed for all synthesized magnetic
chains. Additional resonances were found near zero energy
(for energies of the order of 80 $\mu$eV). It should be borne in
mind that these resonances, due to their proximity in energy,
could modify the MBSs. In view of these complications in
the interpretation of the experimental data, we emphasize
that the identification of MBSs against the background of
topologically trivial low-energy states is still an experimental
problem that is far from solved.

Resonance at zero voltage was also observed in a ferromagnetically ordered chain consisting of Co atoms on a single
crystal of lead \cite{ruby-17}. However, it was detected when the tip
passed over the entire chain, which indicated its delocalized
character. Therefore, such a peak could not be associated
with MMs. It was believed that an even number of Fermi
points were realized in the Co chain when the Fermi level
crossed the electron bands, in contrast to an odd number of
points in the case of an Fe chain, and therefore the cobalt
chain was in a topologically trivial phase and the conductance
peak was determined by other mechanisms.

The results of the experimental studies cited above did not
allow unambiguously interpreting the observed conductance
peak as the one related to MBSs existing in the chain for the
following reasons:
(1) the peak was not quantized; (2) it was
asymmetric with respect to the voltage reversal; (3) it was
observed only for some chains.

In more accurate experiments carried out at a temperature
of 20 mK \cite{feldman-17}, it was possible to separate the presumed
MBSs from the Yu-Shiba-Rusinov impurity states and
demonstrate a symmetric Majorana peak. Its magnitude
increased, but it was still an order of magnitude less than the
predicted conductance value of $2G_0 = 2e^2/h$. Additional
difficulties in the interpretation also arose, because the
influence of defects in the chain on the realization of a zero-voltage resonance remained insufficiently studied.

We note that, in practice, for a number of reasons, it is also
difficult to identify the helicoidal ordering of spin moments in
atomic chains. In \cite{nadj-perge-14}, ferromagnetic ordering was assumed
to occur in the chain of Fe atoms. In this case, the presence of
MBSs was associated with the spin-orbit coupling in lead.
However, as noted in \cite{pawlak-16}, even the formation of a spin
helicoid with a small angle in the chain can be sufficient for
MMs to be partially generated by this mechanism.

Later, 120-degree ordering of the spin moments of Fe
atoms was realized when they were placed on a superconducting rhenium substrate Re(0001) \cite{kim-18}. However, an
experiment using scanning tunneling spectroscopy did not
confirm the presence of MMs in such a structure. Notably, in
the study of differential conductance at zero voltage with the
tip located near the boundaries of the chain, the presence of
only local maxima, and not peaks, was demonstrated.
Experimental studies of an Mn monolayer on Re(0001) \cite{spethmann-20} have led to the discovery of a complex spin structure (the so-called $3Q$ state). Such spin ordering can also induce
topological superconductivity and MMs \cite{bedow-20}. We note that
several hybrid superconducting structures with magnetic
ordering allowing the realization of MBSs have been
proposed recently \cite{crawford-20, rex-20}.

Note that MBSs can be realized in chains with antiferromagnetic ordering of magnetic atoms brought into contact
with a superconductor. In that case, the underlying mechanism is the modification of the initial impurity states by an
external magnetic field oriented in the plane of the superconductor and by the supercurrent running along the
chain axis~\cite{heimes-14}. Additional information on the possibility of
realizing MBSs in spin chains is contained in recent reviews \cite{choi-19, pawlak-19}.

A transformation similar to \eqref{Trnsf}, \eqref{Trnsf2} can also take place
in 2D systems. In this case, as shown in \cite{martin-12}, the helicoidal
spin ordering corresponds to a certain combination of
Rashba and Dresselhaus spin-orbit couplings. This magnetic ordering does not set in due to external conditions but
is induced by internal mechanisms. Hence, the realization of
MMs can be expected in spin-singlet superconductors with a
noncollinear magnetic order.

Evidence of the formation of the phase of coexisting
superconductivity and helicoidal or noncollinear magnetic
order (SC+NCM) has been found in ternary rare-earth
borides and chalcogenides, for example, HoMo$_6$S$_8$ and ErRh$_4$B$_4$~\cite{buzdin-84},
and recently in EuRbFe$_4$As$_4$ \cite{iida-19, devizorova-19, kim-20}.
In HoMo$_6$S$_8$ and ErRh$_4$B$_4$, superconductivity first appears as
the temperature decreases, and then a tendency toward
ferromagnetic ordering reveals itself. But because of superconductivity, nonuniform magnetic ordering forms in a
narrow temperature range~\cite{buzdin-84}. With a further decrease in
temperature, the ternary compounds pass into the normal
state with ferromagnetic ordering.

Promising materials with the possible realization of the
SC+NCM phase include superconductors with a triangular
lattice, such as Na$_x$CoO$_2 \cdot y$H$_2$O~\cite{takada-03}, Ir$_{1-x}$Pt$_x$Te$_2$~\cite{pyon-12}, $\kappa$-(BEDT-TTF)$_2$X \cite{mckenzie-98}.
They are conducive to the
appearance of the SC+NCM phase, because the frustration
of the triangular lattice facilitates the realization of a noncollinear structure during the formation of magnetic ordering
\cite{golosov-89, ritchey-90, chubukov-91, barabanov-92, chubukov-94, chubukov-94b}.
We note that competition between the nearest-neighbor and subsequent exchanges can also lead to helicoidal magnetic ordering \cite{mikheenkov-18}.

It was shown in~\cite{lu-13} that, if a stripe magnetic order is
formed in a triangular lattice with chiral $d_1+id_2$ superconductivity, then MMs arise in such a system. We emphasize that the existence of magnetic ordering and Cooper
instability was ensured in~\cite{lu-13} by adding the mean-field
terms that generate the corresponding averages to the
quadratic Hamiltonian. Sodium cobaltates are also interesting, because topologically nontrivial surface states were
recently discovered in their normal phase at high doping
levels $x$ \cite{yao-15}.

For most compounds whose phase diagrams contain
the superconductivity and antiferromagnetism coexistence
regions \cite{buzdin-86, pfleiderer-09, VVV-ZAO-12},
the conditions for the formation of
MMs are not satisfied. At the same time, in many theoretical
studies where MMs were considered in systems with antiferromagnetic ordering, their realization mechanism was not
directly related to magnetism. For example, the presence of
zero modes in a state with antiferromagnetic ordering and
triplet superconductivity was demonstrated in~\cite{lu-14} for
weakly doped cuprates. In that case, the zero modes occur
solely due to triplet superconductivity. In iron pnictides and
chalcogenides, the possible realization of topologically nontrivial surface states is due to the multiorbital electron
structure of these compounds and already manifests itself in
the normal phase \cite{lau-13}. In the superconducting phase, such
states can coexist \cite{lau-14} with ABSs induced by the $s_{\pm}$-wave symmetry of the superconducting order parameter \cite{ghaemi-09}. In
the phase of coexistence of superconductivity and a spin
density wave, which is experimentally observed in pnictides,
superconductivity with a spin-triplet order parameter that is
even in orbital angular momentum ($\textbf{p}$-even, s-wave) and
odd in frequency ($T$-odd) \cite{youmans-18}, can form at the boundary of
the material due to the presence of magnetic ordering and
translation symmetry breaking; this superconductivity can be
accompanied by the formation of MMs \cite{balatsky-92, coleman-93, asano-13}.

\subsection{\label{sec.4.2}Concepts of the coexistence phase of superconductivity
	and noncollinear spin ordering}

In this subsection, we consider the formation of MMs in
materials with a triangular lattice when the SC+NCM phase is realized in them. To be specific, we discuss the compound Na$_x$CoO$_2 \cdot y$H$_2$O,  the minimal model for which is the $t-J-V$ model \cite{baskaran-03, VVV-VTA-MVA-15}.
In this model, the electron states
correspond to the upper Hubbard subband. In view of this,
the Hamiltonian in the atomic representation has the form
\begin{eqnarray}
\label{Ham}
H_{tJV} &=& \sum_{f \sigma} \left( \varepsilon - \mu \right) X_f^{\sigma \sigma} + \sum_{f} \left( 2\varepsilon + U - 2\mu \right) X_f^{2 2} +
\nonumber \\
&+& \sum_{fm\sigma} t_{fm} X_f^{2 \bar{\sigma}} X_m^{\bar{\sigma} 2} + \frac{V}{2} \sum_{f \delta} n_f n_{f+\delta} +
\sum_{f m} J_{fm} \left( X_f^{\uparrow\downarrow} X_m^{\downarrow \uparrow} -
X_f^{\uparrow \uparrow} X_m^{\downarrow \downarrow} \right),
\nonumber \\
\end{eqnarray}
where $\varepsilon$ is the bare on-site energy of an electron, $\mu$ is the
chemical potential, $U$ is the on-site Hubbard repulsion, $t_{fm}$ is
the electron hopping rate, $n_f = X_f^{\uparrow \uparrow} + X_f^{\downarrow \downarrow} + 2X_f^{22}$ is the
operator of the number of electrons at a site, and $J_{fm}$ is the
exchange interaction parameter. The intersite Coulomb
interaction is taken into account by the term with the
parameter $V$. The Hubbard operators are defined standardly \cite{VVV-SGO-04}: $X_f^{n p} = \left| f, n \right\rangle \left \langle f, p \right|$, where $\left| f, n \right\rangle$ are the basis electron
states at a site $f$ and $n$ is a single-site state index: the value $n = \sigma$ corresponds to a state with one electron with the spin
projection $\sigma$, and $n = 2$, to a state with two electrons. We note
that Hamiltonian \eqref{Ham} is defined in the Hilbert subspace
from which states that do not contain electrons are excluded.
The Hubbard operators act on the basis states as $X_f^{n p} \left| m, q \right\rangle = \delta_{fm} \delta_{pq}  \left| f, n \right\rangle$, where $\delta_{fm}$ and $\delta_{pq}$ are Kronecker
symbols.

As noted in the Introduction and in Section \ref{sec2.2}, the
possibility of realizing edge states is closely related to a
nontrivial value of the TI for the energy structure of the
material under periodic boundary conditions. Such a TI can
be calculated most easily in the case where the system under
consideration is described by a quadratic form in secondary-quantization operators. Unfortunately, this approach is
ineffective for systems with strong electron correlations.

A more general method for finding the TI is to calculate it
using the Green's functions. In this case, the most convenient
approach is based on the use of the TI~$\tilde{N}_3$.
Let us discuss this in more detail.

The TI $\tilde{N}_3$ was first proposed in \cite{ishikawa-87} in the study of
conductance in the quantum Hall effect and was subsequently
used for various phases of superfluid $^{3}$He \cite{volovik-09, volovik-09b}.
According to \cite{ishikawa-87, volovik-09}, the $\tilde{N}_3$
invariant is applicable to spatially 2D systems
(so-called $(2+1)$-dimensional systems, where the additional
dimension is time), in which the excitation spectrum has a gap
in the entire Brillouin zone. It is essential that this invariant
also preserves its form for interacting systems.

The TI $\tilde{N}_3$ can be calculated based on its relation to the
Green's functions and their derivatives,
\begin{eqnarray}
\label{Z_inv} \tilde{N}_{3} =
\frac{\varepsilon_{\mu\nu\lambda}}{24\pi^{2}}
\int\limits_{-\infty}^{\infty}d\omega\int\limits_{-\pi}^{\pi}
dp_{1}dp_{2}
\text{Sp}\left(\widehat{G}\partial_{\mu}\widehat{G}^{-1}
\widehat{G}\partial_{\nu}\widehat{G}^{-1}
\widehat{G}\partial_{\lambda}\widehat{G}^{-1} \right), \nonumber
\end{eqnarray}
where $\varepsilon_{\mu\nu\lambda}$ is the Levi-Civita symbol, $\partial_{1} \equiv \partial/\partial p_1$,
$\partial_{2} \equiv \partial/\partial p_2$, and $\partial_{3} \equiv
\partial/\partial \omega$. The notation $\widehat{G}(p, i\omega)$ is introduced for a
matrix composed of the propagator parts of the Fermi
normal and anomalous Green's functions in the SC+NCM
phase:
\begin{eqnarray}
\label{Def_G}
&& \widehat{G} =
\nonumber \\
&& \left[
\begin{array}{cccc}
G_{\downarrow 2, \downarrow 2} (p, i\omega) & G_{\downarrow 2, 2 \uparrow} (p, i\omega) & G_{\downarrow 2,\uparrow 2} (p, p-Q; i\omega) & G_{\downarrow 2, 2 \downarrow} (p, p-Q; i\omega) \\
G_{2 \uparrow, \downarrow 2} (p, i\omega) & G_{2 \uparrow, 2 \uparrow} (p, i\omega) & G_{2 \uparrow, \uparrow 2} (p, p-Q; i\omega) & G_{2 \uparrow, 2 \downarrow} (p, p-Q; i\omega) \\
G_{\uparrow 2,\downarrow 2} (p-Q, p; i\omega) & G_{\uparrow 2, 2 \uparrow} (p-Q, p; i\omega) & G_{\uparrow 2,\uparrow 2} (p-Q, i\omega) & G_{\uparrow 2, 2 \downarrow} (p-Q, i\omega) \\
G_{2 \downarrow, \downarrow 2} (p-Q, p; i\omega) & G_{2 \downarrow, 2 \uparrow} (p-Q, p; i\omega) & G_{2 \downarrow, \uparrow 2} (p-Q, i\omega) & G_{2 \downarrow, 2 \downarrow} (p-Q, i\omega)\\
\end{array}
\right].
\nonumber \\
\end{eqnarray}

As is known, the full matrix Green's function in the
atomic representation can be expressed as the product $\widehat{D} = \widehat{G} \cdot \widehat{P}$, where $\widehat{P}$ is the force operator matrix. The matrix $\widehat{D}$ is then made of Fourier transforms of the Green's functions in
the Matsubara representation,
\begin{equation}
\label{FGR_perp}
D_{\nu, \mu} \left(f \tau; f' \tau'\right) = - \left\langle T_{\tau} \tilde{X}_{f}^{\nu}\left(\tau\right) \tilde{X}_{f'}^{-\mu}\left(\tau'\right) \right\rangle,
\end{equation}
where $\nu$, and $\mu$ --- are indices indicating pairs of single-site states.
To calculate the TI $ \tilde{N}_{3}$, we have to move from the discrete
Matsubara frequencies $i\omega_n$ to the continuous frequency $i\omega$.

For noninteracting systems, the Green's functions can be
calculated exactly. But for strongly correlated systems, these
functions can only be found in some approximation. In
describing the Cooper instability in materials with noncollinear spin ordering, the loopless approximation is often used.
In that case, the vanishing of the determinant of the inverse
matrix $\widehat{G}^{-1}$
yields an expression for the spectrum of
elementary Fermi excitations of a system in the SC+NCM
phase,
\begin{eqnarray}
\label{spectr}
\varepsilon_{1,2 \, p} = \left[ \frac{1}{2} \left( \xi_{p}^2 + \xi_{p-Q}^2 + |\Delta_p|^2 + |\Delta_{-p+Q}|^2  \right) + R_{p}R_{p-Q}  \mp \lambda_{p} \right]^{1/2},
\end{eqnarray}
where
\begin{eqnarray}
\lambda_{p} = \left\{ \frac{1}{4} \left( \xi_{p}^2 - \xi_{p-Q}^2 + |\Delta_p|^2 - |\Delta_{-p+Q}|^2 \right)^{2} + R_{p}R_{p-Q} \left[ \left( \xi_{p} + \xi_{p-Q} \right)^2 + |\Delta_p+\Delta_{-p+Q}|^2 \right] \right\}^{1/2}.
\nonumber
\end{eqnarray}
In \eqref{spectr}, we use the following notations:
\begin{equation}
\xi_{p} = \varepsilon + U
-\mu+J_0(1-n/2)+V_0n+nt_{p}/2,
\nonumber
\end{equation}
$n = \left\langle
n_f \right\rangle$ is the on-site concentration of electrons, $J_0$ and $J_{Q}$ are
the values of the Fourier transform of the exchange integral
for the magnetic structure vectors $(0, 0)$ and ${\bf Q}$, $V_0
= 6V$, $R_{p} = M(t_{p} - J_{Q})$, $R_{p - Q} = M(t_{p - Q} -
J_{Q})$, $M$ is the amplitude
of the inhomogeneous magnetic order parameter that defines
the spin structure,
$\left\langle {\bf S}_f \right\rangle = M \left( \cos({\bf Q f}),
-\sin({\bf Q f}), 0 \right)$, and $\Delta_p$ is
the superconducting order parameter.

In most cases, MMs are studied within a simplified
scheme, where the essential parameters of the model vary
arbitrarily. There are a small number of studies in which the
proximity effect in topologically nontrivial heterostructures
or the order parameters are calculated self-consistently when
the nontrivial topology is caused by internal interactions (see,
e.g.,~\cite{stanescu-11, reeg-17, kopasov-18, VVV-ZAO-19, theiler-19}).

In contrast to these simplified approaches, we discuss the
results of solving the problem of finding topologically
nontrivial regions and topological transitions in the SC+NCM phase in the framework of the $t-J-V$ model,
with the order parameters determined from the system of
integral self-consistency equations.

It is essential for what follows that, near the filling $n\simeq1$ the 120-degree spin ordering (${\bf Q} = (2\pi/3, 2\pi/3)$) is more
favorable \cite{chubukov-94, chubukov-94b, jiang-14, pasrija-16}. In the Heisenberg regime, $n=1$, this order is stable up to $J_2 \approx J_1/10$ \cite{barabanov-92}.
It turns out
that, in the SC+NCM phase, the superconducting order
parameter symmetry correlates with the spin ordering structure. For example, for the stripe structure proposed in \cite{lu-13} for $J_2/J_1 \to \infty$, the chiral symmetry of the superconducting
order parameter becomes disadvantageous in comparison to
the modified $d_{x^2-y^2}$ symmetry or the $d_{xy}$ symmetry \cite{VVV-ZAO-16}.
At
the same time, the 120-degree magnetic order preserves the
chiral symmetry type of superconductivity in the SC+NCM
phase. Importantly, in the $d_{x^2-y^2}+id_{xy}$ superconducting
phase without magnetic ordering, topologically protected
edge states \cite{volovik-97} are realized that are not Majorana states,
whereas no such states exist in the superconducting $d_{x^2-y^2}$ or $d_{xy}$ phase. Therefore, it seems relevant to search for MMs in
the phase of coexistence of superconductivity and 120-degree
magnetic ordering (SC+120).

To simplify the calculation, we here consider only the
exchange interaction between Hubbard fermions at the
nearest-neighbor sites. As a result, the quasimomentum
dependence of the superconducting order parameter in the
SC+120 phase takes the form
\begin{equation}\Delta_p = 2\Delta \phi_{21}(p), \, \, \, \phi_{21}(p) = \cos(p_2) + e^{i2\pi/3}\cos(p_1) + e^{i4\pi/3}\cos(p_1+p_2).
\end{equation}
A more general case with the exchange interaction in the
second coordination sphere and the intersite Coulomb
repulsion taken into account was considered in \cite{VVV-ZAO-17}.

From the Green's functions found in considering periodic
boundary conditions, self-consistency equations follow for
the magnetic order parameter $M$ and the amplitude $\Delta$ of the
superconducting order parameter \cite{VVV-ZAO-17, VVV-ZAO-19}. The equation for $M$ has the compact form
\begin{eqnarray}
\label{eq_M} && M = \frac{n}{2} + \frac{1}{2} - \sum_q \frac{A_q/2 -
	J_Q}{2\gamma_q} -
\frac{1}{2} \sum_p \left( f_{1p} + f_{2p} \right)
\nonumber \\
&-& M \sum_{p} \frac{J_Q-\left( t_p + t_{p-Q} \right)/2}{\varepsilon^{(0)}_{2p} - \varepsilon^{(0)}_{1p}} \left( f_{1p} - f_{2p} \right),
\nonumber \\
\end{eqnarray}
where $A_{q} = J_{q} + \left( J_{q-Q} + J_{q + Q}
\right)/2$, $J_q$ is, as previously, the
Fourier transform of the exchange integral, and $\gamma_{q}$ is related
to the bare spectrum of spin-wave excitations as
\begin{equation}
\omega_{0 q} = 2M \gamma_{q} = 2M\sqrt{\left( J_{q} - J_Q \right)
	\left[ \frac{J_{q - Q} + J_{q + Q}}{2} - J_Q \right]},
\end{equation}
where $f_{jp} \equiv f(\varepsilon^{(0)}_{jp}/T)$ is the Fermi-Dirac function. In
deriving this equation, the effect of superconducting pairing
on spin ordering was neglected.

The use of this approximation is justified by the following
argument. First, both types of long-range order are induced
by the same exchange interaction. As a result, magnetic
ordering forms at temperatures $T_{\text{N}} \sim J$, often exceeding the Cooper instability temperature $T_{\text{c}}$. Second, we see in what
follows that the SC+120 phase is realized mainly in the
underdoped region of the phase diagram, far from the
optimal doping, where the condition $T_{\text{N}} \gg T_{\text{c}}$ is certainly
satisfied. To describe quantum topological transitions, it
suffices to restrict ourself to the low-temperature limit, and
therefore Eqn \eqref{eq_M} is then written with $T \ll T_{\text{N}}$. As a result,
the branches of Fermi spectrum \eqref{spectr}, without corrections
coming from the superconducting order parameter, are determined by expressions for the spectrum in the normal magnetic
phase with 120-degree ordering:
\begin{eqnarray} \varepsilon^{(0)}_{1,2p} = \frac{\xi_p + \xi_{p-Q}}{2} \mp  \sqrt{\frac{\xi_p - \xi_{p-Q}}{2} + R_{p} R_{p - Q}}.
\end{eqnarray}

The self-consistency equation for $\Delta$ can be expressed as:
\begin{eqnarray}
\label{eq_Del}
&& 1 - \frac{J_1}{N}\sum_{qj}\frac{(-1)^{j} \phi_{21}(q)\text{th}\left( \varepsilon_{jq}/2T\right)}{\varepsilon_{jq} \lambda_{q}} \times
\nonumber \\
& \times & \left\{ \cos(q_2) \left[ \varepsilon_{jq}^2 - \xi_{q-Q}^2 - \left| \Delta(-q+Q) \right|^2 \right] + \cos(q_2-Q_2)R_{q}R_{q-Q} \right\} = 0,
\end{eqnarray}
where $j = 1, 2$.

\begin{figure}[ht]
	\begin{center}
		\includegraphics[width=0.78\textwidth]{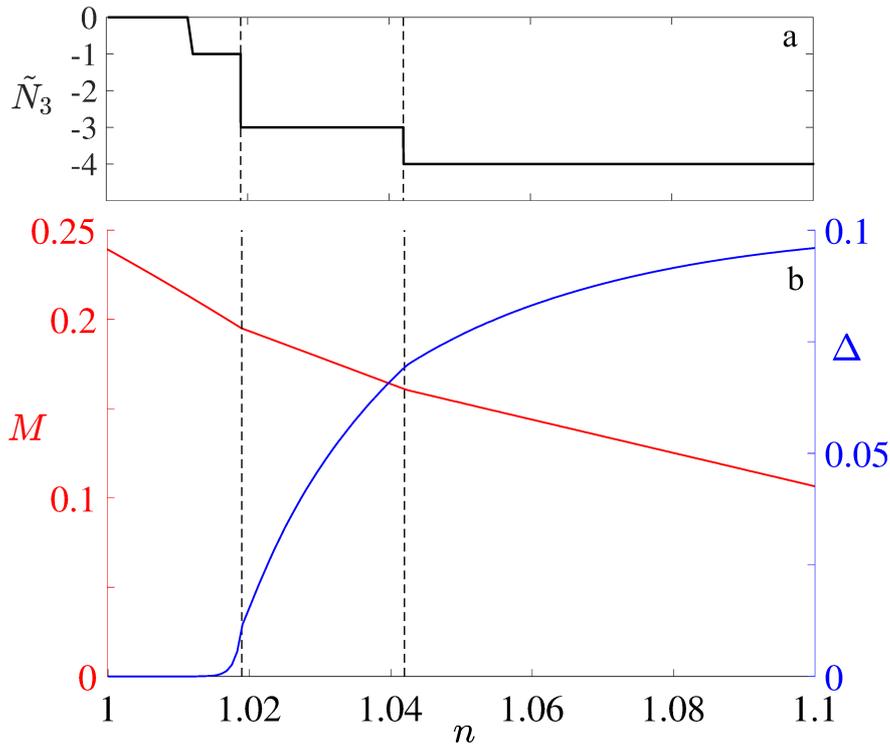}
		\caption{\label{fig_N3_po_n}
			(a) Dependence of topological index $\tilde{N}_3$ on
			electron concentration $n$ for exchange and Coulomb interaction parameters on neighboring sites $J_1 = 0.5 t_1$, $V = 0$. (b) Concentration
			dependence of magnetic and superconducting order parameter amplitudes ($M$ and $\Delta$) at the same parameter values.}
	\end{center}
\end{figure}

In Fig.~\ref{fig_N3_po_n}b, we show the concentration dependences of the
order parameters obtained from the solution of Eqns \eqref{eq_M} and \eqref{eq_Del}, including in the SC+120 phase, for the nearest-neighbor exchange parameter $J_1 = 0.5 t_1$. The dependence of
the TI $\tilde{N}_3$ on the electron concentration corresponding to
these parameters is shown in Fig.~\ref{fig_N3_po_n}a. We can see that a series
of quantum topological transitions with a change in the value
of $\tilde{N}_3$ is realized in the system as the electron concentration
increases. Regions with a nonzero $\tilde{N}_3$ correspond to
topologically nontrivial phases. An even value of $\tilde{N}_3$ then
indicates the possible realization of edge states in a system
with open boundary conditions, and an odd value indicates
the possibility of forming an MBS with zero excitation energy \cite{ghosh-10, VVV-ZAO-ShMS-18}.

In the considered strongly correlated 2D system with a
triangular lattice, as $n$ increases, the onset of the phase of
coexistence of superconductivity and 120-degree spin ordering is accompanied by a transition from the topologically
trivial phase to a topologically nontrivial phase. With a
further increase in the electron concentration, two more
topological quantum transitions occur, separating topologically nontrivial phases with different values of $\tilde{N}_3$. The critical
electron concentrations of such transitions are indicated by
dashed vertical lines.

At the critical electron concentrations, the energy spectrum of the system in the SC+120 phase with periodic
boundary conditions becomes gapless. In this case, for the
first and third transitions, the gap closes at the $K^{\prime}$ point of the
hexagonal Brillouin zone, and for the second, at points $K$ and $\Gamma$. As can be seen from Fig.~\ref{fig_N3_po_n}, the dependence of the
magnetization $M$ on the electron concentration has kinks
at the topological transition points, which is due to the
rearrangement of the electron structure and the realization
of gapless excitations in the SC+120 phase \cite{VVV-ZAO-19}. Despite the
small magnitude of the effects and the difficulties in experimentally detecting them, they are of fundamental interest,
indicating the possibility in principle of realizing a series of
topological quantum transitions in superconductors with
noncollinear spin order.

We note that, in the considered approach, an increase in
the intersite Coulomb repulsion leaves the characteristics of
magnetic ordering unchanged, but strongly suppresses superconductivity. As a result, in particular, Cooper instability
arises at higher values of the electron concentration when the
Coulomb interaction is taken into account. The number of
topological transitions can then decrease \cite{VVV-ZAO-19}. Another
important effect of intersite Coulomb repulsion in the
SC+120 phase is the mixing of triplet superconducting
pairing potentials \cite{VVV-ZAO-17} into the kernel of the integral self-consistency equation for the superconducting order parameter.

We also note that the possibility of realizing the SC+120
phase was investigated previously in the $t-J$ model by the
variational Monte Carlo method \cite{weber-06}. Our method of considering a system with strong electron correlations leads to
similar results at a qualitative level. The formation of MMs in
the SC+120 phase for a quadratic Hamiltonian without
interaction is described in \cite{VVV-ZAO-FAD-ShMS-17, VVV-ZAO-ShMS-18}.

\subsection{Majorana modes in the coexistence phase
	of superconductivity and noncollinear magnetism}

In seeking the MMs, we assume that the sites of a triangular
lattice occupy $N_1$ positions along the direction of the
translation vector ${\bf a}_1$ and correspond to an open geometry,
while the periodic boundary conditions are realized along the
direction ${\bf a}_2$ (cylinder geometry).

For this geometry, the closed system of equations for the
Green's functions can be represented in the block form
\begin{eqnarray}
\label{Ham_kQ}
&&\left( {\begin{array}{*{20}{c}}
	{{i \omega_n \hat{I}_{N_1} - { \hat{\xi} }_{k_2}}}&{\hat{h}_{k_2-Q_2}(Q)}&\hat{0}_{N_1}&{{{ \hat{D} }_{k_2}}}\\
	{{\hat{h}_{k_2}(-Q)}}&{{i \omega_n \hat{I}_{N_1} - { \hat{\xi}}_{k_2 - Q_2}}}&{ - \hat{D}_{ k_2-Q_2}}&\hat{0}_{N_1}\\
	\hat{0}_{N_1}&{ -  \hat{D}_{k_2 - Q_2}^{\dag}}&{ i \omega_n \hat{I}_{N_1} + \hat{\xi}_{ k_2 - Q_2}}&{ -\hat{h}_{k_2}(-Q)}\\
	{ \hat{D}_{k_2}^{\dag} }&\hat{0}_{N_1}&{-\hat{h}_{k_2-Q_2}(Q)}&{i \omega_n \hat{I}_{N_1} + \hat{\xi}_{k_2}}
	\end{array}} \right) \cdot \nonumber \\
&& \cdot \left[ {\begin{array}{*{20}{c}}
	\hat{G}_{\downarrow2,\downarrow2}(k_2,k_2;l^{\prime};i\omega_n)\\
	\hat{G}_{\uparrow2,\downarrow2}(k_2-Q_2,k_2;l^{\prime};i\omega_n)\\
	\hat{G}_{2\downarrow,\downarrow2}(k_2-Q_2,k_2;l^{\prime};i\omega_n)\\
	\hat{G}_{2\uparrow,\downarrow2}(k_2,k_2;l^{\prime};i\omega_n)
	\end{array}} \right]  = \left[ {\begin{array}{*{20}{c}}
	\hat{\delta}_{l^{\prime}}\\
	\hat{0}\\
	\hat{0}\\
	\hat{0}
	\end{array}} \right],
\end{eqnarray}
where $\hat{I}_{N_1}$ and $\hat{0}_{N_1}$ are the unit and zero $N_{1} \times N_{1}$ matrices. The
matrices $\hat{\xi}_{k_2}$, $\hat{D}_{k_2}$, and $\hat{h}_{k_2}(Q)$ are the same size. In the
approximation where the operator averages are independent
of the site number and are calculated within the problem with
periodic boundary conditions, the expressions for $\hat{\xi}_{k_2}$, $\hat{D}_{k_2}$, and $\hat{h}_{k_2}(Q)$ take the form
\begin{eqnarray}
\label{Ham_Matrices}
&& \hat{\xi}_{k_{2}} = \left( {\begin{array}{*{20}{c}}
	{{\xi_0 + F t_{k_{2}}}}&{F T_{k_{2}}}&0&0\\
	{F T_{-k_{2}}^{}}& \ddots & \ddots &0\\
	0& \ddots & \ddots &{F T_{k_{2}}}\\
	0&0&{F T_{-k_{2}}^{}}&{\xi_0 + F t_{k_{2}}}
	\end{array}} \right),
\nonumber \\
&& \hat{D}_{k_{2}} = - \left( {\begin{array}{*{20}{c}}
	{\Delta_{k_{2}}^*}&{\psi_{ - {k_{2}}}^{*}}&0&0\\
	{\psi _{k_{2}}^*}& \ddots & \ddots &0\\
	0& \ddots & \ddots &{\psi _{ - {k_{2}}}^*}\\
	0&0&{\psi _{k_{2}}^{*}}&{\Delta _{k_{2}}^{*}}
	\end{array}} \right),
\\ \nonumber
&& \hat{h}_{k_2}(Q) = -M \times
\nonumber \\
&&\times \left( {\begin{array}{*{20}{c}}
	{{(t_{k_2} - J_Q)e^{iQ_1}}}&{T_{k_{2}}e^{iQ_1}}&0&0\\
	{T_{-k_{2}}e^{i2Q_1}}& \ddots & \ddots &0\\
	0& \ddots & \ddots &{T_{k_{2}}e^{i(N_1-1)Q_1}}\\
	0&0&{T_{-k_{2}}e^{iN_1Q_1}}&{(t_{k_2} - J_Q)e^{iN_1Q_1}}
	\end{array}} \right).
\nonumber
\end{eqnarray}
For the Hubbard renormalization in the noncollinear
ordering phase, we introduce the notation $F = \left\langle n_{ll_2} \right\rangle/2$ and $M = \left\langle S_{ll_2}^{+} \right\rangle e^{i(Q_1 l+Q_2 l_2)}$, where $\left\langle n_{ll_2} \right\rangle$ is the on-site electron concentration. The coordinate of a lattice site
is here given in the two-parameter form: $l$ for the
coordinate along the ${\bf a}_1$ direction and $l_2$ along the ${\bf a}_2$ direction. The additional notation is $t_{k_{2}} = 2t_1\cos(k_{2})$, $\Delta_{k_{2}} = 2\Delta\cos(k_{2})$, ${T_{k_{2}}} = {t_1}\left( 1 + \exp(ik_{2}) \right)$, and
\begin{eqnarray}
\psi_{k_{2}} & = & \Delta \exp(i2\pi/3)\left(1+\exp(i2\pi/3+ik_{2}) \right).
\nonumber \\
\end{eqnarray}

We write $\hat{G}_{\nu,\mu}(k_2,k_2;l^{\prime};i\omega_n)$ for the column vector whose $l$th element is $G_{\nu,\mu}(k_2,k_2;l,l^{\prime};i\omega_n)$, where $l$ is the site number
along the ${\bf a}_1$ direction for the first Hubbard operator in
definition \eqref{FGR_perp} and $l^{\prime}$ is the site number along the same direction for the second Hubbard operator. The quasimomentum dependence on the $k_2$ component appears after
taking the periodic boundary conditions along ${\bf a}_2$ into
account. The nonzero element on the right-hand side of
system of equations \eqref{Ham_Matrices} can be represented in terms of the
Kronecker symbols: $\hat{\delta}_{l'} = F \left[ \delta_{1l^{\prime}}, \delta_{2l^{\prime}}, \dots \delta_{N_1 l^{\prime}} \right]^{T}$.

We write system of equations \eqref{Ham_kQ} in the compact form
\begin{eqnarray}
\left[ i \omega_n \hat{I}_{4N_1} - \hat{M} \right] \hat{V}_G = \hat{C}.
\end{eqnarray}
and multiply this relation from the left with the matrix $\widehat{S}^{\dag}$, such that $\widehat{S}^{\dag} \hat{M} \widehat{S} = \hat{M}_D$, where $\hat{M}_D$ is the diagonal matrix
composed of the positive excitation energies $\varepsilon_{j k_2}$ ($j = 1, 2, \dots, 2N_1$) and the corresponding negative energies. For
definiteness, we assume that the elements on the diagonal of $\hat{M}_D$ are ordered as follows: the first $2N_1$ elements are the
energies $\varepsilon_{j k_2}$ taken in ascending order, and the other $2N_1$ elements are the energies $-\varepsilon_{j k_2}$ taken in descending order. As a
result, we obtain the expression
\begin{eqnarray}
\label{Sys_comp}
\left[ i \omega_n \hat{I}_{4N_1} - \hat{M}_D \right] \widehat{S}^{\dag} \hat{V}_G = \widehat{S}^{\dag}\hat{C}.
\end{eqnarray}
We introduce new `quasiparticle' Green's functions:
\begin{eqnarray}
\label{Gab}
\left[ {\begin{array}{*{20}{c}}
	\hat{G}_{\alpha,\downarrow2}(k_2;l^{\prime};i\omega_n)\\
	\hat{G}_{\beta,\downarrow2}(k_2;l^{\prime};i\omega_n)\\
	\hat{G}_{\alpha^{\dag},\downarrow2}(k_2;l^{\prime};i\omega_n)\\
	\hat{G}_{\beta^{\dag},\downarrow2}(k_2;l^{\prime};i\omega_n)
	\end{array}} \right] = \widehat{S}^{\dag}
\left[ {\begin{array}{*{20}{c}}
	\hat{G}_{\downarrow2,\downarrow2}(k_2,k_2;l^{\prime};i\omega_n)\\
	\hat{G}_{\uparrow2,\downarrow2}(k_2-Q_2,k_2;l^{\prime};i\omega_n)\\
	\hat{G}_{2\downarrow,\downarrow2}(k_2-Q_2,k_2;l^{\prime};i\omega_n)\\
	\hat{G}_{2\uparrow,\downarrow2}(k_2,k_2;l^{\prime};i\omega_n)
	\end{array}} \right].
\end{eqnarray}
It is then easy to find the Green's function of low-energy
quasiparticles with the $j$th energy (here $j = 1, 2, \dots, N_1$):
\begin{eqnarray}
\label{GRFM}
G_{\alpha_j, \downarrow 2} \left( k_2; l^{\prime}; i\omega_n \right) =
F \frac{(\widehat{S}^{\dag})_{jl^{\prime}}}{i \omega_n - \varepsilon_{j k_2} }.
\end{eqnarray}

\begin{figure}[ht]
	\begin{center}
		\includegraphics[width=0.78\textwidth]{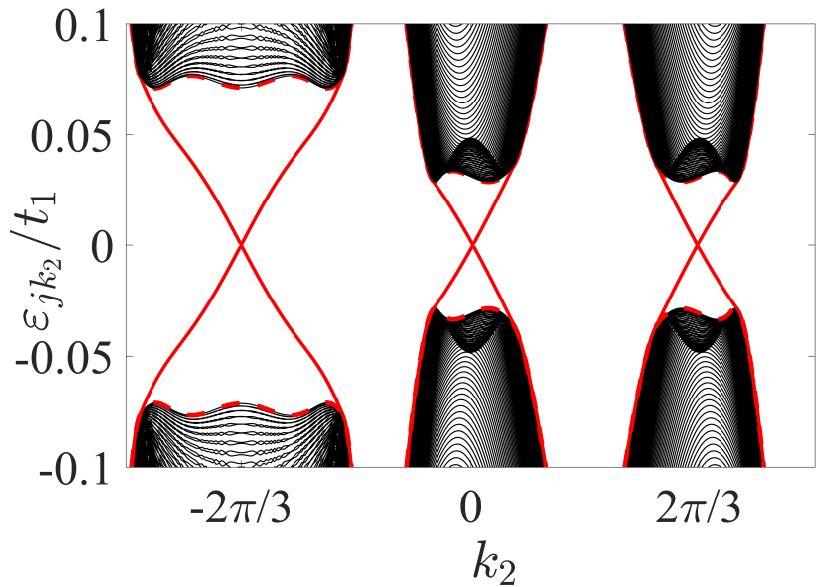}
		\caption{\label{exc_maj_sc_nco}
			Spectrum of Fermi excitations in the phase of
			coexistence of superconductivity and noncollinear magnetism at electron
			concentration $n = 1.025$ corresponding to the TI value $\tilde{N}_3 = -3$, for
			cylinder geometry.}
	\end{center}
\end{figure}

Relation \eqref{Gab} between the Green's function found in (\ref{GRFM}) and the original Green's functions allows finding the
Fermi operators of elementary excitations for the SC+120
phase as linear combinations of the Hubbard operators,
\begin{eqnarray}
\label{DEFUV} \alpha_{j k_2} & = & \sum_{l=1}^{N_1} u_{jl}
X_{k_2,l,\uparrow} + w_{jl} X_{k_2-Q_2,l,\downarrow}
+ z_{jl} X^{\dag}_{-k_2+Q_2,l,\uparrow} + v_{jl}
X^{\dag}_{-k_2,l,\downarrow},
\end{eqnarray}
where the coefficients $u_{jl}$, $w_{jl}$, $z_{jl}$ and $v_{jl}$ are elements of the $j$th row of the matrix $\widehat{S}^{\dag}$.
It follows from this definition and
symmetry considerations~\cite{VVV-ZAO-ShMS-18}
that MMs are realized in the
cylinder geometry at the momentum space point corresponding to the symmetry between particles and holes, the particle-hole invariant momentum point $K_2 = -K_2+Q_2+G$, where $Q_2 = 2\pi/3$ and $G$ is the reciprocal lattice vector, i.e., exactly at
the point $K_2 = -2\pi/3$, for which, as can be seen from Fig.~\ref{exc_maj_sc_nco} the excitation energy is zero.

The dependence of the excitation spectrum of the system
on the quasimomentum $k_2$ at the electron concentration
$n = 1.025$ and $V = 0$ is plotted in Fig. \ref{exc_maj_sc_nco}.
The dashed line
shows the boundaries of the bulk excitation spectrum under
periodic boundary conditions along both directions of the
triangular lattice. We can see that, for the chosen values of the
parameters, a superconducting gap is realized in the excitation spectrum. For the cylinder geometry, edge states appear inside the spectrum gap (thick solid line). Thin lines represent
the excitation branches lying in the bulk spectrum.

\begin{figure}[ht]
	\begin{center}
		\includegraphics[width=0.78\textwidth]{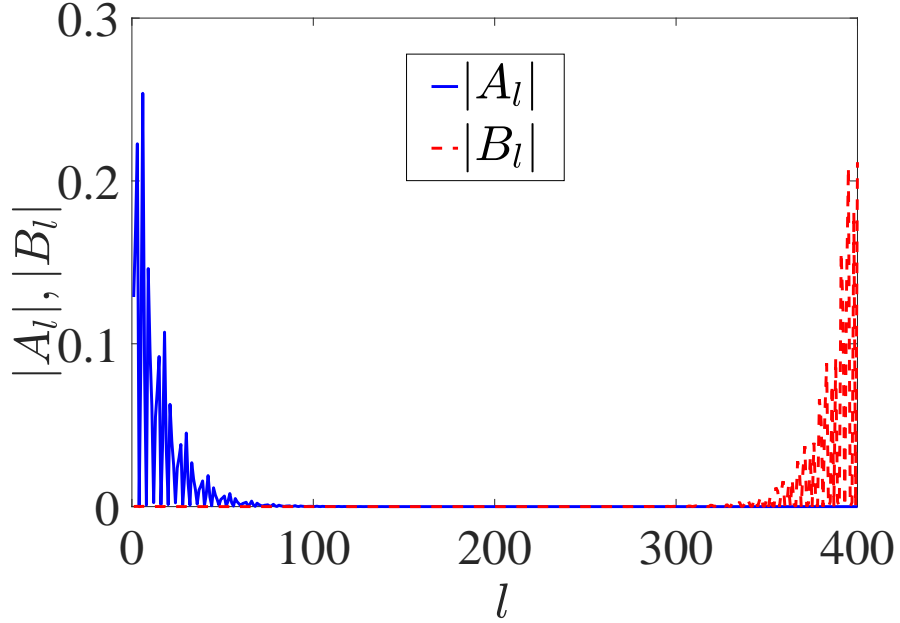}
		\caption{\label{str_maj_sc_nco}
			Spatial structure of MMs at the $K_2 = - 2\pi/3$ point in the superconductivity and magnetism coexistence phase with TI value $\tilde{N}_3 = -3$. Parameters are the same as in Fig. \ref{exc_maj_sc_nco}.}
	\end{center}
\end{figure}

To show the spatial structure of MMs for $K_2 = - 2\pi/3$,  we
use the approach described in \cite{kitaev-01}.
For this, we introduce two
self-adjoint operators, $b_0^{\prime} =
\alpha_{1} + \alpha_1^{\dag}$ and $b_0^{\prime \prime} =
i(\alpha_1^{\dag}-\alpha_1)$. Using expansion
(\ref{DEFUV}), we then express these operators in
terms of Majorana operators in the atomic representation:
\begin{eqnarray}
\gamma_{A l \sigma} = X_{l \sigma} + X^{\dag}_{l \sigma},~~~
\gamma_{B l \sigma} = i\left(X^{\dag}_{l \sigma}-  X_{l
	\sigma}\right).
\end{eqnarray}
As a result, we find
\begin{eqnarray}
\label{DEF_b_prim} b_0^{\prime}=\sum_{l=1}^{N_1}
\left\{{\rm{Re}}(u_{1l}+z_{1l})\gamma_{A l \uparrow}
+{\rm{Re}}(w_{1l}+v_{1l})\gamma_{A l \downarrow}\right\}-
\nonumber \\
\sum_{l=1}^{N_1} \left\{{\rm{Im}}(u_{1l}-z_{1l})\gamma_{B l \uparrow}
+{\rm{Im}}(w_{1l}-v_{1l})\gamma_{B l \downarrow}\right\},
\nonumber
\end{eqnarray}
\begin{eqnarray}
\label{DEF_b_pr_pr} b_0^{\prime\prime}=\sum_{l=1}^{N_1}
\left\{{\rm{Im}}(u_{1l}+z_{1l})\gamma_{A l \uparrow}
+{\rm{Im}}(w_{1l}+v_{1l})\gamma_{A l \downarrow}\right\}+
\nonumber \\
\sum_{l=1}^{N_1} \left\{{\rm{Re}}(u_{1l}-z_{1l})\gamma_{B l \uparrow}
+{\rm{Re}}(w_{1l}-v_{1l})\gamma_{B l \downarrow}\right\}.
\nonumber \\
\end{eqnarray}

In Fig. \ref{str_maj_sc_nco} we show the site-number dependence of the
coefficients $A_l=\rm{Re}(u_{1l}+z_{1l})$ and
$B_l=\rm{Im}(u_{1l}+z_{1l})$ appearing in the decompositions of the operators $b_0^{\prime}$ and
$b_0^{\prime \prime}$ in terms of Majorana operators $\gamma_{A l \sigma}$ and $\gamma_{B l \sigma}$ in the atomic
representation. We can see that the dependence of these
coefficients on the site number exhibits localization near the
opposite edges. Other decomposition coefficients have a
similar localization, and we do not show them in Fig. \ref{str_maj_sc_nco}.

\begin{figure}[ht]
	\begin{center}
		\includegraphics[width=0.78\textwidth]{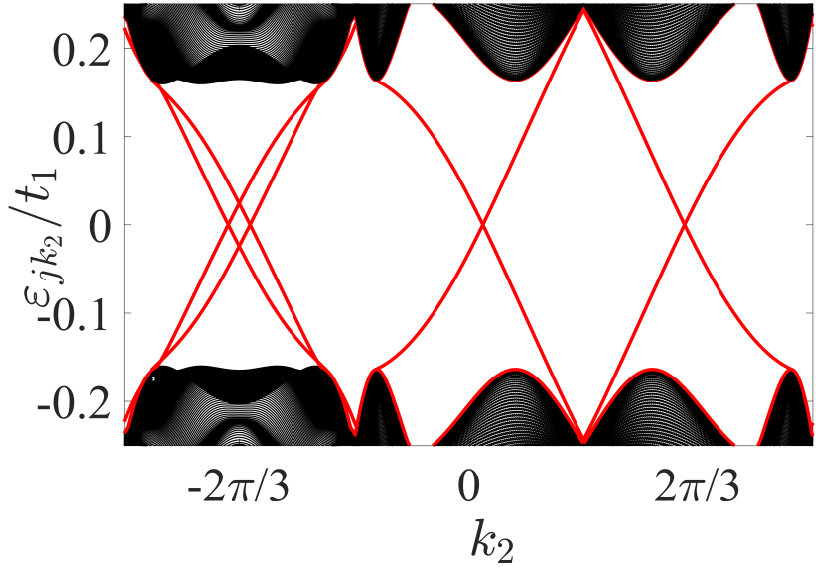}
		\caption{\label{exc_maj_sc_nco2}
			Spectrum of Fermi excitations in the superconductivity and magnetism coexistence phase at electron concentration $n = 1.08$ corresponding to TI value $\tilde{N}_3 = -4$, for the cylinder geometry.}
	\end{center}
\end{figure}

In the region with the TI $\tilde{N}_3 = -4$, after the third
topological quantum transition, the excitation spectrum in
the cylinder geometry is shown in Fig. \ref{exc_maj_sc_nco2}. We can see from the
figure that, although excitation energies that are exponentially close to zero are realized for four values of $k_2$, there are
no MMs at the symmetric point $K_2 = - 2\pi/3$. The positions
of the four points with momenta $k_2$ and zero energy can
change, and therefore such zero modes are sensitive to the
parameters of the system and are not of primary interest. The
distribution of the coefficients $A_l$ and $B_l$ for these zero modes
is localized at the same edge, in contrast to that shown in
Fig. \ref{str_maj_sc_nco}.

In a narrow range of electron concentrations of the
SC+120 phase with the TI $\tilde{N}_3 = -1$, the formation of
MMs is also possible at $K_2 = -2\pi/3$. In this region, in
contrast to that in the region with $\tilde{N}_3 = -3$, a branch of
edge states forms only near $K_2 = -2\pi/3$. However, such a
topologically nontrivial phase is of little practical interest
due to the extremely small values of the superconducting
gap, even if the intersite Coulomb interaction $V$ is disregarded.

Thus, based on a self-consistent calculation of the order
parameters and a TI, we have demonstrated a series of
quantum topological transitions between regions with different nontrivial topologies in the phase of coexistence of
superconductivity and noncollinear 120-degree magnetic
ordering, as well as the formation of MMs in one of the
regions. It is essential that the Majorana states are induced
just by the noncollinear spin order in the presence of
superconductivity. The presented results expand the list of
probable candidates for the experimental detection of
Majorana fermion modes.

\section{Conclusions}

We have analyzed the origination and development of the
problem of realizing and identifying MMs in low-dimensional
condensed matter systems that allow the existence of superconducting phases with a nontrivial topology. Systems in
which the superconducting pairing potential arises as a result
of either the proximity effect or the Cooper instability
initiated by internal interactions are considered as objects
allowing the existence of MMs.

With the example of a Kitaev chain, we discussed the
principles of realization of elementary excitations corresponding to MMs. For clarity, we considered particular
cases where simple analytic expressions exist that reflect the
structure of such modes. A correspondence has been demonstrated between the conditions for the realization of MMs in
an open geometry of the system and those for the formation
of a topologically nontrivial state determined by the value of
the Zak-Berry phase for a closed chain. As we have seen, an
analysis of the properties of the Kitaev Hamiltonian in the
quasimomentum representation and its ground state yields a
simple method for determining the fermion parity (FP) of the
ground state for various parameters of the model and allows
identifying the regions of the realization of a topologically
nontrivial phase in the phase diagram.

We have shown that the hybridization of the MM wave
functions due to the effects of a finite chain size gives rise to a
cascade of quantum transitions in an open system accompanied by a change in the FP of the ground state. Quantum
critical points appear in the parameter region where a
topologically nontrivial phase is realized in a closed system.
Identifying these quantum transitions based on the analysis
of electrocaloric anomalies is important for experimental
applications.

The noted possibility of detecting quantum transitions is
available not only in the Kitaev chain but also in systems that
are more realistic from the practical standpoint. This has
been demonstrated for a semiconducting nanowire with the
Rashba spin-orbit coupling and induced superconductivity
(superconducting nanowire) placed in an external magnetic
field. Of significant importance is the result on the preservation of magneto- and electrocaloric anomalies when the
electron-electron interactions are taken into account. The
detection of caloric anomalies can therefore not only serve as
a test for the manifestation of quantum transitions but also
give grounds to assert the realization of states with MMs.

We have reviewed the experimental results related to the
realization of the conductance peak at zero bias in a finite
range of magnetic fields. Arguments have been scrutinized
that allow regarding the occurrence of this feature as evidence
of the realization of MMs in a superconductor-semiconducting nanowire hybrid structure. We have also discussed
alternative scenarios for the appearance of the conductance
peak, unrelated to a topological phase transition.

We have discussed the properties of MMs and the kinetic
and static characteristics of systems that are currently being
used to identify MBSs. We have shown that the modern
principles of detecting MMs are based on the study of the
nonlocality of such excitations and their spin polarization.
In addition, we discussed the features of the fluctuation
characteristics of the current (shot noise), transport in the
Coulomb blockade regime, and interference processes.

In the framework of the last area, we theoretically
analyzed the conducting characteristics of a device in the
form of an Aharonov-Bohm ring with the arms connected by
a superconducting wire (SW, bridge). The transport properties of such a device were discussed, with the interaction
between the low-energy states of the normal-phase wires that
make up the arm and the bridge taken into account.

In the linear response approximation at low temperatures,
we have shown that a number of symmetric and asymmetric
resonances (respective Breit-Wigner and Fano resonances)
arise in the conductance of the ring when the wire is
transferred by a magnetic field into a topologically nontrivial
phase. We analyzed the dependence of the properties of Fano
resonances on the spatial distribution of the low-energy SW
state (Majorana or Andreev types). It is important that the
type of this state can be effectively tested for a particular case
of the ring, a T-shaped transport scheme.

The predicted Fano effect and its features are of
considerable interest for establishing differences in the
transport properties of the Majorana and Andreev states.
The results obtained can contribute to the further development of experimental methods for detecting MBSs in
coherent quantum transport.

Superconducting systems with inhomogeneous magnetic
ordering were shown to be promising for the detection of
MMs. In such systems, a dense magnetic nanostructure is
applied on a superconductor, a superconductor is placed in
an inhomogeneous magnetic field, or a superconductor is
considered in the phase of coexistence with noncollinear
magnetic ordering. For such magnetic superconductors, it
has been shown that a nontrivial topology of the ground state
can be preserved even in the regime of strong electron
correlations, and MMs can arise. It has been shown that
topological quantum transitions can, in principle, affect the
characteristics of the magnetic ordering of such superconductors. The experimental detection of topologically nontrivial
surface states in such systems is an urgent issue for future
experiments.
We note that the effect of strong electron correlations on
topologically nontrivial phases can be significant and lead to
the induction of new topological transitions and even to a
change in the topological classification of these phases.

\textbf{Acknowledgments.} This review was prepared in the framework of the Russian Foundation for Basic Research (RFBR)
grant 19-12-50087. Part of
the presented results were obtained in the framework of
RFBR projects 19-02-00348, 20-32-70059, and 20-02-00015.
S V A expresses his gratitude for support through a grant
from the President of the Russian Federation, MK-1641.2020.2. A O Z and M S Sh express their gratitude to
the Theoretical Physics and Mathematics Advancement
Foundation ``BASIS''. M Yu K thanks the Program for
Basic Research of the National Research University Higher
School of Economics for support.

\end{document}